\documentclass[11pt]{article}
\usepackage{braket}
\usepackage{amsmath}
\usepackage{amsfonts}
\usepackage{amssymb}
\usepackage{amsthm}
\usepackage{cite}
\usepackage{color}
\usepackage{hyperref}
\usepackage{authblk}
\usepackage{graphicx}
\usepackage{subcaption}

\def\beq#1\eeq{\begin{align}#1\end{align}}
\def\beqa#1\eeqa{\begin{align}#1\end{align}}
\def\bes#1\ees{\begin{split}#1 \end{split}}

\newcommand{\ben}{\begin{enumerate}}  
\newcommand{\een}{\end{enumerate}}  
\newcommand{\timeso}{\overset{\mathrm{out}}{\times}}

\newlength{\dinwidth}
\newlength{\dinmargin}
\makeatletter
\newcommand\bibname{References}%
\def\bibfont{\fontsize{10}{11}\selectfont}
\newdimen\bibindent
\def\@biblabel#1{[#1]}

\makeatother

\renewcommand{\P}{\mathsf P}
\newcommand{\out}{\mathrm{out}}
\newcommand{\inc}{\mathrm{in}}
\newcommand{\m}{a}
\newcommand{\Omeps}{\Omega_{\veps}}

\newcommand{\mcK}{\mathcal K}
\newcommand{\mcF}{\mathcal F}
\newcommand{\B}{Z}
\newcommand{\aloc}{\mathrm{a-loc}}
\newcommand{\Dep}{\De_{\mfh}'}

\newcommand{\mcX}{\mathcal X}
\newcommand{\whGa}{\wh \Ga}
\newcommand{\mcT}{\mathcal T}
\newcommand{\dist}{\mathrm{dist}}
\newcommand{\diam}{\mathrm{diam}}
\newcommand{\fin}{\mathrm{fin}}
\newcommand{\loc}{\mathrm{loc}}
\renewcommand{\i}{\mathrm{i}}
\newcommand{\mbZ}{\mathbb{Z}}
\renewcommand{\xi}{a}

\newcommand{\veps}{\epsilon}

\newcommand{\mcL}{\mathcal L}

\newcommand{\Sp}{\mathrm{Sp}}

\newcommand{\ti}{\tilde}

\newcommand{\Om}{\Omega}

\newcommand{\Si}{\Sigma}
\newcommand{\La}{\Lambda}

\newcommand{\ka}{\kappa}

\newcommand{\wh}{\widehat}
\newcommand{\pa}{\partial}

\newcommand{\Ran}{\mathrm{Ran}}
\newcommand{\ov}{\overline}
\newcommand{\vp}{\varphi}
\newcommand{\mfh}{\mathfrak{h}}

\newcommand{\eps}{\varepsilon}
\newcommand{\de}{\delta}

\newcommand{\De}{\Delta}
\newcommand{\e}{\mathrm{e}}

\newcommand{\nin}{\noindent}
\newcommand{\si}{\sigma}

\newcommand{\h}{\fr{1}{2}}
\newcommand{\nat}{\mathbb{N}}
\newcommand{\hil}{\mathcal{H}}
\newcommand{\om}{\omega}
\newcommand{\mfa}{\mathfrak{A}}
\newcommand{\mco}{\mathcal{O}}

\newcommand{\supp}{\mathrm{supp}}
\newcommand{\fr}[2]{\frac{#1}{#2}}
\newcommand{\al}{\alpha}
\newcommand{\real}{\mathbb{R}}
\newcommand{\complex}{\mathbb{C}}
\newcommand{\la}{\lambda}
\newcommand{\non}{\nonumber}
\newcommand{\Ga}{\Gamma}

\newcommand{\lan}{\langle}
\newcommand{\ran}{\rangle}

\def\proof{\noindent{\bf Proof. }}
\def\qed{$\Box$\medskip}

\newtheorem{theoreme}{Theorem } [section]
\newtheorem{proposition}[theoreme]{Proposition}
\newtheorem{lemma}[theoreme]{Lemma}
\newtheorem{definition}[theoreme]{Definition}
\newtheorem{corollary}[theoreme]{Corollary}
\newtheorem{remark}[theoreme]{Remark}
\newtheorem{example}[theoreme]{Example}
\newtheorem{criterion}[theoreme]{Criterion}

\newcommand{\bex}{\begin{example}}
\newcommand{\eex}{\end{example}}
\newcommand{\ber}{\begin{remark}}
\newcommand{\eer}{\end{remark}}
\newcommand{\bec}{\begin{corollary}}
\newcommand{\eec}{\end{corollary}}
\newcommand{\bep}{\begin{proposition}}
\newcommand{\eep}{\end{proposition}}
\newcommand{\becr}{\begin{criterion}}
\newcommand{\eecr}{\end{criterion}}

\def\bel{\begin{lemma}}
\def\eel{\end{lemma}}
\def\bet{\begin{theoreme}}
\def\eet{\end{theoreme}}
\def\bed{\begin{definition}}
\def\eed{\end{definition}}

\newcommand{\caB}{{\mathcal B}}
\newcommand{\caC}{{\mathcal C}}

\newcommand{\caE}{{\mathcal E}}

\newcommand{\caG}{{\mathcal G}}
\newcommand{\caH}{{\mathcal H}}

\newcommand{\bbN}{{\mathbb N}}

\newcommand{\bbR}{{\mathbb R}}

\newcommand{\bbZ}{{\mathbb Z}}

\title{Lieb-Robinson bounds, Arveson spectrum and Haag-Ruelle scattering theory for gapped quantum spin systems}

\author{Sven Bachmann}
\affil{Mathematisches Institut der Universit{\"a}t M{\"u}nchen, Theresienstrasse 39, D-80333 M{\"u}nchen, Germany, E-mail: \rm{\texttt{sven.bachmann@math.lmu.de}}}

\author{Wojciech Dybalski}
\affil{Zentrum Mathematik, Technische Universit{\"a}t M{\"u}nchen, D-85747 Garching, Germany, E-mail:~\rm{\texttt{dybalski@ma.tum.de}}} 

\author{Pieter Naaijkens}
\affil{Institut f{\"u}r Theoretische Physik, Leibniz Universit{\"a}t Hannover, Appelstrasse 2, D-30167 Hannover, Germany, E-mail: \rm{\texttt{pieter.naaijkens@itp.uni-hannover.de}}}

\begin{document}

\maketitle

\begin{abstract}
We consider translation invariant gapped quantum spin systems satisfying the Lieb-Robinson bound and containing single-particle states in a ground state representation.
Following the Haag-Ruelle approach from relativistic quantum field theory, we construct  states describing collisions of several particles, and define the corresponding  $S$-matrix. We also obtain some general restrictions on the shape of the energy-momentum spectrum. For the purpose of our analysis we adapt the concepts of almost local observables and  energy-momentum transfer (or Arveson spectrum) from relativistic QFT to the lattice setting.  The Lieb-Robinson bound, which is the crucial substitute of strict locality from relativistic QFT,  underlies all our constructions. Our results hold, in particular, in the Ising model in strong transverse magnetic fields.
\end{abstract}

\section{Introduction}

Since many important physical experiments involve collision processes, our understanding of physics relies to a large extent on scattering theory.
The multitude of possible experimental situations is reflected by a host of theoretical descriptions of scattering processes,  see for example~\cite{Reed:1979aa}. One theoretical approach, due to Haag and Ruelle \cite{Ha58,Ru62}, proved to be particularly convenient in the context of quantum systems with infinitely many degrees of freedom. Originally developed in axiomatic relativistic Quantum Field Theories (QFT), Haag-Ruelle (HR) scattering theory was adapted to various non-relativistic QFT
and models of Quantum Statistical Mechanics. However, the  most appealing aspect of HR theory, which 
is the conceptual clarity of its assumptions, typically gets lost in a non-relativistic setting: some authors use physically less transparent 
Euclidean or stochastic assumptions   \cite{BarataFredenhagen,Malyshev:1978}, other resort to model-dependent computations \cite{Al73}. In this paper we show that there is an exception to this rule: For a class of gapped quantum spin systems satisfying the Lieb-Robinson bound
and admitting single-particle states in their ground-state representations, HR scattering theory can be developed in a natural, model-independent manner
parallel to its original relativistic version. 

\begin{figure}
\centering
\begin{subfigure}{.5\textwidth}
  \centering
  \includegraphics[width=9.3cm]{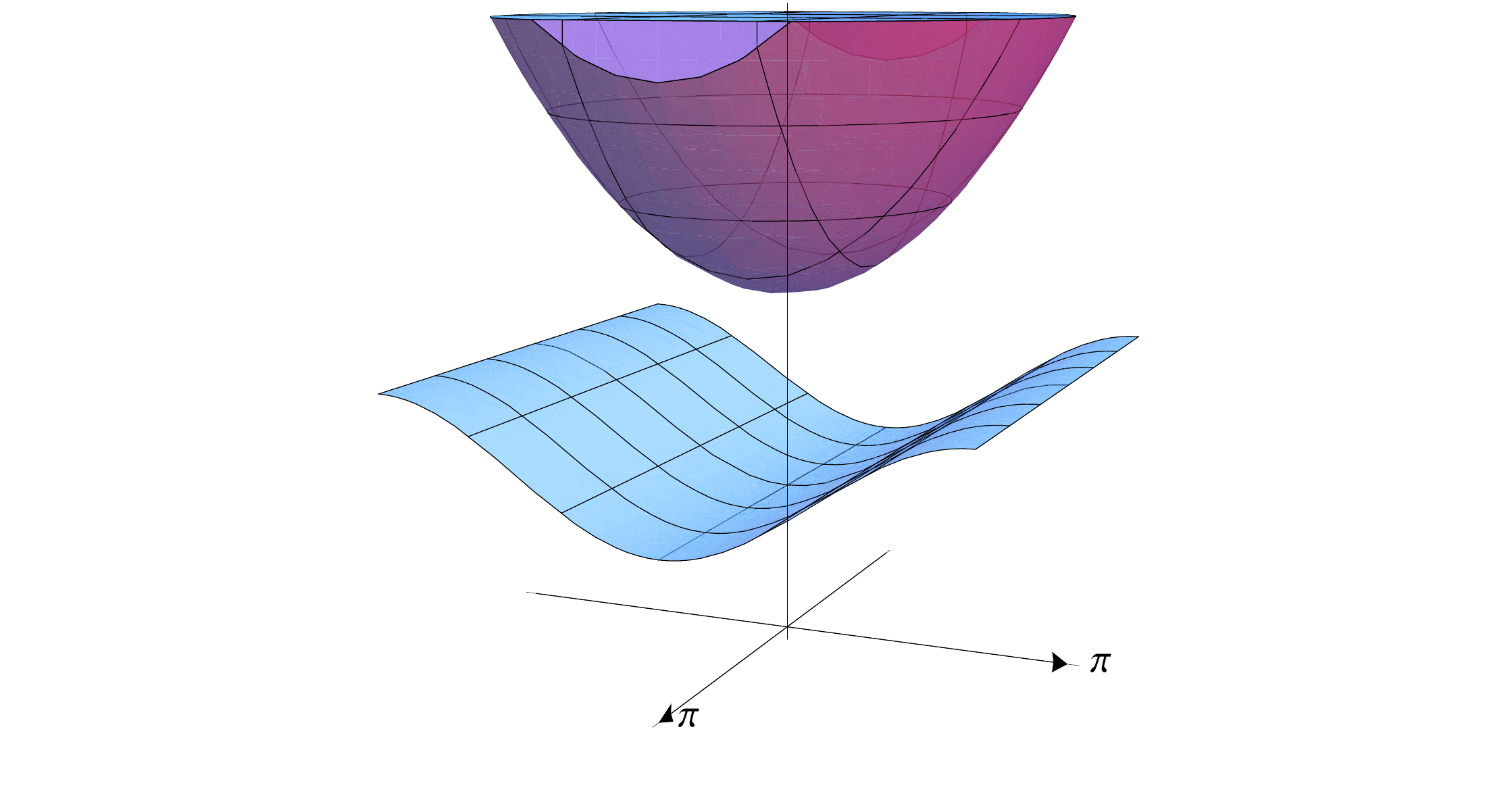}
  \caption{A pseudo-relativistic mass shell.}
  \label{fig:sub1}
\end{subfigure}%
\begin{subfigure}{.5\textwidth}
  \centering
  \includegraphics[width=7cm]{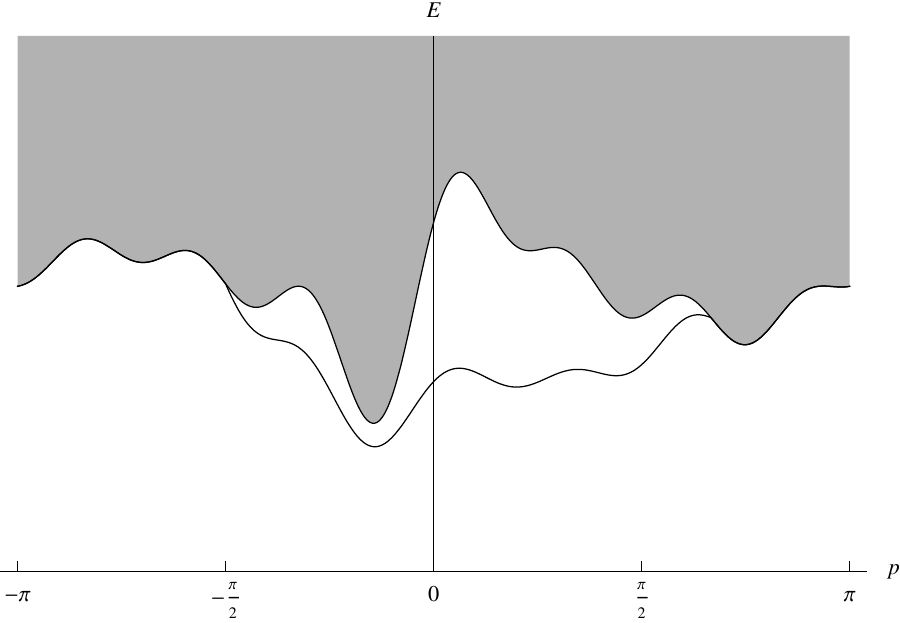}
  \caption{A regular mass shell.}
  \label{fig:sub2}
\end{subfigure}
\caption{Two examples of mass shells within the scope of this paper: (\ref{fig:sub1}) A pseudo-relativistic mass shell which is not regular as it
has a flat direction. The spectrum above the upper shell may be continuous.  (\ref{fig:sub2}) A regular mass shell. (The shaded region is continuous spectrum). }
\label{mass-shell-fig}
\end{figure}

To explain the content of this paper it is convenient to adopt a general framework which
encompasses both local relativistic theories and spin systems:  Let $\Ga$ be the group of space translations
with $\Ga=\real^d$ in the relativistic case and $\Ga=\mathbb{Z}^d$ for lattice theories. We denote by $\wh\Ga$
the Pontryagin dual of $\Ga$ which in the latter case is the torus $S_1^d$.
We consider a $C^*$-dynamical system $(\tau, \mfa)$, where $\tau$ is a representation of the group of space-time translations $\real\times \Ga$ in the automorphism group of a unital $C^*$-algebra $\mfa$. 
 We fix a $\tau$-invariant ground state $\om$ on $\mfa$ (or a vacuum state in the relativistic terminology) and proceed to its GNS representation $(\pi,\hil,\Om)$. Let $U$ be a unitary representation of $\real\times \Ga$ implementing $\tau$ in $\pi$ and $\Sp\,U$  be its spectrum defined via the SNAG theorem, or in more physical terms its energy-momentum spectrum.
We say that $\pi$ contains single-particle states, if there is an isolated mass shell $\mfh\subset \Sp\,U$. $\mfh$ should be
the graph of a smooth function $p\mapsto\Si(p)$ whose Hessian matrix is almost everywhere non-zero. As $\Si$ is the dispersion relation of the particle, the
latter condition says that the group velocity $p\mapsto \nabla\Si(p)$ is non-constant on sets of non-zero measure. This is automatically satisfied   in relativistic theories where dispersion relations  are given by mass hyperboloids $\Si(p)=\sqrt{p^2+m^2}$. Two key properties of such mass hyperboloids enter into various proofs of the relativistic HR theorem: First, the momentum-velocity relation $p\mapsto \nabla\Si(p)$ is invertible. Second, the difference of two vectors on a mass-hyperboloid is spacelike or zero. In the context of spin systems, whose mass shells can a priori have arbitrary shape, we use these general properties to characterize two classes of mass shells which we can handle: 
We call them \emph{regular} and \emph{pseudo-relativistic}, respectively, cf. Figure~\ref{mass-shell-fig} and Definition~\ref{mass-shell-definition}.  We check that such mass shells do appear in the concrete example of the Ising model in strong transverse magnetic fields in various space dimensions. 

Single-particle states in $\hil$ are elements of the spectral subspace of the mass shell $\mfh$  introduced above. To obtain states in $\hil$ describing several particles one needs to `multiply' such single-particle states.  This is performed following a prescription familiar from the Fock space: one identifies
creation operators of individual particles and acts with their product on the vacuum. In our context `creation operators' are elements of $\mfa$, well localized in space and with definite energy-momentum transfer. Thus acting on the ground state $\Om$ they  create single-particle states with prescribed localization in phase-space. We recall that the energy-momentum transfer (or the Arveson spectrum) of $A\in\mfa$, denoted $\Sp_A\tau$, is defined as 
the smallest subset of $\real\times\wh\Ga$ s.t. the expression
\beqa
\tau_{f}(A):= ( 2 \pi)^{-\frac{d+1}{2}} \int dt d\mu(x)\, \tau_{(t,x)}(A) f(t,x), \quad f\in L^1(\real\times \Ga), \label{smearing-intro}
\eeqa
vanishes for any $f$ whose Fourier transform is supported outside of this set. (Here $d\mu$ is the Haar measure 
of $\Ga$). We note that for $\Ga=\real^d$ the Arveson spectrum of $A$  is simply the support of the (inverse)  Fourier transform of the
distribution $(t,x)\mapsto \tau_{(t,x)}(A)$. The utility of this concept derives from the 
energy-momentum transfer relation
\beqa
\pi(A_1)P(\De)\hil\subset P(\ov{\De+\Sp_{A_1}\tau})\hil, 
\eeqa
where $P$ is the spectral measure of $U$ defined on Borel subsets $\De\subset \Sp\, U$. In particular, setting $\De=\{0\}$
and choosing $A_1$ s.t. $\Sp_{A_1}\tau$ is a small neighbourhood of a point $(\Si(p_1), p_1)$ on the mass shell, we obtain a candidate for
a creation operator of a particle whose energy and momentum are close to this point. Such operators with their energy-momentum transfer in a prescribed set
are easy to obtain by setting $A_1=\tau_{f}(A)$ and making use of the fact that $\Sp_{\tau_{f}(A)}\tau\subset \supp\, \widehat f$. However, to ensure  localization
of the created particle in space we need  more structure: A minimal requirement seems to be the existence of 
a norm-dense $*$-subalgebra of \emph{almost local} observables
$\mfa_{\aloc}\subset \mfa$ s.t. for any $A, A_1, A_2\in\mfa_{\aloc}$
\beqa
 &\tau_{(t,x)}(A) \in \mfa_{\aloc},\quad (t,x)\in\real\times \Ga, 
\label{time-invariance-intro}\\
 &\tau_{f}(A)\in \mfa_{\aloc}, \quad f\in S(\real\times \Ga),  \label{Schwartz-class-intro}\\
 &[\tau_{x_1}(A_1), \tau_{x_2}(A_2)]=O( \lan x_1-x_2\ran^{-\infty}), \label{commutators-intro}
\eeqa
where  $S(\real\times \Ga)$ are Schwartz class functions and in (\ref{commutators-intro}) a rapid decay of the norm of the commutator is meant, cf. Appendix~\ref{Notations}. Postponing further discussion of $\mfa_{\aloc}$
to a later part of this introduction, we note that with the above input the HR construction goes through
in the usual way: In view of (\ref{Schwartz-class-intro}), we can set $B_1^*:=\pi(A_1)$, where $A_1$  has its energy-momentum transfer near the point
$(E_1, p_1)$ of the mass shell and in addition is almost local. Next, we pick a positive-energy wave packet
\beqa
g_{1,t}(x)=(2\pi)^{-\fr{d}{2}}\int_{\wh\Ga} dp\, \e^{-\i\Si(p)t+\i px} \wh{g}_{1}(p), \quad \wh{g}_1\in C^{\infty}(\wh\Ga),
\eeqa
describing the free evolution of the particle in question.
(Here $dp$ is the Haar measure on $\wh\Ga$ and $\wh{g}_{1}$ is supported near $p_1$). The HR creation operator,
given by
\beqa
B^*_{1,t}(g_{1,t}) := (2 \pi)^{-\frac{d}{2}} \int d\mu(x)\, B_{1,t}^*(x)g_{1,t}(x), \quad B_{1,t}^*(x)=U(t,x)B_1^*U^*(t,x),
\eeqa
creates from  $\Om$ a time-independent single-particle state $\Psi_1$. Given a collection $\Psi_1, \ldots, \Psi_n$
of such single-particle states, corresponding to distinct points $(\Si(p_1),p_1),\ldots, (\Si(p_n),p_n)$ on the mass shell, s.t. also the velocities 
$\nabla\Si(p_1),\ldots, \nabla\Si(p_n)$ are distinct\footnote{In the relativistic case for the dimension of space $d\geq 3$ the velocities of particles do not have to be distinct \cite{Ha}.}, the
$n$-particle scattering state is constructed as follows
\beq
\Psi_1\timeso\cdots\timeso\Psi_n := \lim_{t \to \infty} B_{1,t}^*(g_{1,t}) \cdots B_{n,t}^*(g_{n,t}) \Omega,
\eeq
where the existence of the limit follows via Cook's method and property~(\ref{commutators-intro}). One
obtains in addition that the scattering states span a subspace in $\hil$ naturally isomorphic to the symmetric
Fock space, and thus they can be interpreted as configurations of independent bosons. This latter step requires
that the mass shell is regular or pseudo-relativistic, and that multiples of $\Om$ are the only translation invariant vectors in $\hil$. 
In the language of the physical spins, these states correspond to collective, strongly interacting excitations above the ground state, which become asymptotically independent from each other for large times.
As we shall see, mass shells of particles with fermionic or anyonic statistics do not appear in the energy-momentum spectrum of
a translation invariant ground state (or a vacuum state). Nevertheless,
variants of HR theory for fermions and anyons have been developed in the relativistic setting. In Section~\ref{future-work} we discuss 
prospects of adapting these constructions to  spin systems on a lattice. The existence of anyonic excitations in such systems is believed to be generically associated to topologically ordered ground states (cf.~\cite{Einarsson,Kitaev:2003ul}).

Let us now come back to the requirement that $\mfa$ should contain a norm-dense $*$-algebra $\mfa_{\aloc}$ of almost local observables,  
whose elements satisfy properties (\ref{time-invariance-intro})--(\ref{commutators-intro}). In relativistic theories this requirement is met as
follows: Recall that in this context $\mfa$ is the $C^*$-inductive limit of a  net of observables $\mco\mapsto \mfa(\mco)$ labelled by open bounded regions
$\mco$ of \emph{space-time} $\real^{d+1}$. Due to the covariance property
\beqa
\tau_{(t,x)}(\mfa(\mco))= \mfa(\mco+(t,x)) \label{covariance-property}
\eeqa  
the $*$-algebra of all local observables $\mfa_{\loc}$ is invariant under the action of $\tau$ and thus satisfies (\ref{time-invariance-intro}). 
Locality of the net, which says that observables localized in spacelike-separated regions commute, implies (\ref{commutators-intro}). However,
property~(\ref{Schwartz-class-intro}) cannot be expected, unless $f$ is compactly supported. To ensure~(\ref{Schwartz-class-intro}), one 
proceeds to a slightly larger $*$-algebra consisting of $A\in\mfa$ s.t.
\beqa
A-A_r=O(r^{-\infty}), \textrm{ for some } A_r\in\mfa(\mco_r), \textrm{ where } \mco_r:=\{\,(t,x)\,|\, |x|+|t|< r\,\}. \label{almost-local-def-intro}
\eeqa
That is, $A$ can be approximated in norm by observables localized in double-cones, centered at zero, of radius $r$, up to
an error vanishing faster than any inverse power of $r$. It is easy to see that this algebra of almost local observables $\mfa_{\aloc}$, introduced for
the first time in \cite{AH67}, satisfies (\ref{time-invariance-intro})-(\ref{commutators-intro}). In particular, invariance under $\tau$ follows
from the invariance of $\mfa_{\loc}$ and isotony of the net.

In spin systems the existence of a $*$-algebra $\mfa_{\aloc}$ of almost local observables, satisfying (\ref{time-invariance-intro})-(\ref{commutators-intro}), is less obvious. We recall that here the algebra $\mfa$ is the $C^*$-inductive limit of a net of local algebras $\La\mapsto\mfa(\La)$ labelled by bounded subsets of \emph{space} (more precisely finite subsets of the lattice $\Ga=\mathbb{Z}^d$). As these time-zero algebras are covariant only under
space translations,
the $*$-algebra $\mfa_{\loc}$ of all local observables is usually not invariant under time evolution\footnote{Except for special cases such as for Hamiltonians consisting of commuting interactions of bounded range.}. Thus the algebra of almost local observables $\mfa_{\aloc}$, defined by replacing $\mco_r$ with a ball of radius $r$ in $\mathbb{Z}^d$ in (\ref{almost-local-def-intro}), has a priori no reason to be time invariant. However, and this is essential in our paper, using the Lieb-Robinson bound one can show that it actually \emph{is} invariant. This bound can be schematically stated as follows
\beqa
\| [\tau_t(A), B] \| \leq C_{A,B} \e^{\lambda(v_{\mathrm{LR}} t - d(A,B))},\quad A,B\in \mfa_{\loc}, \label{LR-intro}
\eeqa
where  $d(A,B)$ is the distance between the localization regions of $A$ and $B$, $v_{\mathrm{LR}}>0$ is called the Lieb-Robinson velocity and $\la>0$ is a constant.
With the help of this estimate we show in Theorem~\ref{smearing-theorem}
that $\mfa_{\aloc}$ satisfies the crucial properties
~(\ref{time-invariance-intro}), (\ref{Schwartz-class-intro}). Property~(\ref{commutators-intro}) follows directly from the definition of $\mfa_{\aloc}$. 
The Lieb-Robinson bound provides us via these results a crucial tool to sufficiently localise single-particle excitations. This opens the route to HR scattering theory for spin systems as described above, where particularly in the construction of multi-particle states, the localisation of single-particle states is essential.

Apart from its relevance to scattering theory,
 almost locality combined with  Arveson spectrum is a powerful technique which appears in many contexts in relativistic QFT (see e.g. \cite{BF82}).
To demonstrate its flexibility we include several results, partially or completely independent of scattering theory, which concern the shape of $\Sp\, U$:  
First, we show that  velocity of a particle, defined as the gradient of its dispersion relation $p\mapsto\nabla\Si(p)$,  is always bounded by the 
Lieb-Robinson velocity $v_{\mathrm{LR}}$ appearing in (\ref{LR-intro}). 
Second, we verify additivity of the spectrum: If $\ti p_1, \ti p_2\in   \Sp\, U$
then also $\ti p_1+\ti p_2\in \Sp\, U$, where the addition is understood in $\real\times \wh\Ga$. This result is a counterpart of a well known fact from the relativistic setting \cite{Ha}. Third, we check that gaps in the spectrum of the finite-volume Hamiltonians cannot close in the thermodynamic limit. We hope that the concepts of almost locality and Arveson spectrum will find further interesting applications in the setting of spin systems.

Having summarized the content of this paper, we proceed now to a more systematic comparison of our work with the literature:  
Definition and properties~(\ref{time-invariance-intro}), (\ref{Schwartz-class-intro}), (\ref{commutators-intro}) of the algebra of almost local observables for spin systems are in fact not new.  Nearing the completion of this work, Klaus Fredenhagen pointed out to us the Diplom thesis of Schmitz~\cite{Sch83}, which contains such results. Also the bound $\nabla\Si(p)\leq v_{\mathrm{LR}}$
is proven in \cite{Sch83}. Yet, since our proofs of these facts are different than
Schmitz', and also the reference \cite{Sch83} is not easily accessible, we decided to keep the complete discussion of these results in our paper.
We also stress that the overlap between our work and Schmitz' does not go
beyond the technical facts mentioned above. In particular, there is no 
discussion of scattering theory in \cite{Sch83}.

Let us now turn to the works which do concern scattering theory for lattice systems:
In the special case of the ferromagnetic Heisenberg model it is possible to develop scattering theory adapting  arguments from many body quantum mechanics
\cite{Hepp:1972,GrafSchenker:1997}. For certain perturbations of non-interacting gapped lattice models collision theory was established by Yarotsky,  exploiting the special form of the finite volume Hamiltonians~\cite{Yarotsky:2004vp}. Malyshev discusses particle excitations in the Ising model in external fields (and more generally, in so-called Markov random fields), implementing HR ideas~\cite{Malyshev:1983}. Barata and Fredenhagen have developed 
HR scattering theory for Euclidean lattice field theories~\cite{BarataFredenhagen} on a $d+1$-dimensional lattice.  Clustering estimates play an important role in the last three references, and in fact they also appear in early proofs of the original HR theorem. In our analysis we avoid clustering estimates altogether (in the case of pseudo-relativistic mass shells) or derive them
from the Lieb-Robinson bound via~\cite{Nachtergaele:2006fd} (in the case of regular mass shells).

Let us now discuss briefly more recent literature, centered around the Lieb-Robinson bound: In~\cite{Haegeman:2013a} certain operators
were constructed which create from the ground state single-particle states of a given momentum, up to controllable errors. 
Although technically quite different, this finite volume result inspired the present investigation as it leaves little doubt that the Lieb-Robinson
bound is a sufficient input to develop HR theory for spin systems containing mass shells. Our counterpart of the main result from \cite{Haegeman:2013a}
is Lemma~\ref{single-particle-density} below. We further note that \cite{Haegeman:2013a}  gives a theoretical justification that a matrix product state ansatz describes single-particle states efficiently, see~\cite{Vanderstraeten:2014a} for more details.

Our paper is built up as follows. In Section~\ref{preliminaries-section} we introduce the standard concepts and tools of the theory of quantum spin systems. In Section~\ref{Arveson+almost-local}, which still has a preliminary character, we introduce two concepts which may be less well known to experts in 
quantum spin systems: almost locality and the Arveson spectrum. (A more general perspective on this latter concept is given 
in Appendix~\ref{Spectral-appendix}).
Section~\ref{scattering-section} is the heart of the paper: there we develop the Haag-Ruelle scattering theory for lattice systems.  In Section~\ref{shape-spectrum} we include some results on shape of the spectrum obtained with the methods of this paper. Examples of concrete systems satisfying our assumptions are given in  Section~\ref{examples-section}. Finally, in Section~\ref{future-work}, we comment  on possible future directions. Our conventions and  some more technical proofs are relegated to the appendices.  

\vspace{0.5cm}

\emph{Acknowledgements:} WD is supported by the DFG with an Emmy Noether grant DY107/2-1, PN acknowledges support by NWO through Rubicon grant 680-50-1118 and partly through the EU project QFTCMPS. We like to thank Kamilla Mamedova for help in translating parts of Ref.~\cite{Malyshev:1983} and Tobias Osborne for helpful discussions on Refs.~\cite{Haegeman:2013a,Vanderstraeten:2014a}.  WD would like to thank Herbert Spohn for useful discussions on
spectra of lattice models, Klaus Fredenhagen for pointing out the reference \cite{Sch83} and Maximilian Duell for a helpful discussion about his proof of the energy-momentum transfer relation from \cite{Du13}. SB and WD wish to thank the organizers of the Warwick EPSRC Symposium on Statistical Mechanics, Daniel Ueltschi and Robert Seiringer, as part of the work presented here was initiated during the meeting.

\section{Framework and preliminaries} \label{preliminaries-section}
\setcounter{equation}{0}

\subsection{Quasilocal algebra $\mfa$ and space translations}

Let $\Ga=\mathbb{Z}^{d}$ be the set 
labelling the sites of the system, which we equip with a metric $|\,\cdot\,|$. 
We define by $\P(\Ga)$ the set of
all subsets and by $\P_{\fin}(\Gamma)$ the set of all finite subsets of $\Ga$. Also, for $\La, \La_1, \La_2\in \P_{\fin}(\Gamma)$
we set
\beqa
& \diam(\La):=\sup\{\, |x_1-x_2| \,|\, x_1,x_2\in \La\,\}, \\
& \dist(\La_1, \La_2):=\inf\{\, |x_1-x_2| \,|\, x_1\in \La_1, x_2\in \La_2\,\}, \\
& \La^r:=\{\,x\in \Ga \,|\, \dist(x, \La)\leq r\,\},
\eeqa
and $|\La|:= \sum_{x\in \La} 1$ is the volume of $\Lambda$. We denote by $\B=\{0\}$ the set consisting of one point at the
origin, so that $\B^r$ is the ball of radius $r$ centered at the origin.

We assume that at each site $x\in \Ga$ there is an $\ell$-dimensional
quantum spin whose observables are elements of $M_{\ell}(\mathbb{C})$, where $\ell$ is independent
of $x$. For $\La\in \P_{\fin}(\Ga)$ we set
\beqa
\mfa(\La):=\bigotimes_{x\in\La} M_{\ell}(\complex).
\eeqa
For $\La_1,\La_2\in \P_{\fin}(\Ga)$ such that $\La_1\subset\La_2$, we have the natural embedding $\mfa(\La_1)\subset\mfa(\La_2)$ given by identifying $A\in\mfa(\La_1)$ with $A\otimes 1_{\La_2\setminus\La_1}\in\mfa(\La_2)$. Thus we obtain a net $\La\mapsto\mfa(\La)$ which is local in the sense
that for $\La_1,\La_2\in \P_{\fin}(\Ga)$, $\La_1\cap\La_2=\emptyset$ we have
\beqa
[\mfa(\La_1), \mfa(\La_2)]=0.
\eeqa
We define
\beqa
\mfa_{\mathrm{loc}}:=\bigcup_{\La\in \P_{\fin}(\Ga) }\mfa(\La),
\eeqa
and the quasilocal algebra $\mfa$ as a $C^*$-completion of $\mfa_{\mathrm{loc}}$. 

We denote by $\Ga\ni x\mapsto \tau_x$ the natural representation of the group of translations from $\Ga$ in automorphisms of $\mfa$. 

\subsection{Interactions}
We define an interaction as a map $\Phi: \P_{\fin}(\Ga)\to \mfa$, 
s.t. $\Phi(\La)$ is self-adjoint and $\Phi(\La)\in \mfa(\La)$. The interaction is said to be translation invariant if
\beqa
\Phi(\La+x)=\tau_x(\Phi(\La)). \label{translation-invariance}
\eeqa
We also introduce a family of local Hamiltonians: For any $\La\in \P_{\fin}(\Ga)$ we set
\beqa
H_{\La}:=\sum_{X\subset\La} \Phi(X).
\eeqa
$H_{\La}$ induce the local dynamics $\{\tau^{\La}_t\}_{t\in\bbR} $ given by
\beqa
\tau^{\La}_t(A):=\e^{\i H_{\La} t}A \e^{-\i H_{\La} t}, \quad A\in \mfa.
\eeqa
As stated in Corollary~\ref{global-dynamics} below, the global dynamics $\tau_t:=\lim_{\La\nearrow \Ga} \tau^{\La}_t$ can be defined for a large class of interactions which we now specify.

 Let $F:[0,\infty)\to(0,\infty)$ be a non-increasing function such that
\begin{align}
\Vert F\Vert &:= \sum_{x\in\Gamma} F(\vert x \vert)<\infty, \\
C&:= \sup_{x,y\in\Gamma} \sum_{z\in\Gamma}\frac{F(\vert z-x \vert)F(\vert y-z \vert)}{F(\vert y-x \vert)} < \infty.
\end{align}
For $\Ga = \mathbb{Z}^d$ one can choose $F(x) = (1+x)^{-d-\varepsilon}$ for any $\varepsilon > 0$ and $C \leq 2^{d + \varepsilon + 1} \sum_{x \in \Ga} (1+|x|)^{d+\varepsilon}$, see~\cite{Nachtergaele:2006bh}.
Then, for any $\lambda>0$, $F_\lambda(r) := \exp(-\lambda r) F(r)$ satisfies the same conditions with $\Vert F_\lambda\Vert \leq \Vert F\Vert$ and $C_\lambda\leq C$. The formula
\begin{equation}
\Vert \Phi \Vert_\lambda := \sup_{x\in\Gamma} \sum_{X\ni x,0} \frac{\Vert \Phi(X) \Vert}{F_\lambda(\vert x\vert)}
\end{equation}
defines a norm on the set of translation invariant interactions. We let $\caB_\lambda$ be the set of interactions such that $\Vert \Phi \Vert_\lambda<\infty$.

\subsection{Lieb-Robinson bounds and the existence of dynamics}
\label{Lieb-Robinson-subsection}
The fast decay of interactions from $\caB_{\lambda}$ implies that the associated dynamics satisfies the Lieb-Robinson bounds
\cite{Lieb:1972ts,Nachtergaele:2006bh} stated in Theorem~\ref{Lieb-Robinson-one} below. Beyond its physically natural interpretation as a finite maximal velocity of propagation of correlations through the system, the Lieb-Robinson bounds have many useful structural corollaries such as the existence of the dynamics in the infinite system. This article will further emphasize that they provide an analogue of the speed of light in the framework of quantum spin systems (cf. Corollary~\ref{gradient-of-mass-shell}). Another instance of that analogy can already be found in the exponential clustering property of~\cite{Hastings:2006gv,Nachtergaele:2006fd} stated in Theorem~\ref{clustering} below.

For interactions $\Phi\in \caB_{\lambda}$ we have the following Lieb-Robinson bounds \cite{Lieb:1972ts,Nachtergaele:2006bh}:
\bet\label{Lieb-Robinson-one} Let $\Lambda_1,\Lambda_2,\Lambda$ be finite subsets of $\Gamma$ such that $\Lambda_1,\Lambda_2\subset\Lambda$, and let $A\in \mfa(\La_1)$, $B\in \mfa(\La_2)$. Moreover, assume that there is $\lambda>0$ such that $\Phi\in \caB_\lambda$. Then
\begin{equation}
\|[\tau_t^{\Lambda}(A), B]\|\leq \frac{2 \|A\| \|B\|}{C_\lambda} \e^{2\Vert\Phi\Vert_\lambda C_\lambda\vert t \vert} \sum_{w\in\Lambda_1}\sum_{z\in\Lambda_2} F_\lambda(\vert w-z\vert)
\end{equation}
for all $t\in\bbR$ and uniformly in $\La$. Thus defining the Lieb-Robinson velocity as $v_{\la}:=\fr{2\|\Phi\|_{\la}C_{\la}}{\la}$ we have
\beqa
\left\Vert \left[\tau_t^\Lambda(A) ,  B\right]\right\Vert
\leq \frac{2 \|A\| \|B\|}{C_\lambda}\min(\vert \La_1 \vert, \vert \La_2 \vert  )\Vert F \Vert \e^{-\lambda (\dist(\La_1,\La_2)-v_\lambda\vert t\vert)}. \label{LRB}
\eeqa
\eet
It is well-known~\cite{Ro76, Nachtergaele:2006bh} that the Lieb-Robinson bound allows for the extension of the dynamics from the local to the quasi-local
algebra by a direct proof of the Cauchy property of the sequence $\tau^{\Lambda_n}_t(A)$ for any increasing and absorbing sequence of finite subsets $\Lambda_n\subset\Gamma$.

\begin{corollary} \label{global-dynamics}
Assume that there is $\lambda> 0$ such that $\Phi\in \caB_\lambda$. Then there exists a strongly continuous one parameter group of automorphisms of $\mfa$, $\{\tau_t\}_{t\in\bbR}$, such that for all $A\in\mfa_{\mathrm{loc}}$,
\begin{align*}
\lim_{n\to\infty}\Vert \tau^{\Lambda_n}_t(A)-\tau_t(A) \Vert = 0,
\end{align*}
independently of the choice of the absorbing and increasing sequence $\Lambda_n$. Moreover, there exists a *-derivation $\delta$ such that $\tau_t = \e^{t\delta}$.
\end{corollary}
\subsection{Ground states and clustering}
\label{ground-state-subsection}

We start with a definition of a ground state of a $C^*$-dynamical system $(\mfa,\tau)$:
\bed\label{ground-state-def} Let $\om$ be a state on $(\mfa,\tau)$ s.t. $\om\circ\tau_t=\om$ for all $t$.
Let $(\pi,\hil,\Om)$ be the corresponding GNS representation and $t\mapsto U(t)$
the unitary representation of time translations implementing $t\mapsto \tau_t$ in $\pi$,
which satisfies $U(t)\Om=\Om$ for all $t$.
Let $H$ be the generator of $U$ (the Hamiltonian) i.e. $U(t)=\e^{\i tH}$. 
\begin{enumerate}
\item We say that $\om$ is a \emph{ground state} and $\pi$ a \emph{ground state representation} if $H$ is positive.
\item We say that $\pi$ has a \emph{lower mass gap} if $\{0\}$ is an isolated eigenvalue of $H$.
\item We say that $\pi$ has a \emph{unique ground state vector} if $\{0\}$ is a simple
eigenvalue of $H$ (corresponding up to phase to the eigenvector $\Om$). 
\item We say that $\om$ is \emph{translation invariant} if $\om\circ \tau_x=\om$ for all
$x\in \Ga$.
\end{enumerate}
\eed

We note here that our definition of a ground state is in fact equivalent to the more algebraic one given by 
the inequality
\begin{equation}\label{GS}
-\i \omega(A^*\delta(A))\geq 0,\qquad A\in D(\delta).
\end{equation}
See~\cite[Prop. 5.3.19]{Bratteli:1997aa}.

Finally, we recall that the approximate locality provided by the Lieb-Robinson bound  yields exponential clustering in the ground state for Hamiltonians with a lower mass gap. 
The statement below follows from Theorem 4.1 of \cite{BS09}. We will use it in the proof of Theorem~\ref{Haag-Ruelle-Fock}.
\bet\label{clustering} Let $\om$ be a ground state whose GNS representation has a lower mass gap and a unique ground state vector. 
Then, under the assumptions of Theorem~\ref{Lieb-Robinson-one}, for any local observables $A\in \mfa(\Lambda_1)$,
$B\in \mfa(\Lambda_2)$, s.t. $\dist(\Lambda_1,\Lambda_2)\geq 1$  we have
\beqa
|\om(AB)-\om(A)\om(B)|\leq C\|A\| \|B\|\min(\vert \Lambda_1\vert, \vert \Lambda_2\vert) \e^{-\mu\dist(\Lambda_1,\Lambda_2)},
\eeqa 
where $C,\mu>0$ are independent of $A,B,\Lambda_1,\Lambda_2$ within the above restrictions.
\eet

In the remainder of this paper we will always consider a lattice system with a translation invariant interaction $\Phi\in \caB_{\la}$, $\la>0$, so that Lieb-Robinson bounds apply,
and a translation invariant ground state with a GNS representation $(\pi, \hil, \Om)$.

\section{Arveson spectrum and almost local observables} \label{Arveson+almost-local}
\setcounter{equation}{0}
In this section, which is still preparatory, we introduce the concepts of Arveson spectrum 
and almost local  observables. 
In Section~\ref{scattering-section} we will show their utility in the context of  scattering theory. The Arveson spectrum is a topic in the spectral analysis of groups of automorphisms. A more extensive account  can be found in~\cite{Pedersen}.

\subsection{Space-time translations}\label{space-time-translations}

We claim that for any $x\in\Ga$, $t\in\bbR$ and $A\in\mfa$,
\begin{equation}
\tau_t\circ\tau_x(A) = \tau_x\circ\tau_t(A) \label{exchange-property}.
\end{equation}
Since $\tau$ is an action of $\Ga$ this is equivalent to $\tau_{-x} \circ \tau_t \circ \tau_x(A) = \tau_t(A)$ for all $A\in\mfa$. By  Corollary~\ref{global-dynamics}  and because $\tau_x$ is an automorphism we have
\begin{equation}
\tau_{-x} \circ \tau_t \circ \tau_{x}(A) = \lim_{n \to \infty} \tau_{-x} \circ \tau^{\Lambda_n}_t \circ \tau_x(A) = \lim_{n \to \infty} \tau^{\Lambda_n-x}_t(A), 
\end{equation}
for $A\in \mfa_{\loc}$, where in the last step the translation invariance of the interaction~(\ref{translation-invariance}) is used, and the limits denote convergence in norm. Since the right hand side converges irrespective of the choice of exhausting sequence, and $\mfa_{\loc}$ is norm-dense in $\mfa$, the claim follows. Hence, we can consistently define
\begin{equation}
\bbR\times \Ga\ni(t,x)\mapsto \tau_{(t,x)}:=\tau_t\circ\tau_x 
\label{spacetime-automorphisms}
\end{equation}
as a strongly-continuous representation of the group of space-time translations in automorphisms of $\mfa$. 
\begin{remark} For the sake of clarity, we will sometimes write $\tau^{(1)}$, $\tau^{(d)}$, $\tau$ to
distinguish the respective groups of automorphisms: 
\beqa
\real\ni t\mapsto \tau_t, \quad \Ga\ni x\mapsto \tau_{x}, \quad \real\times \Ga\ni (t,x)\mapsto \tau_{(t,x)}.
\eeqa
\end{remark}
In the following lemma we construct the unitary representation of space-time translations $U$ in the GNS representation of a
translation invariant ground state $\om$.
\bel \label{lem:FT} Suppose that the interaction is  translation invariant, see (\ref{translation-invariance}), and $\om$ is a translation invariant ground state. Let $\Ga\ni x\mapsto U(x)$ be a unitary representation of translations implementing $\tau_x$ in $\pi$, which satisfies
$U(x)\Om=\Om$, and let $U(t)$ be as in Definition~\ref{ground-state-def}. Then we can consistently define a unitary representation of space-time
translations
\begin{equation}\label{GNS SpacetimeTranslations}
\real\times \Ga\ni (t,x)\mapsto U(t,x)=U(t)U(x)=U(x)U(t).
\end{equation}
This representation implements $(t,x)\mapsto \tau_{(t,x)}$  in $\pi$  
and  satisfies $U(t,x)\Om=\Om$ for all $(t,x)\in \real\times \Ga$.
\eel
\proof We note that for $A,B\in\mfa_{\mathrm{loc}}$, for all $x\in\Ga$ and $t\in\bbR$, 
\begin{align}
\left\langle\pi(A)\Omega,U(x)U(t)\pi(B)\Omega\right\rangle &= \left\langle\pi(A)\Omega,\pi(\tau_x\circ\tau_t(B))\Omega\right\rangle = \left\langle\pi(A)\Omega,\pi(\tau_t\circ\tau_x(B))\Omega\right\rangle \non\\
&=\left\langle\pi(A)\Omega,U(t)U(x)\pi(B)\Omega\right\rangle\!\!,
\end{align}
since we treat translation invariant interactions and hence (\ref{exchange-property}) holds. 
Thus we have shown that  $U(x)U(t) = U(t) U(x)$ on $\caH$. \qed\\ 
Recall that the dual group of $\Ga=\mbZ^d$ 
is the $d$-dimensional torus $\whGa=S_1^d$ (the Brillouin zone).
By the Stone-Neumark-Ambrose-Godement (SNAG) theorem~\cite{Riesz:1955aa}, there exists a spectral measure $dP$ on $\real\times \wh\Ga$ with values in  $\hil$ s.t.    
\beqa
U(t,x)=\int_{\real\times \whGa} \e^{\i Et-\i px}dP(E,p). \label{SNAG}
\eeqa
The following is a special case of Definition~\ref{Arveson-definition} in the case of unitaries, see Subsection~\ref{unitaries-subsection}.
\bed We define $\Sp\, U$ as the support of $dP$. 
\eed
\nin By definition of the ground state we have that $\Sp\, U\subset \real_+\times \whGa$.

\subsection{Arveson spectrum} \label{EM-subsection}
In this subsection we give a very brief overview of basic concepts of spectral 
analysis of automorphisms groups. For a more systematic 
discussion we refer to Appendix~\ref{Spectral-appendix}.  

Let $A\in \mfa$ and $f\in L^1(\real\times\Ga)$, $g\in L^1(\Ga)$, $h\in L^1(\real)$. Then, using the
space-time translation automorphisms $(t,x)\mapsto \tau_{(t,x)}$
we set 
\beqa  
&\tau_f(A):=(2\pi)^{-\fr{d+1}{2}}\sum_{x\in \Ga}\int dt\, \tau_{(t,x)}(A)f(t,x), 
\label{smearing-operation} \\
&\tau^{(d)}_g(A):=(2\pi)^{-\fr{d}{2}}\sum_{x\in \Ga} \tau_{x}(A)g(x),\label{space-smearing}\\
&\tau^{(1)}_h(A):=(2\pi)^{-\fr{1}{2}}\int dt\, \tau_{t}(A)h(t). 
\label{time-smearing}
\eeqa
These are again  elements of $\mfa$ by the strong continuity of $\tau$,
which is a consequence of the Lieb-Robinson bound. 
 Now we  define:
\bed\label{EM-definition}  The \emph{Arveson spectrum}  of $A\in \mfa$ w.r.t. $\tau$, denoted $\Sp_{A}\tau$, is the smallest closed subset of $\real \times\wh\Ga$   with the
property that $\tau_f(A)=0$ for any $f\in  L^1(\real\times \Ga)$ s.t. its Fourier transform $\wh f$
is supported outside of this set. The Arveson spectra of $A$ w.r.t. $\tau^{(1)}$ and  $\tau^{(d)}$, denoted 
$\Sp_{A}\tau^{(1)}$ and $\Sp_A\tau^{(d)}$, are defined
analogously.
\eed
\begin{remark} $\Sp_{A}\tau$  can also be called the energy-momentum transfer of $A$, 
cf. relation~(\ref{EM-transfer-relation}) below.
\end{remark}
\nin Items (\ref{Ar-one})--(\ref{Ar-four})  of the following proposition are easy and well known consequences of   Definition~\ref{EM-definition}. 
Equation (\ref{EM-transfer-relation}) is a special case of Theorem 3.5 of \cite{Ar82}.  For the reader's convenience, we give an elementary argument
in Appendix~\ref{energy-momentum-proof}, which is based on the proof of Theorem 3.26 of \cite{Du13}.
\bep\label{Arveson-elementary} We have for any $A\in \mfa$, $(t,x)\in \real\times\Ga$,  $f\in L^1(\real\times \Ga )$, $g\in L^1(\Ga)$:
\beqa
&\Sp_{A^*}\tau=-\Sp_{A}\tau, \label{Ar-one}\\
&\Sp_{\tau_{(t,x)}(A)}\tau=\Sp_A\tau,\\
&\Sp_{\tau_f(A)}\tau\subset \Sp_{A}\tau\cap \supp\, \wh f, \label{EM-inclusion}\\
&\Sp_{\tau^{(d)}_g(A)}\tau\subset \Sp_{A}\tau\cap (\real\times \, \supp\,\wh g). \label{Ar-four}
\eeqa
Moreover, with $P$ as defined in (\ref{SNAG}), we have the \emph{energy-momentum transfer relation}
\beqa
\pi(A)P(\De)=P(\ov{\De+\Sp_{A}\tau}) \pi(A)P(\De)
\label{EM-transfer-relation}
\eeqa
for any Borel subset $\De\subset \real\times \wh\Ga$.
\eep

By~(\ref{EM-inclusion}), the smearing operation (\ref{smearing-operation}) allows one to construct observables $A$ with energy-momentum transfers contained 
in arbitrarily small sets. Such observables will be needed in Section~\ref{scattering-section} to create single-particle states from the ground state.
Since particles are localized excitations, these observables should have good localization properties. However, in view of Proposition~\ref{space-translations-prop} below, if $A\in \mfa_{\loc}$ is s.t.  $\Sp_A\tau$ is a small neighbourhood of some point $(E,p)\in \real\times \wh\Ga$ then 
$A\in \complex I$ and $(E,p)=(0,0)$. Therefore, in the next subsection we introduce a slightly larger class of \emph{almost local observables} which 
are still essentially localized in bounded regions of space-time but contains observables whose Arveson spectra are in arbitrarily small sets. The class of 
almost local observables is
 invariant under time translations and under the smearing operation~(\ref{smearing-operation}).
\begin{proposition}\label{space-translations-prop}
Let $A \in \mfa_{\loc}$, $A\notin \complex I$. Then $\Sp_A \tau^{(d)} = \wh\Gamma$.
\end{proposition}
\proof
The proof follows a remark in the Introduction of~\cite{Bu90}. Since $\mfa$ is a simple algebra and $A$ is not a multiple of the identity, there is a $B \in \mfa_{\loc}$ such that $[\tau_x(A), B] \neq 0$ for some $x\in \Ga$. Define $T(x) := [\tau_x(A), B]$ which is supported, as an operator valued function of $x$, in a bounded region of $\Ga$ by locality. Its Fourier transform, $\wh T(p) := (2 \pi)^{-\frac{d}{2}} \sum_{x \in \Gamma} \e^{- \i p \cdot x} [\tau_x(A), B]$, which is a function on $\wh\Ga=S_1^d$, is such that $\supp\, \wh T=\wh \Ga$. This can be seen by interpreting the Fourier transform $\wh T$ as a periodic function on $\real^d$ and verifying that it is real-analytic and non-zero. Now suppose that there is  some open $U\subset \wh\Ga$ s.t.  $U\cap \Sp_A \tau^{(d)}=\emptyset$. Then, by definition, for any $g\in L^1(\Ga)$ s.t.
$\supp\, \wh g\subset U$ we have $\tau^{(d)}_g(A)=0$. Since $ [\tau^{(d)}_g(A), B]=(2\pi)^{-\fr{d}{2}}\int \wh T(-p)\wh g(p) dp$, this contradicts $\supp\, \wh T=\wh \Ga$. \qed

\subsection{Almost local observables} \label{almost-local}
In this subsection we introduce a convenient class of \emph{almost local observables}. We recall that in the context
of relativistic QFT almost local observables  were first defined  in \cite{AH67}.
This class is larger than $\mfa_{\loc}$
but its elements have much better localization properties than arbitrary elements of $\mfa$ (see Lemma~\ref{commutator-decay} below). 
The relevance of almost local observables to our investigation comes  from Theorems~\ref{smearing-theorem} and \ref{harmonic-theorem} below.  

As mentioned in the Introduction, the class of almost local observables  has been studied before in the context of lattice systems  in the Diplom thesis of Schmitz~\cite{Sch83}. Counterparts of Lemma~\ref{commutator-decay} and Theorem~\ref{smearing-theorem} below
can be found in this work. Since our proofs are different, and also the reference  \cite{Sch83} is not readily accesible, we present these results here in detail.

\bed We say that $A\in \mfa$ is \emph{almost local} if there exists
a sequence $\real_+\ni r\mapsto A_r\in \mfa(\B^r)$ s.t. 
\beqa
A-A_r=O( r^{-\infty}).
\eeqa
The $*$-algebra of almost local observables is denoted by $\mfa_{\aloc}$. 
\eed
\nin In contrast to local observables,  the commutator of two elements of $\mfa_{\aloc}$
need not identically vanish if one of them is translated sufficiently far from the other. Instead
we have a rapid decay of commutators:
\bel\label{commutator-decay} Let  $A_i\in \mfa_{\aloc}$, $i=1,2$. 
Then, for $y\in \Gamma$,
\beqa
[A_1, \tau_{y}(A_2) ]=O(\lan y\ran^{-\infty}),
\eeqa
where the notation $\langle y \rangle^{-\infty}$ is defined in Appendix~\ref{Notations}.
\eel
\proof
Making use of almost locality of $A_i$, we find $A_{i,r}\in \mfa(\B^r)$ 
s.t. $A_i-A_{i,r}=O(r^{-\infty})$.
Thus we get
\beqa
[A_1, \tau_{y}(A_2)]= [A_{1, r}, \tau_y(A_{2,r})]+O(r^{-\infty}).
\eeqa
Setting $r=\eps\lan y \ran$, we obtain that  for sufficiently small $\eps>0$ and $\lan y\ran $ sufficiently large
\beqa
Z^r\cap(Z^r+y)=\emptyset,
\eeqa
which concludes the proof. \qed\\
The following theorem gives important invariance properties of $\mfa_{\aloc}$. We note that they are in general not true for $\mfa_{\loc}$.
For  precise definitions of the Schwartz classes of functions on a lattice, $S(\Ga)$ and $S(\real\times\Ga)$, we refer to Appendix~\ref{Notations}.
\bet\label{smearing-theorem}  Let $A\in \mfa_{\aloc}$. Then
\begin{enumerate}
\item[(a)] $\tau_{(t,x)}(A)\in \mfa_{\aloc}$ for all $(t,x)\in \real\times \Ga$, and
\item[(b)] $\tau_f(A),\tau^{(d)}_g(A),\tau^{(1)}_h(A)\in \mfa_{\aloc}$ for all $f\in S(\real\times\Ga)$, $g\in S(\Ga)$ and $h\in S(\real)$.
\end{enumerate}
 \eet
\nin In contrast to relativistic QFT, for lattice systems this result is not automatic.  
We give a proof in Appendix~\ref{Lieb-Robinson}. 

To conclude this section we give a lattice variant of an important result of Buchholz \cite{Bu90}, which will be
helpful in Section~\ref{scattering-section} (cf. Lemma~\ref{HR-auxiliary} (c)). This result illustrates how the Arveson
spectrum and almost locality can be used in combination.
\bet\label{harmonic-theorem} Let $\De$ be a compact subset of $\Sp\,U$, $A\in \mfa$
and $\Sp_A\tau$ be a compact subset of $(-\infty,0)\times \whGa$.
Then, for any $\La\in \P_{\fin}(\Ga)$,
\beqa
\|P(\De)\sum_{x\in \La} \pi(\tau_{x}(A^*A))P(\De)\|\leq C
\sum_{x\in (\La-\La)}
\|[A^*, \tau_x(A)]\|,\label{Harmonic-analysis} 
\eeqa
where $C$ is independent of $\La$ and $\La-\La=\{\,x_1-x_2\,|\,x_1,x_2\in\La \, \}\subset \Ga$. If in addition $A$ is almost local, we can conclude that
\beqa
\|P(\De)\sum_{x\in \Ga} \pi(\tau_{x}(A^*A))P(\De)\|<\infty. 
\eeqa
 \eet
\proof First, we note that by positivity of the spectrum of $H$,
our assumption on $\Sp_A\tau$  and the compactness of $\De$,
 relation~(\ref{EM-transfer-relation}) gives that $A^nP(\De)=0$
for $n\in \nat$ sufficiently large. Now to obtain (\ref{Harmonic-analysis}) we apply Lemma~2.2
of \cite{Bu90}, noting that its proof remains valid if integrals
over compact subsets of $\real^d$ are replaced with sums over
finite subsets of $\Ga$. 

If $A$ is almost local, by Lemma~\ref{commutator-decay}
 the sum on the r.h.s. of (\ref{Harmonic-analysis}) can be extended from  $\La-\La$ to $\Ga$.
As a consequence also the sum on the l.h.s. can be extended from $\La$ to $\Ga$ and it still defines a bounded operator 
(as a strong limit of an increasing net of bounded operators which is uniformly bounded).
\qed\\
We remark that  relation~(\ref{Harmonic-analysis}) actually holds in any representation
in which space-time translations are unitarily implemented and the Hamiltonian is positive (not necessarily a ground state representation).

\section{Scattering theory} \label{scattering-section} 
\setcounter{equation}{0}
A procedure to construct wave operators and the $S$-matrix, which we adapt to lattice systems in this section, is known in local relativistic QFT  
as the Haag-Ruelle theory \cite{Ha58, Ru62}. 
Our presentation here  is close to \cite{Dy05,DG13} which in turn profited from \cite{He65, BF82, Ar99}.

Let us start with stating our {\bf standing assumptions} for this section:
\begin{enumerate}
\item We consider a  lattice system given by a quasi-local algebra $\mfa$ and a translation invariant interaction $\Phi\in \caB_{\la}$, $\la>0$. Thus the
space-time translations act on $\mfa$ by the group of automorphisms 
$\real\times \Ga\ni (t,x)\mapsto \tau_{(t,x)}$ of equation (\ref{spacetime-automorphisms}).

\item We consider a  translation invariant ground state $\om$ of $\mfa$ and
its GNS representation $(\pi, \hil, \Om)$. 
The space-time translation 
automorphisms $\tau$ are implemented in $\pi$ by the group of unitaries~$U$ constructed in Lemma~\ref{lem:FT}.

\item $\Sp\, U$ contains an isolated simple eigenvalue $\{0\}$ (with eigenvector $\Om$)
and an isolated mass shell~$\mfh$ (see 
Definition~
\ref{mass-shell-definition} below).

\item The mass shell is either pseudo-relativistic or regular (see Definition~\ref{mass-shell-definition} below).

\end{enumerate}

There is no general criterion that ensures the validity of the assumptions in a given quantum lattice system. In fact, proving the very existence of an isolated mass shell is in general a difficult problem. However, there are models where our standing assumptions can be checked in full rigour, see Section~\ref{examples-section}. In other models isolated mass shells have been shown to exist numerically~\cite{Haegeman:2013a}.

\subsection{Single-particle subspace}
A  class of localized excitations of the ground state (particles) is carried by \emph{mass shells} in $\Sp\, U$. We define a mass shell as follows:
\bed\label{mass-shell-definition} Let $\om$ be a translation invariant ground state of 
a system $(\mfa,\tau)$ with translation invariant interaction.
We say that $\mfh\subset \real\times \whGa$   is a \emph{mass shell} if  
\begin{enumerate} 
\item[(a)]  $\mfh\subset \Sp\,U$, $P(\mfh)\neq 0$, where $P$ is the spectral projection of~\eqref{SNAG}.

\item[(b)] There is an open subset  $\De_{\mfh}\subset \whGa$ and a real-valued  function $\Si\in C^{\infty}(\De_{\mfh})$
 such that 
\begin{equation*}
\mfh=\{\,(\Si(p), p)\,|\, p\in \De_{\mfh}  \}.
\end{equation*}
 We shall call $\Si$ the \emph{dispersion relation} and denote by $D^2\Si(p):=[\pa_i\pa_j\Si(p)]_{i,j=1,\ldots d}$ its Hessian.

\item[(c)] The set $\mcT:=\{\, p\in \De_\mfh\,|\, D^2 \Si(p) =0  \, \}$ has Lebesgue measure zero. 

\end{enumerate}
Moreover, we say that:
\begin{enumerate}

	\item  A mass shell is \emph{isolated}, if  for any $p\in \De_{\mfh}$  there is $\eps>0$ such that\footnote{To cover mass shells
which dip into the continuous spectrum, as in Figure~\ref{fig:sub2}, we allow for a different $\eps>0$ for each $p\in \De_{\mfh}$. }
\begin{equation}
	\big([\Si(p)-\eps,\Si(p)+\eps] \times \{p\}\big) \cap \Sp\, U=\{(\Si(p),p)\}.
\end{equation}
\item  A mass shell is \emph{regular}, if the set  $\mcX:=\{\, p\in \De_\mfh\,|\,\det D^2 \Si(p) =0  \, \}$ has Lebesgue measure zero. 

\item  A mass shell is \emph{pseudo-relativistic}, if $(\mfh-\mfh)\cap \Sp\, U=\{0\}$.

\end{enumerate}
Finally, we define $\Dep:=\De_{\mfh}\backslash \mcX$.
\eed
Given a mass shell, we define the corresponding single-particle subspace in a natural way:
\bed\label{sp-subspace-def} Let $\mfh$ be a mass shell in the sense of Definition~\ref{mass-shell-definition}.
The corresponding single-particle subspace is given by $\hil_{\mfh}:=P(\mfh)\hil$.
\eed

Let us add some comments on Definition~\ref{mass-shell-definition}: 
Requirement (c) 
prevents the \emph{group velocity} $\nabla\Si$ from being constant on 
a set of momenta of non-zero Lebesgue measure. This condition ensures
that configurations of particles with distinct velocities form a dense subspace. We will make use of this fact in Subsection~\ref{wave-operators-construction}, see Lemma~\ref{velocity-supports-disjointness}.
We note that for $\De_{\mfh}=\whGa$, and $\Si$ a real-analytic function, condition (c) holds  if and only if $\Si$ is non-constant.
In fact, if (c) is violated, then, by analyticity, the second derivative of $\Si$
vanishes identically on  $\whGa$. Thus interpreting $\Si$ as a function on
$\real^d$, it is periodic in all variables and linear. This is only possible if 
$\Sigma$ is constant. We will use this observation in 
 the proof of Proposition~\ref{examples-proposition}.

For the construction of scattering states in Theorem~\ref{Haag-Ruelle} we only need to assume that the mass shell
is isolated. However, to show the Fock space structure of scattering states in Theorem~\ref{Haag-Ruelle-Fock} we need to assume in addition
that it is either regular or pseudo-relativistic. Regularity means that the momentum-velocity relation $p\mapsto \nabla\Si(p)$
is invertible almost everywhere, in particular $\Si$ has no flat directions.
In $d=1$ obviously every mass shell is regular, but for $d>1$ it is not
easy to verify this condition in models. Therefore we introduced the alternative class of pseudo-relativistic mass shells described in 3. 
of Definition~\ref{mass-shell-definition}. Property 3. can always be ensured at the cost of shrinking the set $\De_{\mfh}$ and is  compatible with flat directions of $\Si$.  We note that mass shells in relativistic theories with lower mass gaps always satisfy this property, which justifies the name
`pseudo-relativistic'.

To conclude this subsection, we discuss briefly
positive-energy wave packets describing the free evolution of a single particle with a dispersion relation 
$\De_{\mfh}\ni p \mapsto \Si(p)$ introduced in Definition \ref{mass-shell-definition}. Such a wave packet is given by
\begin{equation}
g_t(x):=(2\pi)^{-\fr{d}{2}}\int_{\whGa} dp\, \e^{-\i\Si(p)t+\i px} \wh g(p), \label{wave-packet}
\end{equation}
where $\wh g\in C^{\infty}(\whGa)$. Its velocity support is defined as
\begin{equation}
V(g):=\{\, \nabla\Si(p)\, |\, p\in\supp\,\wh g\,\}.
\end{equation}
Some parts of the following  proposition  have appeared already in \cite{Yarotsky:2004vp}.
It is a generalization of the (non)-stationary phase method to wave packets on a lattice.
\bep\label{norm-corollary} Let $\chi_+\in C_0^{\infty}(\real^d)$ be equal to one on $V(g)$ and vanish outside of a slightly larger set and let $\chi_-=1-\chi_+$. We write $\chi_{\pm,t}(x):=\chi_{\pm}(x/t)$. 
Then we have
\begin{align}
\|\chi_{+,t}g_t\|_1=O( t^d),\quad
\|\chi_{-,t}g_t\|_1=O(t^{-\infty}),\quad
\|g_t\|_1=O(t^d). \label{non-stationary}
\end{align}
Assuming in addition that $\supp\,\wh g\subset \Dep$ (i.e. $\det D^2\Si\neq 0$ on the support of $\wh g$) we have
\beqa
\sup_{x\in \Ga}|g_t(x)|=O( t^{-d/2} ), \quad \|g_t\|_1=O(t^{d/2}). \label{stationary}
\eeqa
\eep
\proof To prove~(\ref{non-stationary}) we use  decomposition~(\ref{measure-decomposition})
to express $g_t$ as a finite sum of functions to which the non-stationary phase method
(the Corollary of Theorem~XI.14 of \cite{Reed:1979aa}) applies. Thus
\beqa
|(\chi_{-,t}g_t)(x)|=O(\lan |x|+|t|\ran^{-\infty}), 
\eeqa
either because of the corollary or because $\chi_{-,t}$ vanishes in the cases the corollary does not apply.
It follows that $\|\chi_{-,t}g_t\|_1=O(t^{-\infty})$. Now since $\chi_+$ is compactly supported, we obtain
$\|\chi_{+,t} g_t\|_1=O(t^d)$. 

To prove~(\ref{stationary}) we use again  decomposition~(\ref{measure-decomposition}) in order
to apply the stationary phase method (Corollary of Theorem~XI.15 of \cite{Reed:1979aa}).
This gives $\sup_{x\in \Ga}|g_t(x)|=O(t^{-d/2} )$. By decomposing $g_t(x)=\chi_{+,t}(x)g_t(x)+\chi_{-,t}(x)g_t(x)$, 
  using $\|\chi_{-,t}g_t\|_1=O(t^{-\infty})$, $\sup_{x\in \Ga}|g_t(x)|=O( t^{-d/2} )$ 
and the fact that $\chi_+$ is compactly supported we get $ \|g_t\|_1=O(t^{d/2})$. \qed

\subsection{Haag-Ruelle creation operators}

In this subsection we will define and study properties of  certain `creation operators'  from $\pi(\mfa_{\aloc})$ 
which create elements of the single-particle subspace $\hil_{\mfh}=P(\mfh)\hil$ from the ground state vector $\Om$:
\bed\label{HR-creation-operators} Let $A^*\in \mfa_{\aloc}$ be  s.t. $\Sp_{A^*}\tau\subset (0,\infty)\times \wh\Ga$ is compact and $\Sp_{A^*}\tau\cap \Sp\, U\subset \mfh$.
Let $g_t$ be a wave packet given by (\ref{wave-packet}). 
We say that:
\begin{enumerate}
\item $B^*:=\pi(A^*)$  is a \emph{creation operator}.  $\Sp_{A^*}\tau=\Sp_{\pi^{-1}(B^*)}\tau$ will be called
the energy-momentum transfer of $B^*$.
\item $B^*_t(g_t):=\pi(\tau_t\circ \tau^{(d)}_{g_t}(A^*))$ is the \emph{Haag-Ruelle (HR) creation operator}.
\end{enumerate}
\eed
\nin Recall that $\mfa$ is simple so that $\pi$ is faithful, and thus 1. is well-defined. Furthermore, setting $B^*_t(x):=U(t,x)B^*U(t,x)^*$, we can equivalently write
\begin{equation}
B^*_t(g_t)=(2\pi)^{-\fr{d}{2}}\sum_{x\in \Ga}B^*_t(x)g_t(x),
\end{equation}
consistently with the notation $\pi(A^*)(g):=\pi(\tau_g^{(d)}(A^*))$ from Appendix~\ref{Notations}.

We also note that the map $t\mapsto B_t^*$ is smooth in the norm topology for any creation operator and the respective derivatives are again creation operators. Indeed, since $\Sp_{A^*}\tau$ is compact, we have that $A^* = \tau_f(A^*)$ for any $f\in S(\bbR\times\Gamma)$ such that $\widehat f$ is equal to $1$ on $\Sp_{A^*}\tau$. It follows that $\tau_t(A^*) = \tau_{f_t}(A^*)$, where $f_t(s,x) = f(s-t,x)$. Hence,
\begin{equation}
\pa^n_t\tau_t(A^*) = (-1)^n \tau_{\partial_s^n f_t}(A^*),
\end{equation}
and $t\mapsto \tau_t(A^*)$ is $C^\infty$. The claim follows, since $\pi$ is continuous.

By Proposition~\ref{Arveson-elementary},  $\Sp_{\tau_t\circ \tau^{(d)}_{g_t}(A^*)}\tau \subset \Sp_{A^*}\tau$
thus by the energy-momentum transfer relation (\ref{EM-transfer-relation}) 
\beqa
 B^*_t(g_t)\Om\in \hil_{\mfh}, \quad (B^*_t(g_t))^*\Om=0. \label{HR-EM-transfer}
\eeqa
On the other hand it follows from Theorem~\ref{smearing-theorem} that $\tau_t\circ \tau_{g_t}^{(d)}(A^*)\in \mfa_{\aloc}$,
thus  $B^*_t(g_t)$ creates a well-localized excitation (particle) from the vacuum. The expression $ \tau_t\circ \tau^{(d)}_{g_t}$
amounts to comparing the interacting forward evolution $\tau_t$ with the free backward evolution $\tau_{g_t}^{(d)}$. 
The goal of scattering theory is to show that they match at asymptotic times. In part (a) of the next lemma we show that 
at the single-particle level these two evolutions match already at finite times. 
\bel\label{HR-auxiliary} Let $B^*_t(g_t)$ be a HR creation operator and $\chi_{\pm}$ be as in Proposition~\ref{norm-corollary}. We have
\begin{enumerate}
\item[(a)] $B^*_t(g_t)\Om=B^*(g)\Om=P(\mfh) B^*(g)\Om$,
\item[(b)] $\pa_t B^*_t(g_t)= \dot{B}^*_t(g_t)+B^*_t(\dot{g}_t)$,
\item[(c)] $\|B^*_t(g_t) P(\De)\|=O(1)$ for any compact $\De\subset \Sp\, U$,
\item[(d)] $\|B^*_t(g_t)\|, \|B^*_t(\chi_{+,t}g_t)\|=O(t^d)$, $\|B^*_t(\chi_{-,t}g_t)\|=O(t^{-\infty})$,
\end{enumerate}
where  in (b) $\dot{B}^*$ (resp. $\dot{g}_t$) is again a creation operator
(resp. a wave packet).
\eel
\proof 
Let us show (a): Making use of (\ref{SNAG}) and (\ref{HR-EM-transfer}) we get
\beqa
B^*_t(g_t)\Om
&=(2\pi)^{-\fr{d}{2}}\sum_{x\in \Ga}  g_t(x) U(t,x)B^*\Om\non\\
&=(2\pi)^{-\fr{d}{2}}\sum_{x\in  \Ga}  g_t(x) \int_{\mfh} \e^{\i E t-\i px} dP(E,p) P(\mfh)B^*\Om
\non\\
&= \int_{\mfh}  \e^{\i (E-\Si(p) )t }\wh g(p) dP(E,p) P(\mfh)B^*\Om=B^*(g)\Om. \label{creation-on-vacuum}
\eeqa
Here in the third step we applied the Fourier inversion formula (cf. Appendix~\ref{Notations}). 
In the last step we used that $E=\Si(p)$ in the region of integration and thus the expression
is $t$-independent. Then we reversed the steps to conclude the proof of (a).
 
Part (b) is an  obvious computation, which is legitimate since  $t\mapsto B_t^*$ is smooth in norm as we argued above.  

Part (c) is a consequence of Theorem~\ref{harmonic-theorem}: Let $\Psi,\Phi\in \hil$ be unit vectors, with $\Phi\in \Ran P(\De)$.
Then by the Cauchy-Schwarz inequality
\beqa
|\lan \Psi, B_t^*(g_t)\Phi\ran|\leq (2\pi)^{-\fr{d}{2}}\bigg(\sum_{x\in \Ga}|\lan \Psi, B_t^*(x)\Phi\ran|^2\bigg)^\h 
 \bigg( \sum_{x\in \Ga} |g_t(x)|^2 \bigg)^\h.
\eeqa
By Parseval's theorem, the second factor on the r.h.s. above  is a time-independent constant. As for the first factor, we have
\beqa
\sum_{x\in \Ga}|\lan \Psi, B_t^*(x)\Phi\ran|^2\leq \sum_{x\in \Ga}\lan \Psi, P(\De')B_t^*(x)B_t(x)P(\De')\Psi\ran
\leq \| P(\De')\sum_{x\in \Ga}B^*(x)B(x)P(\De')\|, \label{finite-expression}
\eeqa
where in the first step we made use of the fact that $B^*=\pi(A^*)$, where $\Sp_{A^*}\tau$ is compact, 
and of the energy-momentum transfer
relation~(\ref{EM-transfer-relation}), to introduce the projection on a compact set $\De'\supset (\Delta + \Sp_{A^*}\tau)\cap\Sp\, U$.  Since $A^*$ is almost local, and 
\beqa
\Sp_{A}\tau=-\Sp_{A^*}\tau\subset  (-\infty, 0)\times \wh\Ga,
\eeqa
the expression on the r.h.s. of (\ref{finite-expression}) is finite by Theorem~\ref{harmonic-theorem}.  

Part (d) of the lemma follows from Proposition~\ref{norm-corollary} and the obvious estimate 
$\|B^*(g')\|\leq \|B^*\|\|g'\|_1$, $g'\in L^1(\Ga)$. \qed\\
The next lemma concerns the decay of commutators of HR creation operators associated with 
wave packets with disjoint velocity supports. 
\bel\label{commutators-decay} Let $B_{1,t}^*(g_{1,t})$, $B_{2,t}^*(g_{2,t})$, $B_{3,t}^*(g_{3,t})$ be HR creation operators
s.t. $V(g_1)\cap V(g_2)=\emptyset$ and $V(g_3)$ arbitrary. Then
\begin{enumerate}
\item[(a)] $[ B_{1,t}^*(g_{1,t}), B_{2,t}^*(g_{2,t})]=O(t^{-\infty})$,
\item[(b)] $[ B_{1,t}^*(g_{1,t}), [ B_{2,t}^*(g_{2,t}), B_{3,t}^*(g_{3,t})]]=O(t^{-\infty})$.
\end{enumerate}
The statements remains valid if some of the HR creation operators are replaced with
their adjoints.
\eel
\proof To prove (a) we decompose $g_{i,t}=\chi_{i,+}g_{i,t}+\chi_{i,-} g_{i,t}$, $i=1,2$, where $\chi_{i,\pm}$ appeared in 
Proposition~\ref{norm-corollary}. Then, by Lemma~\ref{HR-auxiliary}, we get
\beqa
[ B_{1,t}^*(g_{1,t}), B_{2,t}^*(g_{2,t})]=[ B_{1,t}^*(\chi_{1,+,t}g_{1,t}), B_{2,t}^*(\chi_{2,+,t}g_{2,t})]+O(t^{-\infty}).
\eeqa
Now for each $n$ there is a constant $c_n > 0$ such that the following estimate holds,
\beqa
\|[B^{*}_{1,t}(\chi_{1,+,t}g_{1,t}), B^{*}_{2,t}&(\chi_{2,+,t}g_{2,t})]\| = \left\|
\sum_{x_1, x_2 \in \Gamma} \chi_{1,+,t}g_{1,t}(x_1) \chi_{2,+,t}g_{2,t}(x_2) [ B_{1,t}^*(x_1), B_{2,t}^*(x_2) ] \right\| \non\\
&\leq \sum_{x_1,x_2\in \Ga} |\chi_{1,+}(x_1/t)| |\chi_{2,+}(x_2/t)|\fr{c_n}{\lan x_1-x_2\ran^n}
, \label{commutator-bound}
\eeqa 
where in the second line the existence of $c_n$ follows from almost locality, Lemma~\ref{commutator-decay}, and the fact that $|g_{i}(t,x)|=O(1)$ uniformly in $x$.
Since $\chi_{1,+},\chi_{2,+}$ are approximate characteristic functions of $V(g_1), V(g_2)$, they may be chosen with compact, disjoint
supports. Therefore, by changing variables to $y_1:=x_1/t, y_2:=x_2/t$ we obtain that (\ref{commutator-bound}) is $O(t^{2d-n})$ and hence
$O(t^{-\infty})$ since $n\in \nat$ is arbitrary.

To prove (b) we decompose $\wh g_3(p)=\wh g_{3,1}(p)+\wh g_{3,2}(p)$, using a smooth partition of unity, s.t. $V(g_{3,1})\cap V(g_1)=\emptyset$
and $ V(g_{3,2})\cap V(g_2)=\emptyset$. Now the statement follows from part (a) and the Jacobi identity. \qed\\
The next lemma is a counterpart of the relation $a(f)a^*(f)\Om=\Om\lan \Om, a(f)a^*(f)\Om\ran$ which holds for the usual creation
and annihilation operators on the Fock space. We will need it to establish the Fock space structure of scattering states. This is the only result of this
subsection which requires that the mass shell is pseudo-relativistic or
regular.
\bel\label{clustering-lemma} Let $B^*_{1,t}(g_{1,t})$, $B^*_{2,t}(g_{2,t})$ be HR creation operators. 
\begin{enumerate}
\item[(a)] If $\mfh$ is pseudo-relativistic, and the energy-momentum transfers of $B_1^*$, $B_2^*$ are contained in a sufficiently  
small neighbourhood of $\mfh$ then
\beqa
(B^*_{1,t}(g_{1,t}))^*B^*_{2,t}(g_{2,t})\Om=\Om\lan \Om, (B^*_{1,t}(g_{1,t}))^*B^*_{2,t}(g_{2,t})\Om\ran.
\eeqa

\item[(b)] If $\mfh$ is regular and $\supp\, \wh g_1, \supp\, \wh g_2\subset \Dep$, then
\beqa
(B^*_{1,t}(g_{1,t}))^*B^*_{2,t}(g_{2,t})\Om=\Om\lan \Om, (B^*_{1,t}(g_{1,t}))^*B^*_{2,t}(g_{2,t})\Om\ran+O(t^{-d/2}). \label{clustering-b}
\eeqa \end{enumerate}
\eel
\proof Part (a) follows immediately from $(\mfh-\mfh)\cap \Sp\, U=\{0\}$ and the energy-momentum transfer relation~(\ref{EM-transfer-relation}).

To prove (b) we will use the clustering property stated in Theorem~\ref{clustering} and a similar strategy as in \cite[p.169-171]{Bu77}:
Let $A_i\in \mfa_{\aloc}$, $i=1,2,3,4$, $A_i(x_i):=\tau_{x_i}(A_i)$ and $A_{i,r}\in \mfa(\B^r)$ be s.t. $\|A_i-A_{i,r}\|=O(r^{-\infty})$.
Now we write $P(\{0\})^{\bot}=1-|\Om\ran\lan\Om|$ and consider the function
\beqa
F(x_1,x_2,x_3,x_4):=\lan\Om, \pi([A_1(x_1), A_2(x_2)])P(\{0\})^{\bot} \pi([A_3(x_3), A_4(x_4)])\Om\ran. 
\eeqa 
We will show that this function is rapidly decreasing in relative variables, i.e.
\beqa
F(x_1,x_2,x_3,x_4)=O(\lan x_1-x_2\ran^{-\infty}) O(\lan x_3-x_4\ran^{-\infty}) O(\lan x_2-x_3\ran^{-\infty}).  \label{full-decay}
\eeqa
By almost locality of $A_i$ and Lemma~\ref{commutator-decay}, we obtain rapid decay in $x_1-x_2$ and $x_3-x_4$. Thus it suffices
to show 
\beqa
F(x_1,x_2,x_3,x_4)=O(\lan x_2-x_3\ran^{-\infty}). \label{required-decay}
\eeqa
 To this end, we first write
\beqa
F(x_1,x_2,x_3,x_4):=\lan\Om, \pi([A_{1,r}(x_1), A_{2,r}(x_2)])P(\{0\})^{\bot} \pi([A_{3,r}(x_3), A_{4,r}(x_4)])\Om\ran+O(r^{-\infty}). \label{cluster-prep}
\eeqa
Denoting the first term on the r.h.s. of (\ref{cluster-prep}) by $F_r(x_1,x_2,x_3,x_4)$, we obtain from Theorem~\ref{clustering}:
\beqa
|F_r(x_1,x_2,x_3,x_4)|\leq C\chi(X_1^r\cap X_2^r\neq\emptyset)\chi(X_3^r\cap X_4^r\neq \emptyset)
\max_{i=1,\ldots,4}| X^r_i| 
\e^{-\mu\dist(X_1^r\cup X_2^r ,X_3^r\cup X_{4}^r)},
\eeqa
where we set $X_{i}^r:=\B^r+x_i=\{x_i\}^r$ and $\chi(\mathrm{K})$ is equal to one if the condition $\mathrm{K}$ is satisfied and zero otherwise.
$X_{1}^r\cap X_2^r\neq \emptyset$ and $X_3^r\cap X_4^r\neq \emptyset$ imply that 
\beqa
|x_1-x_2|\leq 2r, \quad  |x_3-x_4|\leq 2r.  \label{trivial-inequalities}
\eeqa
Moreover, 
\beqa
\dist(X_1^r\cup X_2^r ,X_3^r\cup X_{4}^r)&=\inf\{\,|w_a-w_b|\,|\, w_a\in X_1^r\cup X_2^r, w_b\in X_3^r\cup X_{4}^r\,\}\non\\
&=\inf_{\substack{m\in 1,2 \\ n\in 3,4}} \inf\{\,|w_a-w_b|\,|\, w_a\in  X_m^r, w_b\in X_n^r\,\}\non\\
&\geq \inf_{\substack{m\in 1,2 \\ n\in 3,4}} |x_m-x_n| -2r\non\\
&\geq |x_2-x_3|-6r,
\eeqa
where in the last step we made use of (\ref{trivial-inequalities}). 
Thus noting that $|X^r_i|=|\B^r|\leq (2r)^d$ we can write
\beqa
|F_r(x_1,x_2,x_3,x_4)|\leq C (2r)^d \e^{6\mu r} \e^{-\mu |x_2-x_3|}.
\eeqa 
Setting $r=\eps\lan x_2-x_3\ran$, $\eps>0$ sufficiently small, and making use (\ref{cluster-prep}) we obtain (\ref{required-decay})
and therefore (\ref{full-decay}).

Now we are ready to prove (\ref{clustering-b}). Since $(B^*_{1,t}(g_{1,t}))^*\Om=0$, we have
\beqa
\|P(\{0\})^{\bot}(B^*_{1,t}(g_{1,t}))^*B^*_{2,t}(g_{2,t})\Om\|^2\leq \sum_{x_1,\ldots,x_4\in \Ga} |F(x_1,x_2,x_3,x_4)| 
|g_{1,t}(x_1)g_{2,t}(x_2)g_{1,t}(x_3)g_{2,t}(x_4)|.
\eeqa
By introducing relative variables $x_{i,j}=x_i-x_j$, and using (\ref{full-decay}), we get
\beqa
\|P(\{0\})^{\bot}(B^*_{1,t}(g_{1,t}))^*B^*_{2,t}(g_{2,t})\Om\|^2\leq& t^{-d}\sum_{x_{1,2},x_{2,3},x_{3,4}\in \Ga} O(\lan x_{1,2}\ran^{-\infty})
O(\lan x_{2,3}\ran^{-\infty})O(\lan x_{3,4}\ran^{-\infty}).
\eeqa
Here we exploited  regularity of the mass shell
which gives $\|g_{i,t}\|_1=O(t^{d/2})$ and  $|g_{i,t}(x)|=O( t^{-d/2} )$ uniformly in $x\in \Ga$. (We applied the former estimate
to $g_{1,t}(x_1)$, and the latter to the remaining three wave packets).  
This concludes the proof. \qed
\subsection{Scattering states and their Fock space  structure} \label{Scattering-states-existence}

In this subsection we prove the  main results of this paper which are the existence of Haag-Ruelle scattering states
for lattice systems (Theorem~\ref{Haag-Ruelle}) and their Fock space structure (Theorem~\ref{Haag-Ruelle-Fock}). 
Using these two results we will define the wave operators and the scattering matrix in the next subsection.
The method of proof of the next theorem follows \cite{Ar99}.
\bet\label{Haag-Ruelle} Let $B_1^*,\ldots, B^*_{n}$  be creation operators and  $g_1,\ldots, g_n$ be
positive energy wave packets with disjoint velocity supports. Then  there exists the  $n${\em -particle scattering state} given by
\beqa
\Psi^{\out}=\lim_{t\to\infty} B^{*}_{1,t}(g_{1,t})\ldots B^{*}_{n,t}(g_{n,t})\Om. \label{scattering-state}
\eeqa
Moreover,  let $S_n$ be the set of permutations of an $n$-element set.  Then, for any $\sigma\in S_n$, 
\begin{equation}\label{Asymptotic symmetry}
\Psi^{\out} = \lim_{t\to\infty} B^{*}_{\sigma(1),t}(g_{\sigma(1),t})\ldots B^{*}_{\sigma(n),t}(g_{\sigma(n),t})\Om.
\end{equation}
Finally, let $\ti B^*_{i,t}(\ti g_i)$, $i=1,\ldots, n$ be HR creation operators  satisfying
the same assumptions and let $\ti \Psi^{\out}$ be the corresponding scattering state. If $\ti B^*_{i,t}(\ti g_i)\Om=B^{*}_{i,t}(g_{i,t})\Om$, for all $i=1,\ldots, n$,  and $V(\tilde g_i)\cap V(g_j) = \emptyset$ for all $i\neq j$, then $\ti \Psi^{\out} = \Psi^{\out}$.
\eet

\proof  To prove the existence of scattering states by Cook's method it suffices to
show that the derivative
\begin{align}
 \pa_{t}(B_{1,t}^{*}(g_{1,t})\ldots B^*_{n,t}(g_{n,t}))\Omega= \sum_{i=1}^n B_{1,t}^{*}(g_{1,t}) \ldots \pa_t( B_{i,t}^*(g_{i,t}) ) \ldots B^*_{n,t}(g_{n,t})\Omega   
 \end{align}
is integrable in norm. Due to Lemma~\ref{HR-auxiliary}~(a),  $\pa_t( B_{i,t}^*(g_{i,t}) )$ annihilates $\Om$.
Thus we can commute this expression to the right until it acts on $\Om$ and it is enough to show that
 the resulting terms with the commutators are $O(t^{-\infty})$. This follows from  Lemma~\ref{commutators-decay} (a)
and Lemma~\ref{HR-auxiliary} (b), (d).

Since any permutation can be decomposed into a product of adjacent transpositions, the estimate
\begin{equation}
B^{*}_{1,t}(g_{1,t})\ldots B^{*}_{i,t}(g_{i,t})B^{*}_{i+1,t}(g_{i+1,t})\ldots \Om = B^{*}_{1,t}(g_{1,t})\ldots B^{*}_{i+1,t}(g_{i+1,t})B^{*}_{i,t}(g_{i,t})\ldots \Om + O(t^{-\infty}),
\end{equation}
which holds by Lemma~\ref{commutators-decay} (a) and  Lemma~\ref{HR-auxiliary} (d), is sufficient to prove the symmetry of $\Psi^{\out}$.

Finally, by iterating the relation
\beqa
 B^{*}_{1,t}(g_{1,t})\ldots B^{*}_{n,t}(g_{n,t})\Om&= B^{*}_{1,t}(g_{1,t})\ldots  B^{*}_{n-1,t}(g_{n-1,t})\ti B^{*}_{n,t}(\ti g_{n,t})\Om\non\\
&=\ti B^{*}_{n,t}(\ti g_{n,t})B^{*}_{1,t}(g_{1,t})\ldots  B^{*}_{n-1,t}(g_{n-1,t})\Om+O(t^{-\infty}),
\eeqa  
which follows again from  Lemma~\ref{commutators-decay} (a) and Lemma~\ref{HR-auxiliary} (d), and taking the limit $t\to\infty$,
we obtain that $\Psi^{\out}$ coincides with the scattering state $\ti \Psi^{\out}$. \qed

\nin By the last part of Theorem~\ref{Haag-Ruelle}, the scattering state $\Psi^{\out}$ depends only on the time independent single-particle vectors $\Psi_i=B^{*}_{i,t}(g_{i,t})\Om$, and possibly the velocity supports $V(g_i)$. The latter dependence can easily be excluded making use of 
formula~(\ref{scalar-product}) below. It easily follows from formula~(\ref{scalar-product}) below. Anticipating this fact we will write
$\Psi^{\out}=:\Psi_1\timeso\cdots\timeso\Psi_n$.  
\bet\label{Haag-Ruelle-Fock} Suppose that the mass shell is pseudo-relativistic.
Let  $\Psi^{\out}$ and $\tilde \Psi^{\out}$ be two scattering states defined as in Theorem~\ref{Haag-Ruelle} with $n$ and $\ti n$ particles, respectively. Then
\beq
\lan \tilde \Psi^{\out},\Psi^{\out}\ran&=\de_{n,\ti n}\sum_{\si\in S_n}\lan \tilde \Psi_1,  \Psi_{\si_1}\ran\ldots \lan \tilde \Psi_n,  \Psi_{\si_n}\ran,      \label{scalar-product}\\
U(t,x)(\Psi_1\timeso\cdots \timeso\Psi_n)&=(U(t,x)\Psi_1)\timeso\cdots\timeso (U(t,x)\Psi_n), \quad (t,x)\in \real\times \Ga, \label{energy-factorization-relation}
\eeq
where $S_n$ is the set of permutations of an $n$-element set.

If the mass shell is regular, the same conclusion holds if $\supp\, \wh g_i, \supp\,\wh{(\tilde g_i)}$, $i=1,\ldots,n$ are
supported in $\Dep$.
\eet
\begin{remark} We stress that in (\ref{scalar-product}) we only need  $V(g_i)\cap V(g_j)=\emptyset$ and $V(\ti g_i)\cap V(\ti g_j)=\emptyset$
for $i\neq j$, where $g_i, \ti g_i$ enter the definition of $\Psi^\out, \ti\Psi^\out$ as specified in (\ref{scattering-state}).  $V( g_i)$ and
$V(\ti g_j)$ may overlap.
\end{remark}
\proof \def\tB{{\tilde{B}}} We first consider the case $n=\ti n$. 
To prove (\ref{scalar-product}), we set for simplicity of notation $B_{i}(t):= (B_{i,t}^*(g_{i,t}))^*$, $\tilde{B}_{j}(t):= (\tilde{B}^*_{j,t}(\tilde{g}_{j,t} ))^*$
and denote by $\Psi_t, \tilde \Psi_t$ the approximating sequences of $\Psi^{\out}, \tilde\Psi^{\out}$. 
We assume that (\ref{scalar-product}) holds for $n-1$ and compute
\beqa
\label{prod-x}
\lan\tilde{\Psi}_{t}, \Psi_{t}\ran&=\lan\Om,\tB_n(t)\ldots \tB_1(t) B_1(t)^*\ldots B_n(t)^*\Om\ran\non\\ 
&=\sum_{k=1}^n\lan\Om,\tB_n(t)\ldots \tB_2(t) B_1(t)^*\ldots[\tB_1(t),B_k(t)^*]\ldots B_n(t)^*\Om\ran\non\\ 
&=\sum_{k=1}^n\sum_{l=k+1}^n\lan\Om,\tB_n(t)\ldots \tB_2(t) B_1(t)^*\ldots \check{k} \ldots [[\tB_1(t),B_k(t)^*], B_l(t)^*]\ldots B_n(t)^*\Om\ran\non\\ 
& \ \ +\sum_{k=1}^n\lan\Om,\tB_n(t)\ldots \tB_2(t) B_1(t)^*\ldots \check{k} \ldots B_n(t)^*  \tB_1(t)B_k(t)^*\Om\ran,
\eeqa
where $\check{k}$ indicates that $B_k(t)^*$ is omitted from the product.
The terms involving double commutators vanish in the limit $t\to\infty$ by Lemma~\ref{commutators-decay} (b) and Lemma~\ref{HR-auxiliary}~(d).
To treat the last term on the r.h.s. of (\ref{prod-x}) we have to consider two cases.

For the first case, suppose that the mass shell is pseudo-relativistic. Recall that $B^*=\pi(A^*)$ and set $B^*_f:=\pi(\tau_{f}(A^*))$, $f\in S(\real\times\Ga)$.
Choosing $\wh f$ supported in a small neighbourhood $O\subset \whGa$ of a subset of $\mfh$, we obtain that the energy-momentum transfer of 
$B^*_{f}$ is contained in $O$ (cf. Proposition~\ref{Arveson-elementary}). Demanding in addition that
 $\wh f$ is equal to one on $\mfh\cap \Sp_{A^*}\tau$  we can ensure that $B^*_{f,t}(g_t)\Om=B_t^*(g_t)\Om$ (cf. the proof of Lemma~\ref{HR-auxiliary} (a)).
Thus, in view of the last part of Theorem~\ref{Haag-Ruelle}, we can assume without loss that all the HR creation operators involved in (\ref{prod-x})
satisfy the assumptions of Lemma~\ref{clustering-lemma} (a). Then the last term on the r.h.s. of (\ref{prod-x}) factorizes as follows
\beqa
\sum_{k=1}^n\lan \Om,\tB_n(t)\ldots \tB_2(t) B_1(t)^*\ldots \check{k} \ldots B_n(t)^* \Om\ran\lan\Om, \tB_1(t)B_k(t)^*\Om\ran
\eeqa
by Lemma~\ref{clustering-lemma} (a). Now by the induction hypothesis the expression above factorizes in the limit $t\to\infty$
and gives (\ref{scalar-product}).

For the other case, let us assume that the mass shell is regular. By the energy-momentum transfer relation (\ref{EM-transfer-relation}) 
and Lemma~\ref{HR-auxiliary} (c) we obtain that 
\beqa
|\Om\ran\lan \Om|\tB_n(t)\ldots \tB_2(t) B_1(t)^*\ldots \check{k} \ldots B_n(t)^*=O(1). \label{O-one-estimate}
\eeqa
Consequently, by Lemma~\ref{clustering-lemma} (b), the last term on the r.h.s. of (\ref{prod-x}) equals
\beqa
\sum_{k=1}^n\lan\Om,\tB_n(t)\ldots \tB_2(t) B_1(t)^*\ldots \check{k} \ldots B_n(t)^* \Om\ran\lan\Om,  \tB_1(t)B_k(t)^*\Om\ran +O(t^{-d/2}).
\eeqa
Now the inductive argument is concluded as in the previous case.

If $n\neq \ti n$, a similar argument implies that the scalar product is zero, since the reduction will yield eventually an expectation value of a product of either only creation or only annihilation operators, which vanishes. We omit the details. 

This gives (\ref{scalar-product}). Relation~(\ref{energy-factorization-relation}) follows from  (\ref{scalar-product}) and Lemma~\ref{HR-auxiliary} (a). \qed
\begin{remark} In estimate~(\ref{O-one-estimate}) above we made crucial use of the bound from Lemma~\ref{HR-auxiliary} (c),
which relies on Theorem~\ref{harmonic-theorem}. Weaker bounds from  Lemma~\ref{HR-auxiliary} (d) would clearly not suffice to
conclude the proof along the above lines. Without this sharper bound available, in the regular case  we would have to proceed via cumbersome `truncated vacuum expectation
values' of arbitrary order, which burden all the textbook presentations of the Haag-Ruelle theory known to us  (see e.g. \cite{Ar99,Reed:1979aa}). 
\end{remark}

\subsection{Wave operators and the $S$-matrix}
\label{wave-operators-construction}
Up to now we left aside the question if any of the scattering states $\Psi^{\out}$ is different from 
zero.   In this subsection we will demonstrate that scattering states in fact
span a subspace of $\hil$ which is naturally isomorphic to the symmetric Fock space over $\hil_{\mfh}$. 
This observation allows to define the wave operators and the $S$-matrix. 

We denote by $\Ga(\hil_{\mfh})$ the symmetric Fock space over the 
single-particle subspace $\hil_{\mfh}$ and by $a^*(f), a(f)$,  $f\in  \hil_\mfh$, the
corresponding creation and annihilation operators. We also define the unitary representation
of translations on the single-particle subspace
\beqa
U_{\mfh}(t,x)=U(t,x)|_{\hil_{\mfh}}.
\eeqa    
Its second quantization $(t,x)\mapsto \Ga(U_{\mfh}(t,x))$ is a unitary representation
of space-time translations on $\Ga(\hil_{\mfh})$.
\bed\label{wave-definition} 
We say that an isometry $W^{\out}:\Ga(\hil_{\mfh})\to \hil$ is an \emph{outgoing wave operator} if
\beqa
&W^{\out}\Om=\Om, \\
&W^{\out}(a^*(\Psi_1)\ldots a^*(\Psi_n)\Om)=\Psi_1\timeso\cdots\timeso\Psi_n, \\
&U(t,x)\circ W^{\out}=W^{\out}\circ \Ga(U_{\mfh}(t,x)),
\eeqa
for any collection of $\Psi_1,\ldots, \Psi_n\in \hil_{\mfh}$ as in Theorem~\ref{Haag-Ruelle}  and any $(t,x)\in \real\times \Ga$.

An \emph{incoming wave operator} $W^{\inc}$ is defined analogously, by taking the limit $t\to-\infty$ in Theorem~\ref{Haag-Ruelle}.
The \emph{$S$-matrix} is an isometry on $\Ga(\hil_{\mfh})$ given by
\beqa
S=(W^{\out})^*W^{\inc}.
\eeqa
\eed
\nin The main result of this section is the following theorem:
\bet If the mass shell $\mfh$ is either pseudo-relativistic or  regular then the wave operators and the $S$-matrix exist and are unique.
\label{s-matrix}
\eet
\proof We proceed similarly as in the proof of Proposition 6.6 of \cite{DG13}.
Let $\Ga^{(n)}(\hil_{\mfh})$ be the $n$-particle subspace of $\Ga(\hil_{\mfh})$
and let $\mcF\subset \Ga^{(n)}(\hil_{\mfh})$ be the subspace spanned by vectors $a^*(\Psi_1)\ldots a^*(\Psi_n)\Om$,
$\Psi_i$ as in Theorem~\ref{Haag-Ruelle}. 
Due to (\ref{Asymptotic symmetry},\ref{scalar-product}), there exists a unique isometry $W_n^{\out}:\mcF\to \hil$
s.t. 
\beqa
W_n^{\out}(a^*(\Psi_1)\ldots a^*(\Psi_n)\Om)=\Psi_1\timeso\cdots\timeso\Psi_n.
\eeqa
By (\ref{energy-factorization-relation}) it satisfies
\beqa
U(t,x)\circ W_n^{\out}=W_n^{\out}\circ \Ga^{(n)}(U_{\mfh}(t,x)),
\eeqa
where $\Ga^{(n)}(U_{\mfh}(t,x))$ is the restriction of $\Ga(U_{\mfh}(t,x))$ to $\Ga^{(n)}(\hil_{\mfh})$.
Thus it suffices to prove that $\mcF$ is dense in $\Ga^{(n)}(\hil_{\mfh})$. Indeed, in that case $W_n^{\out}$ extends uniquely to an isometry $W_n^{\out}:\Ga^{(n)}(\hil_{\mfh}) \to \hil$, and $W^{\out}:= \oplus_{n\in \bbN_0}W_n^{\out}$ is the unique outgoing wave operator in the sense of Definition~\ref{wave-definition}. The incoming wave operator $W^{\inc}$ is defined analogously, by taking limits $t\to-\infty$
in Theorems~\ref{Haag-Ruelle} and~\ref{Haag-Ruelle-Fock}.

Let us now show density of $\mcF$. 
We denote by $P_{\mfh}$ the spectral measure of $U_{\mfh}$  and  define
the product spectral measure on $(\real\times \whGa)^{\times n}$ with values in $\hil_{\mfh}^{\otimes n}$ (the non-symmetrized $n$-particle subspace):
\beqa
d\tilde P_{\mfh}((E_1,p_1),\ldots, (E_n,p_n)):=dP_{\mfh}(E_1,p_1)\otimes \cdots \otimes dP_{\mfh}(E_n,p_n).
\eeqa
Clearly, $\tilde P_{\mfh}$ is supported in $\mfh^{\times n}$ and $\ti P_{\mfh}(\mfh^{\times n})\hil_{\mfh}^{\otimes n} = \hil_{\mfh}^{\otimes n}$. Using Lemma~\ref{single-particle-density-two} below, it is easy to see that
\beqa
\ov{\mcF}=\Theta_{\mathrm s}\circ \ti P_{\mfh}(\mfh^{\times n}\backslash D)\hil_{\mfh}^{\otimes n}, \label{expression-for-F}
\eeqa 
where $\Theta_{\mathrm s}: \hil_{\mfh}^{\otimes n} \mapsto \Ga^{(n)}( \hil_{\mfh} )$ is the orthogonal projection
on the subspace of symmetrized $n$-particle vectors and
\beqa
D:=\{\, ((\Si(p_1), p_1),\ldots, (\Si(p_n),p_n)) \in \mfh^{\times n}\,|\, \nabla\Si(p_i)=\nabla\Si(p_j) \textrm{ for some } i\neq j \,\}. 
\eeqa
We have to subtract $D$ in (\ref{expression-for-F}) to account for the disjointness of velocity supports of vectors in $\mcF$.
By Lemma~\ref{velocity-supports-disjointness} below, the auxiliary set
\beqa
D_0:=\{\, p_1,\ldots, p_n \in \De_{\mfh}\,|\, \nabla\Si(p_i)=\nabla\Si(p_j) \textrm{ for some } i\neq j \,\}
\eeqa
has Lebesgue measure zero, which implies finally that $\ti P_{\mfh}(D)=0$ by Lemma~\ref{Lebesgue-absolute-continuity} below. With this and equation~(\ref{expression-for-F}), the proof is complete. \qed

\begin{remark} Although the construction of asymptotic states requires disjoint velocity supports, Theorem~\ref{s-matrix} shows that, up to isometry, their span is dense in the bosonic Fock space. In particular, $n$-particle product states, or condensates, can be obtained as limits of $n$-particle asymptotic states with disjoint velocity supports.
\end{remark}

We now state and prove the lemmas used in the proof of the above theorem.
\bel\label{single-particle-density} Let $\De\subset \mfh$ be an open bounded set and let $\mcK$  be a subspace of $\hil$
spanned by vectors of the form $B^*\Om$,  where $B^*$ is a creation operator in the sense of Definition~\ref{HR-creation-operators}
s.t. $\Sp_{\pi^{-1}(B^*)}\tau\cap\mfh\subset \De$.
Then $\mcK$ is dense in $P(\De)\hil$.
\eel 
\proof  We pick an arbitrary $A^*\in \mfa_{\loc}$ and $f\in S(\real\times \Ga)$ s.t.
$\supp\, \wh f\cap \Sp\, U\subset \De$. Then $B^*=\pi(\tau_f(A^*))$ is a creation
operator  as specified in the lemma and 
\beqa
B^*\Om=\int_{\De} \wh f(E,p) dP(E,p)\pi(A^*)\Om\in P(\De)\hil_{\mfh}.
\eeqa
Approximating with $\wh f$ the characteristic function of $\De$ and exploiting cyclicity of $\Om$ we conclude the proof. \qed
\bel\label{Lebesgue-absolute-continuity} Let $O \subset \De_{\mfh}$ and $\mfh_{O}:=\{\,(\Si(p), p)\in\mfh\,|\, p\in O  \}$.
If $O$ has  Lebesgue measure zero then $P(\mfh_{O})=0$.
\eel
\proof  We proceed similarly as in the first part of the proof of Proposition 2.2 from \cite{BF82}. 
From Lemma~\ref{single-particle-density}  we know that 
vectors of the form $B^*\Om$, where $B^*$ is a creation operator in the sense of Definition~\ref{HR-creation-operators},
span a dense subspace of $\hil_{\mfh}$. By formula~(\ref{SNAG}), we can write
\beqa
\lan B^*\Om,P(\mfh_O)B^*\Om\ran&=
\lim_{n\to\infty} \int_{\mfh} \wh\chi_{n, \mfh_{O}}(E,p)\lan B^*\Om,dP(E,p) B^*\Om\ran\non\\
&=\lim_{n\to\infty} \int \wh\chi_{n, \mfh_{O}}(\Si(p),p)\lan B^*\Om,dP(E,p) B^*\Om\ran,\label{spectral-measure-xx}
\eeqa
where $\chi_{n,\mfh_{O}}\in S(\real\times \Ga)$ and $\wh\chi_{n,\mfh_{O}}$ approximate the characteristic function of
$\mfh_O$ pointwise.  
We set $\wh\chi_{n,O}(p):=\wh\chi_{n, \mfh_{O}}(\Si(p),p)$, which  approximates pointwise
the characteristic function of $O$. We get from (\ref{spectral-measure-xx})
\beqa
\int \wh\chi_{n, \mfh_{O}}(\Si(p),p)\lan B^*\Om,dP(E,p) B^*\Om\ran&=(2\pi)^{-\fr{d}{2}}\sum_{x\in \Ga} \, \chi_{n,O}(x)
\lan B^*\Om,U(x) B^*\Om\ran \non\\
&=(2\pi)^{-\fr{d}{2}}\sum_{x\in \Ga} \, \chi_{n,O}(x)
\lan \Om,[B,  B^*(x)]\Om\ran \non\\
&=(2\pi)^{-\fr{d}{2}}\int_{\whGa} \,\wh\chi_{n,O}(p) \wh{f}(p) dp,
\eeqa
where we used the translation invariance of $\Omega$ and $B \Om = 0$ in the second equality. In the last line,  $\overline{f(x)}:=\lan \Om,[B,  B^*(x)]\Om\ran$ is a rapidly decreasing function by Lemma~\ref{commutator-decay}
and in the last step we made use of Parseval's theorem. Since $\wh\chi_{n,O}$ converges pointwise to a characteristic function of a set of Lebesgue measure zero, we have shown $\Vert P(\mfh_O)B^*\Om \Vert = 0$ and the proof is complete. \qed
\bel\label{single-particle-density-two} Let $\De\subset \mfh$ be an  open bounded set and $\mcL$  be a subspace of $\hil_{\mfh}$
spanned by vectors of the form $B_t^*(g_t)\Om$. Here $B_t^*(g_t)$ is a HR creation operator such that $\supp\, \wh g\subset \De_{\mfh}\backslash\mcX_0$, where $\mcX_0$ is a set of Lebesgue measure zero and  $\{\, (\Si(p),p)\,|\, p\in \supp\,\wh g\,\}\subset \De$. Then  $\mcL$  is dense in $P(\De)\hil_{\mfh}$.
\eel 
\begin{remark} If the mass shell is pseudo-relativistic (resp. regular) we use this lemma with $\mcX_0=\emptyset$ (resp.  $\mcX_0=\mcX$).
\end{remark}
\proof 
First, we note that by Lemma~\ref{Lebesgue-absolute-continuity} the set 
$\mfh_{\mcX_0}:=\{\,(\Si(p), p)\,|\, p\in \mcX_0 \}$ satisfies $P(\mfh_{\mcX_0})=0$,
thus we can assume that $\De$ does not intersect with $\mfh_{\mcX_0}$. 
Now we will argue by contradiction. Suppose that that there is a non-zero vector $\Psi\in  P(\De)\hil_{\mfh}$
which is orthogonal to all elements of $\mcL$.
Now we find $g$, as in the statement of the lemma,  s.t.
\beqa
\int_{\De} \ov{\wh g(p)} dP(E,p)\Psi\neq 0.
\eeqa
To this end we approximate with $\wh g$ the characteristic
function of $\De_0:=\{\, p\in \De_{\mfh}\,|\, (\Si(p),p)\in \De\,\}$ from inside.
This is possible since we assumed that $\mfh_{\mcX}$ does not intersect with $\De$.
Next, recall that $\mcK$, as defined in Lemma~\ref{single-particle-density}, 
consists of vectors of the form $B^*\Om$, where  $B^*$ is a creation operator s.t.  $\Sp_{\pi^{-1}(B^*)}\tau\cap\mfh\subset \De$.
Since $\mcK$ is dense in $P(\De)\hil_{\mfh}$, we can find  $B^*$, within the above restrictions, s.t.
\beqa
0\neq \lan B^*\Om ,\int_{\De} \ov{\wh g(p)} dP(E,p)\Psi\ran=\lan B^*_t(g_t)\Om,\Psi\ran.
\eeqa
But the last expression is zero by definition of $\Psi$. \qed

\nin Note that the claim of the lemma holds a fortiori if $\supp\, \wh g\subset \De_{\mfh}$.
\bel\label{velocity-supports-disjointness} Let $\Si$ be a dispersion relation of a mass shell (not necessarily pseudo-relativistic or regular, see Definition~\ref{mass-shell-definition}).
 Then the set 
\beqa
D_0:=\{\, p_1,\ldots, p_n \in \De_{\mfh}\,|\, \nabla\Si(p_i)=\nabla\Si(p_j) \textrm{ for some } i\neq j \,\}
\eeqa
has Lebesgue measure zero in $\wh\Ga^{\times n}$.
\eel
\proof Note that $D_0$ is a union of sets
\beqa
D_{0,i,j}:=\{\, p_1,\ldots, p_n \in \De_{\mfh}\,|\, \nabla\Si(p_i)=\nabla\Si(p_j)\,\}, \quad 1\leq i<j\leq n.
\eeqa
Suppose, by contradiction, that  $D_{0,1,2}$ has non-zero Lebesgue measure. Then, for some $v\in \real^d$, the set
\beqa
D_{0,v}:=\{\, p_1,\ldots, p_n \in \De_{\mfh}\,|\, \nabla\Si(p_1)=v\,\}
\eeqa 
also has non-zero Lebesgue measure. Otherwise computing the measure of $D_{0,1,2}$ as an iterated integral
of its characteristic function we would get zero. Clearly, we have
\beqa
D_{0,v}\subset \{\, p_1\in \De_{\mfh}\,|\, D^2\Si(p_1)=0\}\times \{\, p_2,\ldots, p_n\in \De_{\mfh}\,\},
\eeqa
hence $\mcT= \{\, p\in \De_{\mfh}\,|\, D^2\Si(p)=0\,\}$ has non-zero Lebesgue measure, which contradicts property
(c) from Definition~\ref{mass-shell-definition}. \qed

\section{Shape of the spectrum} \label{shape-spectrum}
\setcounter{equation}{0}

In this section we prove three results concerning the shape of $\Sp\, U$. In the following, $v_{\la}$ is the Lieb-Robinson velocity defined in Theorem~\ref{Lieb-Robinson-one}.

Proposition~\ref{gradient-of-mass-shell},
whose proof uses the scattering theory developed in the previous section, says that the velocity of a particle computed as
$\nabla\Si(p)$ cannot be larger that the Lieb-Robinson velocity $v_{\la}$. This shows that $v_{\la}$ plays a similar
role in lattice systems as the velocity of light in relativistic theories. A different proof of this result, adopting methods from the proof of Proposition 2.1 of
\cite{BF82}, appeared in \cite{Sch83}.  In~\cite{Radin:1978tr}, $v_{\la}$ was also shown to be a propagation bound for `signals'.

Proposition~\ref{additivity-of-spectrum}
says that $\Sp\, U$ is a semi-group in $\real\times \whGa$ i.e. if $q_1, q_2\in  \Sp\, U$ then also $q_1+q_2\in \Sp\, U$.
A special case of this result actually follows from Theorem~\ref{Haag-Ruelle-Fock} and Lemma~\ref{single-particle-density-two}:
if $q_1,\ldots, q_n\in \mfh$ then $q_1+\cdots+q_n\in \Sp\, U$ (at least if $\mcX=\mcT=\emptyset$). However, the general proof below does not use scattering
theory but follows directly from the clustering property (Theorem~\ref{clustering}) and the energy-momentum transfer 
relation~(\ref{EM-transfer-relation}).
This argument follows closely the proof from relativistic QFT, see Theorem~5.4.1 of \cite{Ha}.

Proposition~\ref{Prop: TD limit of gaps} relates the spectrum of
the finite volume Hamiltonians $H_{\La}$ to the spectrum of the Hamiltonian $H$ in the thermodynamic limit. More precisely,
$H$ is the GNS Hamiltonian 
of the ground state which is the limit of the ground states of $H_{\La}$. Our result says that gaps cannot close in the thermodynamic limit. This is of important practical interest, since quantum spin systems are usually defined in finite volume where explicit spectral estimates can be obtained. And indeed, it will be
useful in the next section where we discuss concrete examples,
to which the scattering theory of the previous section can be applied.

We start with the following lemma from \cite{Ro76}, which we need for the proof 
of Proposition~\ref{gradient-of-mass-shell}. It says that quantum spin systems satisfying the Lieb-Robinson bounds are asymptotically abelian  
w.r.t. space-time translations in the cone $\{|x|>v_{\la}t \}$ (cf. \cite[Satz II.7]{Sch83}).
\begin{lemma}\label{cor:AA}
Under the assumptions of Theorem~\ref{Lieb-Robinson-one}
and for any $0<\epsilon<v_\lambda^{-1}$, we have
\begin{equation}
\Vert [\tau_t \circ \tau_x(A),B] \Vert \leq C(A,B,\lambda) \e^{-\lambda (1-\epsilon v_\lambda) \vert x \vert} \label{asymptotic-abelianess}
\end{equation}
in the cone $\caC_\epsilon:=\{\, (t,x)\,|\, \vert t\vert\leq \epsilon\vert x\vert\}$, where
\begin{equation}
C(A,B,\lambda) = 2 \|A\| \|B\|C_\lambda^{-1}\Vert F \Vert\min\{\vert\Lambda_1\vert,\vert\Lambda_2\vert\}\e^{\lambda\mathrm{diam}(\Lambda_1\cup\Lambda_2)}.
\end{equation}
\end{lemma}
\proof By Theorem~\ref{Lieb-Robinson-one} and the $\tau_x$ invariance of the norm, 
\begin{align}
\Vert [\tau_t \circ \tau_x(A),B] \Vert 
&\leq \frac{2 \|A\| \|B\|}{C_\lambda} \e^{2\Vert\Phi\Vert_\lambda C_\lambda\vert t \vert} \e^{-\lambda\vert x \vert }\sum_{w\in\Lambda_1}\sum_{z\in\Lambda_2} \e^{\lambda\vert z-w \vert}F(\vert z-(w-x)\vert) \non\\ 
&\leq \frac{2 \|A\| \|B\|}{C_\lambda}\Vert F \Vert\min\{\vert\Lambda_1\vert,\vert\Lambda_2\vert\}\e^{\lambda\mathrm{diam}(\Lambda_1\cup\Lambda_2)}\cdot \e^{-\lambda\left(\vert x\vert -v_\lambda \vert t\vert\right)}
\end{align}
and the lemma follows from the condition $\vert t\vert\leq \epsilon\vert x\vert$.
\qed

\bep\label{gradient-of-mass-shell} Under the standing assumptions of Section~\ref{scattering-section}, we have the following: 
$|\nabla \Si(p)|\leq v_{\la}$
for any $p\in \De_\mfh$.
\eep
\proof We prove this by contradiction. Suppose there is  $p_0\in \De_\mfh$ s.t. $|\nabla \Si(p_0)| > v_{\la}$. Then, by smoothness
of $\Si$, there is a neighbourhood $O_{p_0}$ of $p_0$, that is compactly contained in $\De_\mfh$,
and $0<\epsilon<v_\lambda^{-1}$
s.t. $|\nabla \Si(p)| \geq  \epsilon^{-1}$ for all $p\in  O_{p_0}$. We consider
the subset of the mass shell
\beqa
\mfh_{O_{p_0}}:=\{\,(\Si(p), p)\, |\, p\in O_{p_0}\,\},
\eeqa
which is open in $\mfh$ because the dependence $\De_{\mfh}\ni p \mapsto (\Si(p), p)\in \mfh$ is diffeomorphic and bounded since $O_{p_0}$ is away from the boundary of $\De_\mfh$ \footnote{A priori $\Si(p)$ may tend to infinity when $p$ approaches the boundary of the open set $\De_{\mfh}$. Once proven, Proposition~\ref{gradient-of-mass-shell} excludes such behaviour, however.}.
To arrive at a contradiction we will show that the projection $P(\mfh_{O_{p_0}})$ is zero.

By Lemma~\ref{single-particle-density-two},  vectors of the form
\beqa
B^*(g)\Om=B_t^*(g_t)\Om, \quad \supp\,\wh g\subset O_{p_0},
\eeqa
span a dense subspace in  $ P(\mfh_{O_{p_0}})\hil$.  Thus it is enough to show that all these vectors are zero.

Since $(B^*( g))^*\Om=0$,   we can write for any given $A'\in\mfa(\La')$, with $\La'$ arbitrary but finite,
\beqa
\lan\pi(A')\Om, B^*(g)\Om\ran=
\lan\Om,[\pi(A')^*, B_t^*(g_t)]\Om\ran
=\lan\Om, [\pi(A')^{*},B_t^*(\chi_{+,t}g_t)]\Om\ran+O(t^{-\infty}), \label{first-step-gradient}
\eeqa
where we applied Lemma~\ref{HR-auxiliary} (d). Recall that 
$\chi_{+,t}(x)=\chi_+(x/t)$ where $\chi_+\in C_0^{\infty}(\real^d)$ is supported in  a small neighbourhood $V(g)^{\de}$ of the velocity support $V(g)$. Clearly, we can 
assume that
\beqa
V(g)^{\de}\subset \{\, \nabla \Sigma(p)\,|\, p\in O_{p_0}\},
\eeqa
since $\wh g$ is supported inside of $O_{p_0}$. 

Recall that $B^*=\pi(A^*)$ for some $A^*\in \mfa_{\aloc}$. Thus 
we can find $A_r\in \mfa(\B^r)$ s.t. $\|A-A_r\|=O(r^{-\infty})$. Hence
\beqa
\lan\Om, [\pi(A')^{*},B_t^*(\chi_{+,t}g_t)]\Om\ran=
\lan\Om, [\pi(A')^{*},\pi(\tau_t(A_{r}^*))(\chi_{+,t}g_t)]\Om\ran
+O(r^{-\infty} t^d), \label{intermediate-gradient}
\eeqa
where we used that $\|g_t\|_1=O(t^d)$. Now we note that the sum
\beqa
\tau_t(A^*_{r})(\chi_{+,t}g_t)=(2\pi)^{-\fr{d}{2}}\sum_{x\in \Ga}\, \tau_{(t,x)}(A^*_r)\chi_{+}(x/t)g_t(x)
\eeqa
extends over $x$ in the set $X_t:=\{\, x\in \Ga\,|\,   |x/t|\geq \epsilon^{-1}\,\}$. Clearly  $\{\,(t,x)\,|\,x\in X_t\}$ is contained in the  cone  $\caC_\epsilon:=\{\, (t,x)\,|\, \vert t\vert\leq \epsilon\vert x\vert\}$. Thus, by Lemma~\ref{cor:AA},
we have
\beqa
|\lan\Om, [\pi(A')^{*},\pi(\tau_t(A_{r}^*))(\chi_{+,t}g_t)]\Om\ran|&\leq \|\wh g\|_1C(A',A_r^*,\lambda)\sum_{x\in \Ga}\, 
\e^{-\lambda (1-\epsilon v_\lambda) \vert x \vert}|\chi_{+}(x/t)|\non\\
&\leq \|\wh g\|_1 C(A',A_r^*,\lambda) c_{\epsilon, v_{\lambda}} \e^{-\h\lambda (1-\epsilon v_\lambda)\epsilon^{-1} t}, \label{final-gradient}
\eeqa
where
\beqa
C(A',A_r^*,\lambda) = 2 \|A'\| \|A_r\|C_\lambda^{-1}\Vert F \Vert\min\{\vert\Lambda'\vert,\vert\B^r\vert\}\e^{\lambda\mathrm{diam}(\Lambda'\cup\B^r)},
\eeqa
and we note that $\|A_r\|\leq \|A\|+O(r^{-\infty})$. Thus, setting $r(t)=t^{\eps}$, $\eps>0$ sufficiently small, we obtain that
the r.h.s. of   (\ref{final-gradient}) and error terms in (\ref{intermediate-gradient}), (\ref{first-step-gradient}) tend to zero as $t\to \infty$.
Since the l.h.s. of (\ref{first-step-gradient}) is independent of $t$, we conclude that
\beqa
\lan\pi(A')\Om, B^*(g)\Om\ran=0.
\eeqa
As this is true for any $A'\in \mfa_{\loc}$ we obtain that $B^*(g)\Om=0$ and the proof is complete. \qed

\bep\label{additivity-of-spectrum} Under the assumptions of Theorem~\ref{clustering}, the following additivity property holds:
Suppose that $q_1, q_2\in \Sp\, U$. Then $q_1+q_2\in  \Sp\, U$ where $+$ denotes the group
operation in $\real\times \whGa$.
\eep
\proof Let $\De$ be a (bounded) neighbourhood of $q_1+q_2$. Let $\De_1, \De_2$ be (bounded) neighbourhoods
of $q_1,q_2$ s.t. $\ov{\De_1+\De_2}\subset \De$. We pick $A_1, A_2\in \mfa_{\aloc}$ s.t. $\Sp_{A_1}\tau\subset \De_1$
and $\Sp_{A_2}\tau\subset \De_2$. Then, by the energy-momentum transfer relation~(\ref{EM-transfer-relation}), we have
for any $x\in \Ga$
\beqa
\Psi_x:=\pi(\tau_{x}({A_1}))\pi(A_2)\Om\in P(\De)\hil, \label{additivity-zero}
\eeqa 
thus it suffices to show that some of these vectors are  non-zero.

By almost locality, we can find $A_{i,r}\in \mfa(\B^r)$ s.t. $(A_i-A_{i,r})=O(r^{-\infty})$ in norm for $i=1,2$. Thus we have
\beqa
\|\Psi_x\|=\|\Psi_{r,x}\|+O(r^{-\infty}), \textrm{ where } \Psi_{r,x}:=\pi(\tau_{x}({A_{1,r}}))\pi(A_{2,r})\Om, \label{additivity-one}
\eeqa
uniformly in $x$. By Theorem~\ref{clustering}, 
\beqa
\lim_{|x|\to \infty}\|\Psi_{r,x}\|=\|\pi(A_{1,r})\Om\|\|\pi(A_{2,r})\Om\|=\|\pi(A_{1})\Om\|\|\pi(A_{2})\Om\|+O(r^{-\infty}).
\label{additivity-two}
\eeqa
By choosing $r$ sufficiently large,  we can conclude from (\ref{additivity-one}), (\ref{additivity-two}), that (\ref{additivity-zero}) is non-zero for
sufficiently large $|x|$, provided that $\pi(A_{1})\Om, \pi(A_{2})\Om$ are non-zero.

To justify this last point, we pick $A_{0,i}\in \mfa_{\loc}$, $f_i\in S(\real\times \whGa)$, s.t. $\supp \wh f_i\subset \De_i$ and
set $A_i=\tau_{f_i}(A_{0,i})$. Then we have
\beqa
\pi(A_i)\Om=\int \wh f(E,p)dP(E,p)\pi(A_{0,i})\Om.
\eeqa 
Since with the spectral integrals above we can approximate  $P(\De_i)$   (and $P(\De_i)\neq 0$  since $q_i\in \Sp\, U$), by cyclicity of the ground state vector we can find $A_{0,i}$ s.t. $\pi(A_i)\Om$ are non-zero. \qed\\
As a preparation to the proof of Proposition~\ref{Prop: TD limit of gaps}, we discuss briefly finite-volume Hamiltonians and their ground states.

We first show how infinite volume ground states can be obtained from ground states of finite dimensional systems. We essentially follow the argument of~\cite[Prop. 5.3.25]{Bratteli:1997aa}. Let $\psi_\Lambda$ be a ground state vector of $H_\Lambda$ for the eigenvalue $E_\Lambda$, and define $\omega_\Lambda(A) := \langle\psi_\Lambda, A \psi_\Lambda\rangle$ for any $A\in\mfa(\Lambda)$, which can be extended to the whole algebra. Banach-Alaoglu's theorem then ensures that there are weak-* limit points of the net $\omega_\Lambda$ as $\La\nearrow \Ga$, that we shall generically call $\omega$. First, 
\begin{equation}
\langle\psi_\Lambda, A^{*} [H_\Lambda,A] \psi_\Lambda\rangle 
= \langle\psi_\Lambda, A^{*} H_\Lambda A \psi_\Lambda\rangle - \langle\psi_\Lambda, A^{*} A H_\Lambda \psi_\Lambda\rangle \geq 0,
\end{equation}
where we used in the first term that $H_\Lambda\geq E_\Lambda$ and in the second that $H_\Lambda \psi_\Lambda = E_\Lambda \psi_\Lambda$. Hence, $\omega_\Lambda$ is a ground state in the sense of~(\ref{GS}) since $-\mathrm{i}\delta_\Lambda = [H_\Lambda,\,\cdot\,]$ is the generator of the finite-volume dynamics. Furthermore, for any $A\in D(\delta)$ and any convergent (sub)sequence\footnote{A subsequence (rather than a subnet) can be found by exploiting the fact that the net converges in norm on any local algebra.} $\omega_{\Lambda_{n}}$, 
there is a sequence $A_n\in \mfa(\Lambda_n)$ such that $A_n\to A$ and $\delta_{\Lambda_n}(A_n) \to \delta(A)$. Hence,
\begin{equation}
-\mathrm{i} \omega(A^{*} \delta(A)) = \lim_{n\to\infty}-\mathrm{i} \omega_{\Lambda_{n}}(A_n^{*} \delta_{\Lambda_{n}}(A_n)) \geq 0,
\end{equation}
so that the limiting state $\omega$ is a ground state in the general sense.

 In this paper we are interested in models with  isolated mass shells (cf. Definition~\ref{mass-shell-definition}). 
 As we will see in the next section
in an example, in lattice systems
such mass shells may correspond to intervals in the
spectrum of the GNS Hamiltonian, isolated from the rest of the spectrum
by gaps\footnote{Note that this is not possible in relativistic theories.}.
Here again, such gaps can be deduced from the finite volume properties. The following proposition shows that a uniform spectral gap in finite volume cannot abruptly close in the GNS representation.
\begin{proposition}\label{Prop: TD limit of gaps}
Let $E\in\real$ be such that for some $\epsilon>0$, $\La_0\in \P_{\fin}(\Ga)$,
\begin{equation*}
\big(E_\Lambda+ (E-\epsilon,E+\epsilon)\big) \cap \mathrm{Sp}\left(H_\Lambda\right) = \emptyset
\end{equation*}
for all $\Lambda\supset\Lambda_0$. Then $E\notin\mathrm{Sp}(H)$, where $H$ is the GNS Hamiltonian of a weak-$*$ limit $\omega$ as above.
\end{proposition}
\proof
Let $(\mathcal{H},\pi,\Omega)$ be the GNS triplet of the limiting state $\omega$. Let $f\in S(\real)$ be such that $\widehat f$ is supported in $(E-\epsilon,E+\epsilon)$. To prove the claim, it suffices to show that for any $A,B\in\mfa_{\mathrm{loc}}$,
\begin{equation}\label{ClaimSpectrum}
\left\langle \pi(A)\Omega, \widehat f(H) \pi(B)\Omega\right\rangle = 0.
\end{equation}
First, we write
\begin{equation}
\tau_f(B) = \frac{1}{\sqrt{2\pi}} \int dt\, f(t) \tau_t(B), \label{tau-f}
\end{equation}
analogously to (\ref{time-smearing}). (We skip the superscript $(1)$ here as there is no danger of confusion). Next,
we use the invariance of $\Omega$ to get
\begin{equation}
\wh f(H)\pi(B)\Omega = \frac{1}{\sqrt{2\pi}} \int dt\, f(t) \e^{\i H t}\pi(B)\e^{-\i H t}\Omega = \frac{1}{\sqrt{2\pi}} \int dt\, f(t) \pi(\tau_t(B))\Omega = \pi(\tau_f(B))\Omega.
\end{equation}
We further note that 
\begin{equation}\label{Oo}
\left\langle \pi(A)\Omega, \widehat f(H) \pi(B)\Omega\right\rangle = \omega(A^*\tau_f(B)).
\end{equation}
Defining $\tau_f^\Lambda(B)$ analogously to (\ref{tau-f}), and recalling that $\Vert \tau^\Lambda_t(B) - \tau_t(B)\Vert\to0$ as $\Lambda\nearrow\Ga$, we obtain
\begin{equation}
\lim_{\Lambda\to\Ga} \left\Vert \tau_f^\Lambda(B) - \tau_f(B) \right\Vert = 0.
\end{equation}
This, and the weak$^*$ convergence $\omega_\Lambda\to\omega$ imply that for any $\delta>0$, there is $\Lambda_0$ such that for all $\Lambda\supset\Lambda_0$,
\beqa
&\left\Vert A^*\tau_f(B) - A^*\tau^\Lambda_f(B) \right\Vert < \delta / 2, \non\\
&\left\vert\omega(A^*\tau^\Lambda_f(B)) - \omega_\Lambda (A^*\tau_f^\Lambda(B))\right\vert < \delta / 2.
\eeqa
Hence, by~(\ref{Oo}),
\begin{equation}
\left\vert\left\langle \pi(A)\Omega, \widehat f(H) \pi(B)\Omega\right\rangle - \omega_\Lambda(A^*\tau^\Lambda_f(B))\right\vert <\delta.
\end{equation}
But using the fact that $H_\Lambda \psi_\Lambda = E_\Lambda \psi_\Lambda$, 
\begin{equation}
\omega_\Lambda(A^*\tau^\Lambda_f(B)) = \frac{1}{\sqrt{2\pi}} \int dt\, f(t) \omega_\Lambda\left(A^*\e^{\i t (H_\Lambda-E_\Lambda)}B\right) = \omega_\Lambda\left(A^*\wh f (H_\Lambda-E_\Lambda) B \right) = 0,
\end{equation}
since $\wh f$ is supported in $(E-\epsilon,E+\epsilon)$ while, by assumption, this interval is outside the spectrum of $H_\Lambda-E_\Lambda$. This proves~(\ref{ClaimSpectrum}) and therefore the proposition. \qed

\section{Examples}  \label{examples-section}
\setcounter{equation}{0} 
In this section we show that the Ising model in a strong magnetic field $\epsilon^{-1}$ (given by (\ref{Ising}) below) 
satisfies the standing assumptions of Section~\ref{scattering-section} and thus admits scattering theory
developed in this paper. This model belongs to a larger family of systems which are perturbations of `classical' models  studied e.g. in~\cite{Kirkwood:1983tu,Albanese:1989go,Kennedy:1992ta,Datta:2002ff}. Here, we shall follow mostly~\cite{Yarotsky:2005dk} and the closely related~\cite{Yarotsky:2004ga}, and apply concrete results of~\cite{Yarotsky:2004vp} and~\cite{Pokorny:1993wt}.
We conjecture that many other models of this family  fit into the framework of Section~\ref{scattering-section}, but complete proofs are missing. 
We will come back to this point in the later part of this section. 

The Ising model (with open boundary conditions) is given by the following family of finite volume Hamiltonians
\begin{equation}\label{Ising}
H^I_{\Lambda,\epsilon} := -\frac{1}{2} \sum_{x\in\Lambda}\left(\sigma_x^{(3)}-1\right) -\epsilon\sum_{(x,y) \in\caE(\Lambda)}\sigma_x^{(1)}\sigma_y^{(1)},
\end{equation}
where $\caE(\Lambda) := \{(x,y)\in\Lambda\times \Lambda \,|\,  \vert y-x \vert = 1\}$ is the set of undirected edges of the lattice. When $\epsilon = 0$, the Hamiltonian is diagonal in the tensor product basis of eigenvectors of $\sigma^{(3)}$, and $\mathrm{Sp}(H^I_{\Lambda,0}) = \{n\in\bbN \,|\, n\leq\vert\Lambda\vert\}$. The unique ground state is given by the vector $\otimes_{x\in\Lambda}\vert + \rangle$, with ground state energy $E^I_\Lambda = 0$ for all $\Lambda\in \P_{\fin}(\Ga)$. It is easy to see that this model satisfies the assumptions of Theorem~\ref{Lieb-Robinson-one} 
and thus the Lieb-Robinson bound holds. 

In the course of our analysis in this section we will also need a version of the Ising model with periodic boundary conditions:
for any $L\in \nat$ we write $[-L,L]:=\{-L, -L+1,\ldots L\}$ for an interval and $[-L, L]^{\times d}$ for a hypercube of side $2L$
centred at zero. The finite volume Hamiltonian
of the hypercube $\La=[-L, L]^{\times d}$ is given by 
\begin{equation}\label{Ising-periodic}
\ti{H}^I_{\Lambda,\epsilon} :=H^I_{\Lambda,\epsilon}-\epsilon\sum_{  (x,y) \in \mathcal{F} ( \pa\Lambda) }\sigma_x^{(1)}\sigma_y^{(1)},
\end{equation}
where the boundary $\pa\La$ consists of all $x\in\La$ which have nearest neighbours outside of $\La$, 
$\mathcal{F} ( \pa\Lambda) := \{(x,y)\in\pa\Lambda\times \pa\Lambda\,|\,  y-x \in \{\pm 2Le_1,\ldots, \pm 2Le_{d}\, \}\, \}$ and
$\{e_i\}_{i=1,\ldots, d}$ are primitive vectors of the lattice.

We note that due to the non-local character of the boundary terms, $\ti{H}^I_{\Lambda,\epsilon}$  does not satisfy the assumptions of Theorem~\ref{Lieb-Robinson-one}, and we cannot conclude the existence of its
global dynamics $\ti\tau_{\veps,t}$ directly from Corollary~\ref{global-dynamics}. Nevertheless, we have:
\bel\label{Boundary-independence} The global dynamics $\tilde\tau_{\veps,t}$ of the Ising model (\ref{Ising-periodic}) with periodic boundary conditions 
exists and coincides with the
global dynamics $\tau_{\veps,t}$ of the Ising model (\ref{Ising}) with open
boundary conditions. Hence, also the corresponding families of infinite volume ground states
coincide.
\eel
\proof For any bounded $X$ and $A\in\mfa(X)$, let $\Lambda_n$ be the smallest hypercube such that $\{x\in\bbZ^d \,|\,\dist(x,X)\leq n\}\subset\Lambda_n$. Note that by locality it follows that $\delta_{\Lambda_1}(A) = \tilde{\delta}_{\Lambda_1}(A)$, where $\delta_{\Lambda_1}(A) := \i [H^{I}_{\Lambda_1,\epsilon}, A]$ and $\tilde{\delta}$ is defined similarly. Note that by construction $\delta_{\Lambda_1}(\mfa(X)) \subset \mfa(\Lambda_1)$ and the same is true for $\tilde{\delta}_{\Lambda_n}$.

Repeating the argument above it follows that $\delta^m_{\Lambda_n}(A) = \tilde{\delta}^{m}_{\Lambda_n}(A)$ for all $m\leq n$ and $A \in \mfa(X)$. Hence,
\begin{equation}
\tau^{\Lambda_n}_{\epsilon,t}(A) - \tilde\tau^{\Lambda_n}_{\epsilon,t}(A) = \frac{t^n}{n!}\sum_{m>0}\frac{t^m n!}{(m+n)!}\left(\delta_{\Lambda_n}^{m+n} - \tilde\delta_{\Lambda_n}^{m+n}\right)(A).
\end{equation}
Now,
\begin{equation}
\delta^N_{\Lambda_n}(A) = \sum_{x_N,\ldots, x_0}[\Phi(x_N,x_{N-1}),[\Phi(x_{N-1},x_{N-2}),\cdots,[\Phi(x_1,x_{0}),A]\cdots]],
\end{equation}
where the sum is over sequences such that $x_0\in X$ and $\vert x_i- x_{i-1} \vert \leq 1$ for $1\leq i\leq N$, and $\Phi(x_i,x_{j})$ is either of the interaction terms of~(\ref{Ising}) and $\Phi(x_i,x_{i})=\Phi(x_i)$ is understood. Since  $\Vert\Phi(x_i,x_{j})\Vert \leq \kappa$ for $\kappa = \max(1, \epsilon)$, it follows that 
\begin{equation}
\Vert \delta_{\Lambda_n}^{m+n}(A) \Vert \leq (4d\kappa)^{m+n}\Vert A \Vert\vert X \vert,
\end{equation}
and similarly for $\tilde\delta_{\Lambda_n}^{m+n}(A)$, with the exception that the paths can ``wrap around'' the hypercube.  All in all,
\begin{equation}
\Vert \tau^{\Lambda_n}_{\epsilon,t}(A) - \tilde\tau^{\Lambda_n}_{\epsilon,t}(A)\Vert 
\leq 2\Vert A \Vert\vert X \vert \frac{(4d\ka)^n\vert t\vert ^n }{n!}\e^{4d\ka\vert t\vert}
\end{equation}
where we used that $n! / (n+m)! \leq 1/m!$. 

 By Corollary~\ref{global-dynamics} we know that $\tau_{\epsilon,t}^{\Lambda_n}(A)$ converges to $\tau_{\epsilon,t}(A)$. With the above estimate it then follows that $\tilde{\tau}_{\epsilon,t}^{\Lambda_n}(A)$ also converges to 
$\tau_{\epsilon,t}(A)$ for all $t$ in a compact interval. By the group property we conclude that the dynamics of the Ising model with periodic and open boundary conditions coincide in the thermodynamic limit.

It follows from the algebraic characterization of a ground state that also the sets of infinite volume ground states are equal. \qed

In order to make this section more self-contained, we recall some known facts in the following three theorems
before stating the main result of this section in Proposition~\ref{examples-proposition}.
\begin{theoreme}[\!\!\!\cite{Yarotsky:2004vp,Matsui:1990vg}]\label{examples_knownThms_I} 
Let $H^I_{\Lambda,\epsilon}$ be the Ising Hamiltonian. There is $\epsilon_0>0$ such that for any $0<\epsilon<\epsilon_0$,
\begin{enumerate}
\renewcommand*\labelenumi{(\roman{enumi})}
\item $H^I_{\Lambda,\epsilon}$ has a unique ground state $\psi^I_{\Lambda,\epsilon}$ with ground state energy $E^I_{\Lambda,\epsilon}$,
\item there is a constant $\gamma_\epsilon>0$ such that $H^I_{\Lambda,\epsilon} \geq  (E^I_{\Lambda,\epsilon} + \gamma_\epsilon) \left(1 - \vert \psi^I_{\Lambda,\epsilon}\rangle\langle \psi^I_{\Lambda,\epsilon}\vert\right)$, uniformly in $\Lambda$,
\item there is a positive constant $c$ such that
\begin{equation}\label{BandsYarotski}
\mathrm{Sp}(H^I_{\Lambda,\epsilon}-E^I_{\Lambda,\epsilon}) \subset\bigcup_{n\in \mathrm{Sp}(H^I_{\Lambda,0})}\left\{n_\epsilon\,|\, \vert n_\epsilon - n\vert\leq c n \epsilon \right\},
\end{equation}
\item there is a unique translation-invariant ground state $\omega_{\epsilon}$ and $\omega_{\epsilon}(A) = \lim_{\Lambda\nearrow\Gamma}\langle \psi^I_{\Lambda,\epsilon},A \psi^I_{\Lambda,\epsilon}\rangle$ for all $A\in\mfa_{\mathrm{loc}}$.
Moreover, $\om_{\veps}$ is pure.
\end{enumerate}
\end{theoreme}
(i-iii) are Theorem~1 of~\cite{Yarotsky:2004vp}. Theorem~2 there establishes the convergence of finite volume ground states, $\bar\omega_{\epsilon} = \lim_{\Lambda\nearrow\Gamma}\langle \psi^I_{\Lambda,\epsilon},\cdot\, \psi^I_{\Lambda,\epsilon}\rangle$. We further claim that $\bar\omega_{\epsilon}$ is translation invariant. Indeed, the limiting state is unique and independent of the choice of increasing and absorbing sequence $\Lambda_n\nearrow\Gamma$. Now, for any such sequence and any $x\in\bbZ^d$, the sequence $\Lambda_n-x$ also converges to $\Gamma$ so that, making use of translation invariance of the interaction,
\begin{equation}
\bar\omega_{\epsilon}(\tau_x(A)) = \lim_{n\to\infty}\langle \psi^I_{\Lambda_n,\epsilon},\tau_x(A) \psi^I_{\Lambda_n,\epsilon}\rangle =  \lim_{n\to\infty}\langle \psi^I_{\Lambda_n-x,\epsilon},A \psi^I_{\Lambda_n-x,\epsilon}\rangle = \bar\omega_{\epsilon}(A),
\end{equation}
for any local $A$, which proves the claim. This, and the general uniqueness and purity of the translation-invariant ground state (Theorem~3 and Remark 4.8 of~\cite{Matsui:1990vg}) implies that $\bar\omega_{\epsilon} = \omega_{\epsilon}$, hence we obtain (iv). The GNS triple of 
$\omega_{\epsilon}$ will be denoted by $(\pi_{\veps}, \hil_{\epsilon},
\Omeps)$ and the unitary action of space-time translations
by $U_{\veps}$. The GNS Hamiltonian of the Ising model will be called
$H_{\epsilon} $

The notion of a `one-particle subspace' of~\cite{Yarotsky:2004vp} is defined as a subspace $\caH_1\subset \hil_{\veps}$, that is invariant under space-time translations~(\ref{GNS SpacetimeTranslations}) and such that there is a unitary operator $V:\caH_1\to L^2(\widehat{\Gamma})$ satisfying:
\begin{align}
	(V U_\epsilon(x) V^* \widehat{f}\,) (p) &= \e^{\i px}\widehat{f}(p), \label{m_spatial}\\
(V U_\epsilon(t) V^* \widehat{f}\,) (p) &= \e^{\i m_{\veps}(p) t}\widehat{f}(p), \label{m_MassShell}
\end{align}
where we identified $L^2(\widehat{\Gamma})$ with the image of the Fourier transform~(\ref{FT}). Note that this says that the group of unitaries $(t,x) \mapsto U_{\veps}(t,x)$ affords a spectral decomposition on $L^2(\widehat{\Gamma})$ in which the space-time translation operators act as multiplication by $e^{\i (m_{\veps}(p) t + px)}$. Up to the sign of the momentum this corresponds to the characterization of a mass shell as a subset of the spectrum $\Sp\, U_{\veps}$ in Definition~\ref{mass-shell-definition}~(a) as we will see in the proof of Proposition~\ref{examples-proposition} below. 

From (\ref{BandsYarotski}), Proposition~\ref{Prop: TD limit of gaps} and the explicit form of $\mathrm{Sp}(H^I_{\Lambda,0})$ stated above,
we conclude that
\beqa
\Sp\, H_{\veps}\subset
\bigcup_{n\in \nat_0  }\left\{n_\epsilon \,|\, \vert n_\epsilon - n\vert\leq c n \epsilon \right\}. \label{GNS-limit-subsets}
\eeqa
(Similar inclusion is obtained in Theorem 3 of \cite{Yarotsky:2004vp} by
different methods).  Theorem 4 of \cite{Yarotsky:2004vp} ensures
that the $n=1$ component of union on the r.h.s. of (\ref{GNS-limit-subsets})  contains a  non-empty isolated subset of $\Sp\, H_{\veps}$ which gives rise to a one-particle subspace:
\begin{theoreme}[\!\!\!\cite{Yarotsky:2004vp}]\label{examples_knownThms_II} 
Let $H_{\epsilon}$ be the  GNS Hamiltonian of the Ising model, and let $0<\epsilon<\epsilon_0$ as above. Let 
\begin{equation}
P^1_{\epsilon} = -\frac{1}{2\pi\i}\int_{\caC}(H_{\epsilon}-z)^{-1}dz
\end{equation}
where $\caC$ is a circle centered at $1$ and of radius $c\epsilon < r < 1-2c\epsilon$, and let $\caG^1_{\epsilon}$ be its range. Then,
\begin{enumerate}
\renewcommand*\labelenumi{(\roman{enumi})}
\item $\caG^1_{\epsilon}$ is a one-particle subspace,
\item the function $p\mapsto m_{\veps}(p)$ of~(\ref{m_MassShell}) is real-analytic.
\end{enumerate}
\end{theoreme}

Finally, let us define
\begin{equation}\label{f beta}
f_{\Lambda,\epsilon}^\beta(t):= \tilde\omega_{\Lambda,\epsilon}^\beta\left(\sigma_0^{(1)} \e^{-t \tilde H^I_{\Lambda,\epsilon}}\sigma_0^{(1)}\e^{t \tilde H^I_{\Lambda,\epsilon}}\right),
\end{equation}
where $\tilde\omega_{\Lambda,\epsilon}^\beta$ is the Gibbs state over $\mfa(\Lambda)$ associated to the Hamiltonian $\tilde H^I_{\Lambda,\epsilon}$ with periodic boundary conditions. Theorem~1 of~\cite{Pokorny:1993wt} reads:
\begin{theoreme}[\!\!\!\cite{Pokorny:1993wt}]\label{thm:Pokorny}
There exists a Baire measure $\mu_\epsilon$ depending analytically on $\epsilon$, s.t. for $\veps$ small enough
\begin{equation}\label{Pokorny's convergence}
\lim_{\Lambda\nearrow\Ga}\lim_{\beta\to\infty} f_{\Lambda,\epsilon}^\beta(t) =  \int_0^\infty \e^{-t\lambda} d\mu_\epsilon(\lambda).
\end{equation}
Moreover, there are two reals $0<\m_{0,\epsilon}\leq \m_{1,\epsilon}$ (and the latter inequality is strict for $\veps>0$)  such that $\m_{1,\epsilon}-\m_{0,\epsilon} = O(\epsilon)$ 
and $\mu_\epsilon$ has no support in $(0,\m_{0,\epsilon})$, while it is absolutely continuous in $[\m_{0,\epsilon},\m_{1,\epsilon})$. The dependence $\veps\mapsto \m_{0,\veps}$ is real analytic.
\end{theoreme}

The following proposition is now an application of Theorems~\ref{examples_knownThms_I}--\ref{thm:Pokorny}
 to the present case.

\begin{proposition}\label{examples-proposition} 
There is $\epsilon_0>0$ such that for any $0<\epsilon<\epsilon_0$, the infinite volume Ising model~(\ref{Ising}) 
satisfies the standing assumptions of Section~\ref{scattering-section}. In particular, it
has a unique ground state and an isolated mass shell in the sense of Definition~\ref{mass-shell-definition} with $\De_{\mfh}=S_1^{d}$.
For any $d\geq 1$ this mass shell is pseudo-relativistic. For $d=1$ it is also regular. 
\end{proposition}
\begin{remark} In order to ensure that the mass shell is pseudo-relativistic, $\epsilon_0$
may depend on $d$. The question of regularity of the mass shell in the case $d>1$ is not settled.
\end{remark}
\proof
First of all, we claim that the uniqueness and purity of the translation invariant ground state of Theorem~\ref{examples_knownThms_I} (iv) implies that $\pi_{\epsilon}$ has a unique (up to phase) normalized ground state vector. Indeed, suppose that $\Omega'_{\epsilon}\in\caH_{\epsilon}$ is a unit vector such that $H_\epsilon\Omega'_{\epsilon} = 0$, while $\Omega'_{\epsilon}\neq \lambda \Omega_{\epsilon}$ for  $\lambda \in S_1$. Then, the vector state $\omega'_{\epsilon}$ associated to $\Omega'_{\epsilon}$ satisfies
\begin{equation}
-\i \omega_{\epsilon}'(A^*\delta_\epsilon(A)) = \left\langle \pi_{\epsilon}(A)\Omega'_{\epsilon}, [H_\epsilon, \pi_{\epsilon}(A)]\Omega'_{\epsilon}\right\rangle = \left\langle \pi_{\epsilon}(A)\Omega'_{\epsilon}, H_\epsilon \pi_{\epsilon}(A) \Omega'_{\epsilon}\right\rangle \geq 0,
\end{equation}
since $H_{\veps}$ is a positive operator on the GNS Hilbert space. Hence $\omega'_{\veps}$ is a ground state, and, by purity of $\om_{\veps}$, we have $\omega'_{\veps}\neq \om_{\veps}$ which is a contradiction. Furthermore, (ii) of Theorem~\ref{examples_knownThms_I} and Proposition~\ref{Prop: TD limit of gaps} ensure that $0$ is an isolated eigenvalue of the GNS Hamiltonian separated from the rest of the spectrum by a gap of at least $\gamma_\epsilon$.

Let $V$ be the unitary operator as in equations~(\ref{m_spatial}-\ref{m_MassShell}), corresponding to the one-particle subspace of Theorem~\ref{examples_knownThms_II}. With Lemma~\ref{lem:FT}, we obtain
\begin{equation}
(V U_{\epsilon}(t,x) V^* \widehat{f}\,) (p) = (V U_{\epsilon}(t,0) V^* V U_{\epsilon}(0,x)  V^*\widehat{f}\,) (p) = \e^{\i m_\epsilon(p) t + \i px}\widehat{f}(p).
\end{equation}
Since $V$ is unitary,
this shows that $\mathfrak{h}_\epsilon:=\{(m_\epsilon(-p),p)\,|\, p\in \widehat{\Gamma}\}\subset\mathrm{Sp}(U_{\epsilon})$. By construction, $\mathfrak{h}_\epsilon$ has non-zero spectral measure. In other words, $\mathfrak{h}_\epsilon$ fulfills conditions~(a), (b) of Definition~\ref{mass-shell-definition}.  $\mathfrak{h}_\epsilon$ is isolated because it arose from an isolated part of $\Sp\, H_{\veps}$,
see the discussion below (\ref{GNS-limit-subsets}). 

In order to check condition~(c), we identify the dispersion relation $\Sigma_\epsilon$ with the function $p\mapsto m_\epsilon(-p)$, and we will  ensure 
that this function is non-constant. Given that, (c) follows as explained below Definition~\ref{sp-subspace-def}. For $d=1$ (c) is
equivalent to regularity of the mass shell. For any $d\geq 1$, (\ref{GNS-limit-subsets}) implies that $\{0\}$ is isolated from the rest of the spectrum by a  gap of the size at least $1-c\epsilon$, while the energy component of any vector in $\mathfrak h_\epsilon - \mathfrak h_\epsilon$ is at most $2c\epsilon$, so that the mass shell is pseudo-relativistic for $\epsilon<(3c)^{-1}$.

The fact that the mass shell is non-constant follows from Theorem~\ref{thm:Pokorny} as follows. As always, a tilde $\tilde\cdot$ will refer to objects obtained using periodic boundary conditions. Since $\tilde\omega_{\Lambda,\epsilon}^\beta$ is a $(\tilde\tau^\Lambda_\epsilon,\beta)$-KMS state it satisfies the energy-entropy balance inequality
\begin{equation}
-\i \tilde\omega_{\Lambda,\epsilon}^\beta(A^*\tilde\delta_\epsilon(A)) \geq \frac{1}{\beta}\tilde\omega_{\Lambda,\epsilon}^\beta(A^*A) \ln \frac{\tilde\omega_{\Lambda,\epsilon}^\beta(A^*A)}{\tilde\omega_{\Lambda,\epsilon}^\beta(AA^*)},
\end{equation}
and hence, with~(\ref{GS}), the limit $\beta\to\infty$ yields a finite volume ground state $\langle \tilde\psi^I_{\Lambda,\epsilon},\cdot\, \tilde\psi^I_{\Lambda,\epsilon}\rangle$ of the periodic Hamiltonian, see ~\cite[Proposition~5.3.23]{Bratteli:1997aa} and its proof  for details. With this and the spectral decomposition $\tilde H^I_{\Lambda,\epsilon} = \int \lambda d\tilde P^I_{\Lambda,\epsilon}(\lambda)$, the $\beta\to\infty$ limit of~(\ref{f beta}) reads
\begin{equation} \lim_{\beta\to\infty} f_{\Lambda,\epsilon}^\beta(t) = \int_\bbR \e^{-t(\lambda-\tilde E^I_{\Lambda,\epsilon}) } \left\langle \tilde\psi^I_{\Lambda,\epsilon}, \sigma_0^{(1)} d\tilde P^I_{\Lambda,\epsilon}(\lambda) \sigma_0^{(1)} \tilde\psi^I_{\Lambda,\epsilon}\right\rangle,
\end{equation}
and by a change of variables,
\begin{equation}
\lim_{\beta\to\infty} f_{\Lambda,\epsilon}^\beta(t)  = \int_0^\infty \e^{-t \lambda}d\tilde\mu_{\Lambda,\epsilon}(\lambda),
\end{equation}
where $d\tilde\mu_{\Lambda,\epsilon}(\lambda) = \langle \tilde\psi^I_{\Lambda,\epsilon}, \sigma_0^{(1)} d\tilde Q^I_{\Lambda,\epsilon}(\lambda) \sigma_0^{(1)} \tilde\psi^I_{\Lambda,\epsilon}\rangle$ and $d\tilde Q^I_{\Lambda,\epsilon}(\lambda)$ is the spectral measure of $\tilde H^I_{\Lambda,\epsilon} - \tilde E^I_{\Lambda,\epsilon}$. Let us denote $\wh{g_t}(\lambda):= \Theta(\lambda) \exp(-\lambda t)\in L^1(\bbR)$, where $\Theta\in C^\infty(\bbR)$ is such that $\Theta(\lambda) = 1$ for all $\lambda \geq 0$ and $\Theta(\lambda) = 0$ if $\lambda \leq -1$. Since $g_t$ (the inverse Fourier transform of $\wh{g_t}$) decays faster than any polynomial, it is integrable. The right hand side above now reads
\begin{align}
\int_0^\infty \e^{-t \lambda}d\tilde\mu_{\Lambda,\epsilon}(\lambda) 
&= (2\pi)^{-\h}\int_0^\infty   \int_\bbR ds\, \e^{\i s \lambda} g_t(s) \langle \tilde\psi^I_{\Lambda,\epsilon}, \sigma_0^{(1)} d\tilde Q^I_{\Lambda,\epsilon}(\lambda) \sigma_0^{(1)} \tilde\psi^I_{\Lambda,\epsilon}\rangle \nonumber\\
&= (2\pi)^{-\h} \int_\bbR  ds\, g_t(s) \langle \tilde\psi^I_{\Lambda,\epsilon}, \sigma_0^{(1)} \tilde\tau^\Lambda_{\epsilon,s}(\sigma_0^{(1)}) \tilde\psi^I_{\Lambda,\epsilon}\rangle, \label{FVPeriodic}
\end{align}
where we used that $(\tilde H^I_{\Lambda,\epsilon} - \tilde E^I_{\Lambda,\epsilon})\tilde\psi^I_{\Lambda,\epsilon} = 0$. 
We can now consider the infinite volume limit of~(\ref{FVPeriodic}). As shown in Lemma~\ref{Boundary-independence}, the infinite volume dynamics,
the set of algebraic ground states and (by the uniqueness of the algebraic ground state) the GNS Hamiltonian are independent of the boundary
conditions in the Ising model. Thus, by dominated convergence,
the weak-* convergence of the ground states and the existence of the infinite volume dynamics $\tau_{\epsilon}$,
\begin{equation}
\lim_{\Lambda\nearrow\Gamma}\int_0^\infty \e^{-t \lambda}d\tilde\mu_{\Lambda,\epsilon}(\lambda) =  (2\pi)^{-\h} \int_\bbR  ds\, g_t(s) 
\omega_{\epsilon}\left(\sigma_0^{(1)} \tau_{\epsilon,s}(\sigma_0^{(1)})\right) = \int_0^\infty \e^{-t\lambda} d\nu_\epsilon(\la),
\end{equation}
where
\begin{equation}
d\nu_\epsilon(\Lambda) = \left\langle \pi_{\epsilon}(\sigma_0^{(1)})\Omega_\epsilon, d\tilde Q_{\epsilon}(\lambda) \pi_{\epsilon}(\sigma_0^{(1)})\Omega_\epsilon\right\rangle,
\end{equation}
and $d\tilde Q_{\epsilon}$ is the spectral measure of the GNS Hamiltonian. This shows that the moment generating functions of the measure $\mu_\epsilon$ of~(\ref{Pokorny's convergence}), and that of $\nu_\epsilon$ are equal in a neighbourhood of zero, and hence the measures are the same, see e.g.~\cite[Chapter~30]{Billingsley:1995aa}. In particular, $\nu_\epsilon$ is absolutely continuous on $[\m_{0,\epsilon},\m_{1,\epsilon})$, which implies that $d\tilde Q_{\epsilon}(\lambda)$ has an absolutely continuous part throughout that interval. It remains to identify that interval with the spectral band $n=1$ of~(\ref{GNS-limit-subsets}). At $\epsilon = 0$, $\sigma_0^{(1)}$ causes a `spin flip' at the origin and $\sigma_0^{(1)}\psi_{\Lambda,0}= \sigma_0^{(1)}\tilde \psi_{\Lambda,0}$ is an eigenstate with energy equal to $1$, so that $\tilde\mu_{\Lambda,0}$ is a Dirac measure at $\lambda = 1$ for all $\Lambda$ and so is the limit $\mu_0$. Since $\epsilon\mapsto\mu_\epsilon$ and $\epsilon\mapsto \m_{0,\epsilon}$ are  analytic,  $\epsilon\mapsto[\m_{0,\epsilon},\m_{1,\epsilon})$ is the band arising from the isolated point $\lambda = 1$. In particular, the interval $[\m_{0,\epsilon},\m_{1,\epsilon})$ is a subset of the $n=1$ interval given in~(\ref{GNS-limit-subsets}). Finally, its non-vanishing width and the analyticity of $p\mapsto m_\epsilon(p)$ imply that $\Sigma_\epsilon(p) = m_\epsilon(-p)$ is not constant. \qed \\ 
Altogether, this proves that the Ising model satisfies the standing assumptions of Section~\ref{scattering-section} for large enough transverse field, and hence the scattering theory can be constructed.

We also mention that the existence of an isolated mass shell is expected to hold in a large class of models.  In fact it is  proved in some sense for the strongly anisotropic antiferromagnetic spin-$1/2$ Heisenberg Hamiltonian
\begin{equation}
H^H_{\Lambda,\epsilon} = \frac{1}{2}\sum_{(x,y) \in\caE(\Lambda)}\left(\sigma_x^{(3)}\sigma_y^{(3)}+1\right) + \epsilon \left(\sigma_x^{(1)}\sigma_y^{(1)}+\sigma_x^{(2)}\sigma_y^{(2)}\right),
\end{equation}
whose version with periodic boundary conditions will be denoted by $\ti H^H_{\Lambda,\epsilon}$.
In particular, the convergence~(\ref{Pokorny's convergence}) and the properties of the measure obtained for the Ising Hamiltonian also hold for the Heisenberg model for $d\geq 2$~\cite{Pokorny:1993wt}. Moreover, in~\cite{Albanese:1989go}, it is  shown that if $n\in\bbN$ indexes the eigenvalues at $\epsilon = 0$, then for all $\Lambda$ large enough, the spectral projections $\tilde P^H_{\Lambda,\epsilon}(n)$ of $\tilde H^H_{\Lambda,\epsilon}$ depend analytically on $\epsilon$ for $\vert \epsilon \vert$ small enough. Furthermore, the projection $\tilde P^H_{\Lambda,\epsilon}(0)$ is two-dimensional, the two eigenvalues converge to each other as $\Lambda\nearrow\bbZ^d$ for fixed $\epsilon$, while they remain uniformly isolated from the rest of the spectrum. There is however one obstruction: the ground state of the model is degenerate. Hence the analysis above  would have to be adapted for the antiferromagnetic Heisenberg model. We do however believe that a similar program could be carried out to obtain a bona fide mass shell.

In the case of the isotropic ferromagnetic model, the works~\cite{Hepp:1972,GrafSchenker:1997} describe the asymptotic behavior of spin waves, which are the low-lying collective excitations (see~\cite{Correggi:2013tf} for a precise statement). However, the situation is very different since the spectrum is not gapped.


\section{Outlook}\label{future-work}
\setcounter{equation}{0}

In this paper we have developed  Haag-Ruelle scattering theory for gapped systems on a lattice, and have shown that it applies  to the Ising model in a strong transversal magnetic field.  Compared to the conventional quantum-mechanical scattering theory,  the HR approach has two merits: First, there is no  need to identify a `free' Hamiltonian, since the free dynamics enters via the dispersion relation $\Si$ in the wave packets~\eqref{wave-packet}. This is of advantage, because  in the context of lattice systems the free and interacting Hamiltonians could act in disjoint representations of the algebra of observables $\mfa$. Second, HR theory  treats all scattering channels on equal footing: Indeed, a generalization of the discussion from Section~\ref{scattering-section}  to the case of several mass shells in $\Sp\, U$ is straightforward.
These two features (among others) make the HR theory  particularly convenient in the case of infinite quantum systems\footnote{We note, however, that quantum-mechanical scattering theory can also be recast in the HR fashion \cite{Sandhas, BrenigHaag}.} and justify its further study in the context of spin systems
on a  lattice. We discuss several interesting future directions below. We will suggest many times that certain results from
relativistic QFT should be adapted to the lattice setting. However, we never mean that this is automatic. A pertinent obstacle is the breakdown of the Reeh-Schlieder theorem in the context of lattice systems.

\subsection{Non-triviality of the $S$-matrix and the LSZ reduction formulae}

Although Theorem~\ref{s-matrix} guarantees the existence and the uniqueness of the $S$-matrix, it does not automatically follow that it is non-trivial i.e. $S\neq I$. In other words, it is not \emph{a priori} clear if any scattering takes place at all. In local relativistic QFT non-triviality of the $S$-matrix was settled  in certain low-dimensional models \cite{OS76, Lechner2008, Ta14} and proven in an axiomatic setting to be a consequence of  anyonic statistics (see below) \cite{BrosMund:2012}.
 However,  it is an old open problem to exhibit a QFT with $S\neq I$ in physical space-time.
While in the framework of lattice models one can reasonably expect that models with non-trivial $S$-matrix are in abundance for any $d\geq 1$, rigorous results of this sort are not known to us. Clarifying this point, e.g. in the Ising model of Section~\ref{examples-section}, could help to identify general properties of infinite quantum systems  responsible for (non-)triviality of the $S$-matrix. A possible starting point of such an investigation, which  in fact was accomplished in Euclidean lattice field theories \cite{Ba92}, is to establish the LSZ reduction formulae~\cite{LSZ55}. That is, to 
express the $S$-matrix via ground state expectation values of time-ordered products of observables. In a concrete model, these  `Green's functions' should be tractable e.g. using perturbation theory. 

\subsection{Asymptotic completeness}

The conventional property of asymptotic completeness requires that all states in the Hilbert space have an interpretation in terms of particles i.e. belong to the ranges of the wave operators. This property was established in many-body quantum mechanics for particles with quadratic
dispersion relations \cite{Sigal:1987aa,Derezinski:1993aa}. If the latter assumption is dropped, the problem of complete particle interpretation is open beyond the two-body scattering.  It is therefore not a surprise that for lattice systems, whose basic excitations typically have non-quadratic dispersion relations, asymptotic completeness is rather poorly understood. An additional problem which arises in infinite systems is a possible presence of non-equivalent representations of $\mfa$ describing charged particles, so-called `charged sectors': A configuration of charged particles whose total charge is zero 
gives rise to a vector in the ground state Hilbert space $\hil$ which is not in the ranges of the wave operators of neutral particles $W^{\pm}$. Thus in the presence of charged sectors the conventional asymptotic completeness relation $\Ran\, W^{\pm}=\hil$ cannot hold as it stands. One way to save it is to construct wave operators of charged particles as we discuss below. Another possibility is to formulate a weaker notion of asymptotic completeness in the  ground state representation which is compatible with non-trivial superselection structure. In the context of local relativistic
QFT, where analogous problems arise, such a notion was formulated in  \cite{DG13} and verified using methods from many-body quantum mechanics (e.g.  propagation estimates). We are confident that the analysis from  \cite{DG13} could be adapted to the lattice setting and possibly pursued much further in this more tractable framework.  We recall in this context that in the case of the isotropic ferromagnetic Heisenberg model\footnote{This model is not gapped and thus outside of the scope of this paper. Nevertheless it admits a meaningful scattering theory which relies on  a specific form of its Hamiltonian.} a similar programme of transporting methods from quantum-mechanical scattering (propagation estimates, Mourre theory) was
carried  out in \cite{GrafSchenker:1997} and led to a proof of conventional asymptotic completeness of two-magnon scattering.

\subsection{Scattering of charged particles}
\label{charge-carrying-fields}

Charged particles are described by vector states in  representations of $\mfa$ which are not equivalent to the ground state representation. To create such particles from the ground state vector $\Om$, which is the first step of a construction of scattering states,
it is necessary to use charge-carrying fields\footnote{The actual mathematical formulation may involve the field algebra~\cite{DR:1990} or the field bundle~\cite{DHRII} depending on a situation.}. While such fields are not elements of $\mfa$, in massive relativistic theories they can be obtained from
observables by some variant of the DHR construction~\cite{DHRII, BF82}. 
 In general, charged fields have different properties than observables:
First, they may not be localizable in bounded regions, but only in spacelike `fattening' strings. Second, their commutation rules may not be governed by the Bose statistics - in physical space-time one obtains the familiar Bose/Fermi alternative while in lower dimensions more exotic braid group statistics corresponding to anyonic excitations is possible. 
In all these cases scattering theory is available in the abstract framework of local relativistic QFT \cite{BF82,FGR:1996}, but concrete examples
satisfying all the assumptions are scarce (see however~\cite{Plaschke:2013}).

In the setting of lattice systems the opposite situation seems to be the case: there are plenty of tractable models, containing charged \cite{Barata:1991} or even anyonic~\cite{Kitaev:2003ul} excitations, but a general theory of superselection sectors and scattering is not well developed.  First steps towards
such a general theory for  lattice systems satisfying the Lieb-Robinson bounds were taken over three decades ago in the Diplom thesis of Schmitz~\cite{Sch83}, but this direction of research seems to have been abandoned. We stress that nowadays there is strong physical motivation to revisit this subject:  anyonic excitations appear
in \emph{topologically ordered} systems, which are relevant for quantum computing~\cite{Wang:2010}.  Thus a model independent theory of 
superselection sectors and scattering for anyonic systems on a lattice could prove very useful.

\subsection{Scattering theory of anyons in Kitaev's toric code model}\label{Kitaev}

The simplest example of a lattice system exhibiting topological order is Kitaev's toric code model~\cite{Kitaev:2003ul}. This model is defined as
follows:
On finite $\Lambda\subset\bbZ^2$ that we imagine painted as a chessboard, let $\Lambda_B$, resp. $\Lambda_W$ be the subsets of black, respectively white squares, and define local Hamiltonians by
\begin{equation}
H^K_{\Lambda,0} = \sum_{X\in \Lambda_B} \frac{1}{2}\Big(1-\prod_{x\in X}\sigma_x^{(3)}\Big) + \sum_{Y\in\Lambda_W} \frac{1}{2}\Big(1-\prod_{y\in Y}\sigma_y^{(1)}\Big). \label{Kitaev-eq}
\end{equation}
The Hamiltonian is a sum of mutually commuting projections, and it has frustration-free ground states: for any $X\in \Lambda_B$ and $Y\in\Lambda_W$, the ground state projection $P^K_{\Lambda,0}$ is such that 
\beqa
(1-\prod_{x\in X}\sigma_x^{(3)}) P^K_{\Lambda,0} = (1-\prod_{y\in Y}\sigma_y^{(1)}) P^K_{\Lambda,0} = 0. 
\eeqa
Imposing periodic boundary conditions, this model can be seen as living on a torus, and in fact, it can be defined on tori of higher genera. Topological order manifests itself in the ground states by a degeneracy depending on the genus $g$ of the surface, in this particular example $4^g$, while the various ground states are locally indistinguishable, see~\cite{Bravyi:2011ea} for precise definitions. Excited states can be obtained by flipping a pair of either black squares or white squares. Such an excited square has a natural interpretation of an abelian anyon~\cite{Kitaev:2003ul}. Note that it is only possible to create \emph{pairs} of excitations, which can be interpreted as a form of charge conservation. To consider a single excitation, one has to go to the thermodynamic limit, where it is possible to move one excitation of the pair to infinity~\cite{Naaijkens:2011aa}. The anyonic nature of that `particle' is another manifestation of topological order.

The first step to  develop scattering theory for anyons in this model is to obtain charge-carrying fields  discussed in Subsection~\ref{charge-carrying-fields}.
This may be possible by adapting to the lattice setting results from \cite{Rehren:1992,HalvorsonMueger,MuegerTuset:2008}. The second step is to check 
if the mass shells of the anyons satisfy the technical conditions stated in Section~\ref{scattering-section}. In fact this turns out not to be the case: 
the excitations of the Kitaev model have flat `mass shells', thus there is no propagation. A possible solution is to proceed to perturbed
models of the form
\begin{equation}
H^K_{\Lambda,\epsilon} = H^K_{\Lambda,0} + \epsilon\sum_{X\subset\Lambda} \Phi(X),
\end{equation}
where  $\Phi\in\caB_\lambda$ for some $\la>0$. While the shape of the spectrum, in particular of mass shells, should be sensitive to such perturbations,
the superselection structure typically is not, so we can expect $H^K_{\Lambda,\epsilon}$ to describe propagating anyons. 
Since  $H^K_{\Lambda,0}$  belongs to a class of theories whose perturbation theory is well understood~\cite{Bravyi:2011ea, Michalakis:2013gh}, there are good chances to verify this claim proceeding similarly as in Section~\ref{examples-section}.  
In particular a band structure similar to (\ref{BandsYarotski}) is available. Given mass shells and charge-carrying fields of anyons, a construction of 
scattering states should be performed by adapting the relativistic results from \cite{FGR:1996} to the lattice setting.

\appendix

\section{Spectral theory of automorphism groups} 
\label{Spectral-appendix}
\setcounter{equation}{0}

\subsection{Groups of isometries}
 We give a brief account of concepts and results
in spectral analysis of automorphism groups which are used in
this work. For a more extensive introduction to this subject
we refer to \cite{Ar82} and \cite[Section~3.2.3]{Bratteli:1997aa}. 

Let $G$ be a locally compact abelian group.
Let $\wh G$ be its dual group, that is the set of all characters $\{\la\}$ (i.e. continuous group homomorphisms $G\ni g\mapsto \lan \la, g\ran\in S_1$).
Given $f\in L^1(G)$, we define the Fourier transform of $f$ as
\beqa
\wh f(\la)=\int_{G} \ov{\lan\la,g\ran}f(g)d\mu_G(g), \label{Fourier-one}
\eeqa
where $d\mu_G$ is the Haar measure of $G$. In general $\wh f$ is an element of $C_0(\wh G)$. In case $\wh f\in L^1(\wh G)$
we have the Fourier inversion formula
\beqa
 f(g)=\int_{\wh G} \lan \la,g\ran \wh f(\la )d\mu_{\wh G}(\la), \label{Fourier-two}
\eeqa
 provided that the  Haar measure of $\wh G$ is suitably normalized.

Let $G\ni g\mapsto T_g$ be a strongly continuous representation of $G$ in a group of isometries 
of a complex  Banach space $V$. Then, for any $f\in L^1(G)$, there is a unique bounded operator $T_f$ of $V$
defined by
\beqa
T_f v=\int_{G} (T_gv) f(g) d\mu_{G}(g), \quad v\in V,  \label{Bochner-integral}
\eeqa
where the r.h.s. is a Bochner integral. 
\bed\label{Arveson-definition} The Arveson spectrum of $T$ and the Arveson spectrum of $v$ with respect to $T$ are defined by
\beqa
\Sp\, T:=&\{\,\la\in \wh G \,|\, \forall O_{\la} \, \,\exists f\in L^1(G) \ \mathrm{ s.t. } \ \supp\,\wh f\subset O_{\la} \ \mathrm{ and } \ T_f\neq 0 \,  \},\\
\Sp_v\, T:=&\{\,\la\in \wh G \,|\, \forall O_{\la} \, \,\exists f\in L^1(G) \ \mathrm{ s.t. } \ \supp\,\wh f\subset O_{\la} \ \mathrm{ and } \ T_fv\neq 0 \,  \},
\eeqa
respectively, where $O_{\la}$ denotes a neighbourhood of $\la$. Also, for any closed $E\subset \wh G$ we define the spectral subspace
\beqa
M^T(E):=\ov{\{\, v\in V\,|\, \Sp_v\, T\subset E\,\}}, \label{Ar-subspace}
\eeqa
where the closure is taken in the weak topology.
\eed
We note that in the case of $G=\real^n$ the Arveson spectrum of $v$ w.r.t. $T$ can equivalently be defined 
as the support of the inverse Fourier transform of the distribution $g\mapsto T_g v$. The closure of the union
of such supports  (over all $v\in V$) coincides with $\Sp\,T$.  Coming back to general $G$, one can easily show that
\beqa
\Sp_{T_fv} T\subset \Sp_{v} T\cap\supp\,\wh f. \label{restriction}
\eeqa
This relation allows to construct elements of $V$ whose  Arveson spectrum is contained in a prescribed set.

\subsection{Groups of unitaries}\label{unitaries-subsection}

As a first example fitting into the above framework let us consider a strongly continuous abelian group of unitaries $G\ni g\mapsto U(g)$
acting on a Hilbert space $\hil$. In this case the SNAG theorem provides us with a spectral measure $dP$ on
$\wh G$ with values in projections on $\hil$, given by
\beqa
U(g)=\int_{G}\ov{\lan \la,g\ran}dP(\la). \label{spectral-measure}
\eeqa 
$\Sp\, U$ coincides here with the support of $dP$ and $\Sp_{\Psi}U$ for some $\Psi\in \hil$, with the
support of $\lan\Psi, dP(\,\cdot\,)\Psi\ran$. Moreover,
\beqa
M^U(E)=\Ran\, P(E).
\eeqa
Finally we remark that if $G=\real$ then, by the Stone theorem, $\Sp\, U$ is simply the spectrum of the infinitesimal generator
of $U$.

\subsection{Groups of automorphisms of $C^*$-algebras} \label{appendix-automorphisms-two}

As a second example let us consider a strongly continuous group of automorphisms $G\ni g\mapsto \tau_g$
of a $C^*$-algebra $\mfa$.  In the case $G=\real$ one can show that $\Sp\,\tau$ (resp.  $\Sp_{A}\,\tau$ for some $A\in\mfa$) 
 coincides with the operator-theoretic spectrum  of the infinitesimal generator $-\i\de$ of $\tau$ on the entire Banach space $\mfa$ 
(resp. on the weakly closed subspace spanned by  the orbit of $A$ under the action of $\tau$) \cite{Ev76}.
Coming back to the  general $G$, we obtain  from (\ref{restriction}):
\beqa
\Sp_{\tau_f(A)}\tau\subset \Sp_A\tau \cap \supp\,\wh f.
\eeqa

Now let $\pi: \mfa\to B(\hil)$ be a representation of  $\mfa$ in which $\tau$ is unitarily implemented
i.e. there exists a strongly continuous group of unitaries $G\ni g\mapsto U(g)$ s.t. 
\beqa
 \pi(\tau_g(A))=U(g)\pi(A)U(g)^*, \quad g\in G.
\eeqa
This group of unitaries defines a group of automorphisms $\tilde\tau_g(\,\cdot\,)=U(g)\,\cdot\, U(g)^*$
on $\mathcal{B}(\hil)$ and thus for any $B\in\mathcal{B}(\hil)$ we can define its Arveson spectrum\footnote{The representation $g\mapsto \tilde\tau_g$ is usually not strongly continuous, 
and (\ref{Bochner-integral}) can only be defined as a weak integral. By the Riesz theorem the weak integral
defines a unique element of $\mathcal{B}(\hil)$ and thus the Arveson spectrum of $\tilde\tau$ is well defined. The closure in (\ref{Ar-subspace}) should then be taken in the weak-$^*$ topology.} 
$\Sp_{B}\tilde\tau$ w.r.t $\tilde\tau$. It follows from Definition~\ref{Arveson-definition} that 
\beqa
 \Sp_{\pi(A)}\tilde\tau\subset \Sp_{A}\tau
\eeqa
and for faithful $\pi$ this inclusion is an equality. In the following theorem we summarize the relations between 
the spectra of $\tau$, $\tilde\tau$  and $U$. This is
 a special case of Theorem 3.5 of \cite{Ar82}:
\bet\label{Energy-momentum-transfer} Let $\De$ be a closed subset 
of $\wh G$ and $A\in \mfa$. Then 
\begin{equation}
\pi(A)P(\De)=P(\ov{\De+\Sp_{\pi(A)}\tilde\tau}) \pi(A)P(\De). \label{transfer-relation}
\end{equation}
If $\pi$ is faithful then $\Sp_{\pi(A)}\tilde\tau=\Sp_{A}\tau$.
\eet
To conclude this section, we point out an important difference between the 
spectral theory of groups and unitaries and of groups of automorphisms: in the
latter case no obvious counterpart of the spectral measure  (\ref{spectral-measure}) exists.
Therefore, the finer structure of the Arveson spectrum of automorphism groups (e.g. its division into pure-point, absolutely continuous and
singular continuous parts)  remains to a large extent unexplored. Some steps to close this
gap, with physical applications in mind, have been taken in \cite{Bu90, Dy10, He14}.

\subsection{Examples of groups and their Haar measures}
In relativistic quantum field theory we have $G=\real^{d+1}$, $\wh G=\real^{d+1}$,
the respective Haar measures are the Lebesgue measures with normalization factors $(2\pi)^{-(d+1)/2}$, and relations   (\ref{Fourier-one}), (\ref{Fourier-two})
give the usual Fourier transforms of functions on Minkowski space-time. The Arveson spectrum
$\Sp_A\tau$ is often called the \emph{energy-momentum transfer} of $A$ due to relation~(\ref{transfer-relation}).

In the case of lattice systems considered in the present paper, we consider the locally compact abelian group $G=\real\times \Ga$, where $\Ga:=\mathbb{Z}^d$ 
is the group of lattice translations. The  Haar measures of $\Ga$ and $G$ are fixed by  the relations
\beqa
\int f(x) d\mu_{\Ga}(x)&=(2\pi)^{-\fr{d}{2}}\sum_{x\in \Ga } f(x), \quad\quad\quad\quad\quad f\in L^1(\Ga), \label{Ga-measure}\\
\int f(t,x) d\mu_{G}(t,x)&=(2\pi)^{-\fr{d+1}{2} }\sum_{x\in \Ga }\int_{\real} dt \, f(t,x),\quad f\in L^1(G).
\eeqa

The dual group of $G$  is $\wh G=\real\times \wh \Ga$, where $\wh\Ga=S_1^d$ is the $d$-dimensional
torus. For $p\in S_1^d$ the corresponding characters are given by
\beqa
\lan p, x\ran_{\Ga}&=\e^{\i px},\quad\quad\quad\quad\quad  p\in \wh\Ga, \ \ x\in \Ga, \\
\lan (E, p), (t, x)\ran_{G}&=\e^{-\i Et+\i px}, \quad (E,p)\in \wh G, \  (t,x)\in G. \label{G-character}
\eeqa 
With definitions (\ref{Ga-measure})--(\ref{G-character}) relation~(\ref{Fourier-one}) reproduces our formulas
for the Fourier transform in (\ref{spatialFT}), (\ref{FT}).

To construct the Haar measure on $\wh G$ and $\wh\Ga$ we proceed as follows: Let us consider two distinct points $p_{\pm}\in S_1$
and open intervals $I_{\pm}$  identified with $S_1\backslash \{p_{\pm}\}$ via 
\beqa
I_{\pm}\ni p^1\mapsto \e^{\i p^1}\in S_1.
\eeqa
For example, we can choose $I_+=(0,2\pi)$, $I_-=(-\pi,\pi)$.
Now let $\vp_{\pm}\in C^{\infty}(S_1)$ be a smooth partition of unity on $S_1$  s.t.
$ \supp\,\vp_{\pm}\subset I_{\pm}$.  We can express the Haar measure on $S_1$ by the relation
\beqa
\int_{S_1} f(p^1) d\mu_{S_1}(p^1)=(2\pi)^{-\h}\sum_{i\in \{\pm\}}\int_{I_i}\,\vp_i(p^1) f(p^1) dp^1, \quad  f\in L^1(S_1).
\eeqa
By the uniqueness of the Haar measure (up to normalization) the r.h.s. above does not depend on the details of the construction.
Thus the Haar measure on the torus $\wh \Ga=S_1^d$ can be expressed as
\beqa
\int_{\wh\Ga} f(p) d\mu_{\wh \Ga}(p)= (2\pi)^{-\fr{d}{2}}\sum_{i_1,\ldots, i_d \in \{\pm\}}\int_{I_{i_1}\times\cdots \times I_{i_d}}\,
\vp_{i_1}(p^1)\ldots \vp_{i_d}(p^d)f(p^1,\ldots, p^d ) dp, \quad  f\in L^1(\wh\Ga), \label{measure-decomposition}
\eeqa
where $dp=dp^1\ldots dp^d$. This careful construction of $d\mu_{\wh\Ga }$ is important in the proof of Proposition~\ref{norm-corollary}. Otherwise, due to the close relation between  $d\mu_{\wh\Ga }$ and the Lebesgue measure we use 
the short-hand notation
\beqa
\int_{\wh\Ga} f(p) d\mu_{\wh \Ga}(p)=:(2\pi)^{-\fr{d}{2}}\int_{\wh\Ga} f(p) dp, \quad  f\in L^1(\wh \Ga). \label{short-hand-measure}
\eeqa
In case $f$ is supported on a subset of $\wh\Ga$ which is diffeomorphic with $\real^d$, (\ref{short-hand-measure}) makes sense also literally.
Using this short-hand notation, the Haar measure on $\wh G$ is given by
\beqa
\int_{\wh G} f(E,p) d\mu_{\wh G}(E,p)=(2\pi)^{-\fr{d+1}{2}}\int_{\real\times \wh\Ga} f(E,p) dEdp, \quad f\in L^1(\wh G). \label{dual-measure-4d}
\eeqa 
Given  expressions (\ref{short-hand-measure}), (\ref{dual-measure-4d}), relation~(\ref{Fourier-two}) reproduces formulas~(\ref{FT-inv}), (\ref{spatialFT-inv}) for the inverse Fourier transforms.

Finally, we note that partial derivatives of a function
$f\in C^{\infty}(\De)$, where $\De$ is an open subset of $\whGa$, 
are independent of a chosen parametrization of the torus, an observation which is implicitly used e.g. in Definition~\ref{mass-shell-definition}.

\section{The energy-momentum transfer lemma}\label{energy-momentum-proof}
\setcounter{equation}{0}
Here we present a proof of the energy-momentum transfer relation, equation~\eqref{EM-transfer-relation}, in the context of lattice systems.
This is a special case of Theorem 3.5 of~\cite{Ar82}, but for the benefit of the reader we here adapt an elementary argument given in the context of relativistic QFT in \cite{Du13}.
\begin{lemma}
	Let $A\in \mfa$ and $\tau$ be the group of space-time translation automorphisms, and $U$ its implementing group of unitaries as in Subsection~\ref{space-time-translations}.
Write $P$ for the associated spectral measure defined in (\ref{SNAG}). We then have the \emph{energy-momentum transfer relation}
\beqa
\pi(A)P(\De)=P(\ov{\De+\Sp_{A}\tau}) \pi(A)P(\De) \label{App-em-transfer}
\eeqa
for any  Borel subset $\De\subset \real\times \wh\Ga$.
\end{lemma}
\proof By countable additivity of the spectral measure, we can assume without loss of generality that $\De$ is bounded. 
Now for any $g\in S(\bbR\times \Gamma)$ let
\begin{equation}
P(g) :=  (2\pi)^{-\frac{d+1}{2}} \sum_{x\in\Gamma} \int_\bbR  dt\, U(t,x) g(t,x) 
\end{equation}
which has the following properties: Firstly,  $P(g) = P(K) P(g)$ for any closed $K\supset\mathrm{supp}(\widehat{g})$. Secondly,  $P(\Delta) = P(g)P(\Delta)$ for any function $g$ such that $\widehat g\upharpoonright_\Delta = 1$.

Now, let $f,g\in S(\bbR\times \Gamma)$. We have that
\begin{align}
 \pi(\tau_f(A))P(g) &= (2\pi)^{-(d+1)}\sum_{x,y\in\Gamma}\int_{\bbR^2}dtds\,  f(t,x) g(s,y) U(t,x)\pi(A)
U(t-s,x-y)^*  \non\\
&=(2\pi)^{-(d+1)}\sum_{z,y\in\Gamma}\int_{\bbR^2}drds\,  f(r+s,z+y) g(s,y) U(s,y) \pi(\tau_{(r,z)}(A)) \non\\
&=(2\pi)^{-\frac{d+1}{2}} \sum_{z\in\Gamma} \int_\bbR  dr \, P(h_{(r,z)}) \pi(\tau_{(r,z)}(A)),
\end{align}
where the function $h_{(r,z)}(s,y) = f(r+s,z+y) g(s,y)$ clearly satisfies
\begin{equation}
\mathrm{supp}(\widehat{h_{(r,z)}}) \subset \mathrm{supp}(\widehat{f}) + \mathrm{supp}(\widehat{g}).
\end{equation}
Hence
\begin{equation}\label{SmoothTransfer}
 \pi(\tau_f(A)) P(g) = P\big(\mathrm{supp}(\widehat{f}) + \mathrm{supp}(\widehat{g})\big)\pi(\tau_f(A))P(g).
\end{equation}

Let now $\varphi\in C^\infty(\bbR\times\widehat\Gamma)$ be bounded (as well as all its derivatives) and such that  $\varphi\upharpoonright_{\mathrm{Sp}_{A} \tau} = 1$ and $\varphi\upharpoonright_{(\mathrm{Sp}_{A} \tau)_\epsilon^c} = 0$, where for any $\Omega\subset\bbR\times\widehat\Gamma$ and $\epsilon>0$, we define $\Omega_\epsilon := \{x\in\bbR\times\widehat\Gamma : \mathrm{dist}(x,\Omega)\leq\epsilon\}$, and $\Omega_\epsilon^c$ is its complement. Such $\varphi$ can be obtained by convoluting the characteristic function of $(\mathrm{Sp}_{A} \tau)_{\epsilon/2}$ with a suitable
function from $C_0^\infty(\bbR\times\widehat\Gamma)$.
This yields a decomposition $\widehat f = \widehat f_1 + \widehat f_2 = \varphi \widehat f + (1-\varphi) \widehat f$, where both $f_1,f_2$ are Schwartz functions, and further
\begin{equation}
\tau_{f}(A) = \tau_{f_1}(A) + \tau_{f_2}(A).
\end{equation}
By definition of the Arveson spectrum, $\tau_{f_2}(A) = 0$.

Let $K$ be a compact set such that $K\cap  (\ov{\Delta + \mathrm{Sp}_{A} \tau}) = \emptyset$, and consider $g\in S(\bbR\times\Gamma)$ such that $\widehat g\upharpoonright_\Delta = 1$ and $\widehat g\upharpoonright_{\Delta_{\epsilon}^c} = 0$. Then,
\begin{align}
P(K)\pi(\tau_{f}(A))P(\Delta) &= P(K)\pi(\tau_{f_1}(A))P(g)P(\Delta) \non\\
&= P(K) P\big(\mathrm{supp}(\widehat{f}_1) + \mathrm{supp}(\widehat{g})\big)\pi(\tau_{f_1}(A))P(g) P(\Delta).
\end{align}
Since $K$ is disjoint from the compact set $\ov{\Delta + \mathrm{Sp}_{A} \tau}$, there is $\epsilon$ small enough such that
\begin{equation}
\mathrm{dist}\left(K,\mathrm{supp}(\widehat{f}_1) + \mathrm{supp}(\widehat{g})\right) 
\geq \mathrm{dist}\left(K,(\mathrm{Sp}_{A}\tau)_\epsilon + \Delta_\epsilon\right) >0,
\end{equation}
and hence $P(K) P\big(\mathrm{supp}(\widehat{f}_1) + \mathrm{supp}(\widehat{g})\big) = 0$. Therefore, $P(K)\pi(\tau_{f}(A))P(\Delta)=0$.

Finally, let $(f_n)_{n\in\bbN}$ be defined by
\begin{equation}
f_{n}(t,x) = \tilde f_n(t) \delta_0(x),
\end{equation}
where $\tilde f_n(t)=(2\pi)^{\frac{d+1}{2}} {(4\pi n^{-1})}^{-\h} \e^{-nt^2/4}$ is a sequence converging to the Dirac $\delta$ (multiplied by $(2\pi)^{\frac{d+1}{2}}$), and $\delta_0$ is Kronecker's delta on $\bbZ^d$. For any $\psi,\psi'\in\caH$,
\begin{equation}
\left\langle \psi', \pi(\tau_{f_n}(A)) \psi \right\rangle = (2\pi)^{-\frac{d+1}{2}}\sum_{x\in\Gamma}\int_{\bbR}dt\, \left\langle U(t,x)\psi', \pi(A) U(t,x) \psi \right\rangle f_n(t,x)\to \left\langle \psi', \pi(A) \psi \right\rangle,
\end{equation}
by the strong continuity of $(t,\,\cdot\,)\mapsto U(t,\,\cdot\,)$,  $U(0,0) = 1$ and the dominated convergence theorem. Hence, for $\Delta,K$ as above,
\begin{equation}
	P(K) \pi(A)P(\Delta) = \mathrm{w-}\lim_n P(K)\pi(\tau_{f_n}(A))P(\Delta) = 0.
\end{equation}
The restriction of $K$ being compact can be lifted by considering a countable partition of the complement of $\ov{\Delta + \mathrm{Sp}_{A} \tau}$ into bounded sets, so that the statement above extends to any $K$ such that $K\cap(\ov{\Delta + \mathrm{Sp}_{A} \tau}) = \emptyset$. It follows that
\begin{equation}
\pi(A)P(\Delta) \caH \subset \left(P\left((\bbR\times\widehat\Gamma)\setminus (\ov{\Delta + \mathrm{Sp}_{A} \tau})\right)\caH\right )^\perp = P\left(\ov{\Delta + \mathrm{Sp}_{A} \tau}\right)\caH,
\end{equation}
i.e. $\pi(A)P(\Delta) = P(\ov{\Delta + \mathrm{Sp}_{A} \tau})\pi(A)P(\Delta)$. Since the reverse inclusion is trivial, this proves the result.
\qed

\section{Proof of Theorem~\ref{smearing-theorem}} 
\label{Lieb-Robinson}
 
\setcounter{equation}{0} 
In this section we give a proof that space-time translations leave the algebra $\mfa_{\aloc}$ invariant, and that the same is true if we smear against Schwartz functions. To this end, we first recall Lemma~3.2 of~\cite{Bachmann:2011kw}:
\bel\label{Sven-lemma} If $\caH_1$ and $\caH_2$ are Hilbert spaces, and if $A\in\caB(\caH_1\otimes\caH_2)$ is such that
\begin{equation*}
\Vert [A,1\otimes B]\Vert<\epsilon\Vert B\Vert
\end{equation*}
for all $B\in\caB(\caH_2)$, then for any normal state $\omega$ on $\caB(\caH_2)$, the map $\Pi_\omega=\mathrm{id}\otimes \omega$ satisfies the bound
\begin{equation*}
\Vert \Pi_\omega(A) - A \Vert \leq 2\epsilon.
\end{equation*}
\eel
\nin The key step in our proof of Theorem~\ref{smearing-theorem} is the following consequence of the Lieb-Robinson bound:
\bep\label{cor:localization}
 For any $Y\in \P_{\fin}(\Gamma)$, $A\in \mfa(Y)$,  $\delta>0$ and $t\in\real$ there exists $A_{t,\delta}\in\mfa\left(Y^{(v_\lambda \vert t \vert +\delta)}\right)$ such that 
\begin{equation}
\left\Vert \tau_t(A) - A_{t,\delta} \right\Vert \leq C(A,\lambda) \e^{-\lambda \delta}, \label{cor:localization-formula}
\end{equation}
and $\la>0$ is as in Theorem~\ref{Lieb-Robinson-one}.
Moreover, the dependence $t\mapsto A_{t,\delta}$ is piecewise continuous w.r.t. the norm topology.
\eep
\proof In order to apply  Lemma~\ref{Sven-lemma} in the present case, let $\rho$ be an arbitrary state on $\mfa$. It is necessarily locally normal with density matrices $\rho_\Lambda$, that is $\rho(A) = \mathrm{Tr}(\rho_\Lambda A)$ for $A\in\mfa(\Lambda)$. For any finite $X\subset\Lambda\subset \Gamma$, we consider the projection $\Pi_{\Lambda,X}:\mfa(\Lambda) \to \mfa(X) \otimes 1_{\Lambda\setminus X} \simeq \mfa(X)$ onto the subalgebra $\mfa(X)$ given by $\Pi_{\Lambda,X}: = \Pi_{\rho_{\Lambda\setminus X}} = \mathrm{id_{\mfa(X)}}\otimes \rho_{\Lambda\setminus X}$. It satisfies the consistency condition $\Pi_{\Lambda,X}(A) = \Pi_{\Lambda',X}(A)$ whenever $A\in\mfa(\Lambda)$ with $X\subset\Lambda\subset\Lambda'\subset\Gamma$. We can therefore define $\Pi_X:\mfa \to \mfa(X)$ as follows: for any $A\in\mfa_\mathrm{loc}$, for any $\Lambda$ such that $A\in\mfa(\Lambda)$ we let $\Pi_X(A) = \Pi_{\Lambda,X}(A)$. Since, moreover, $\Vert \Pi_{\Lambda,X} (A) \Vert \leq \Vert A \Vert$, the densely defined linear map $\Pi_X$ can be extended to a bounded operator on the whole algebra $\mfa$ with the same bound.

Let now $A\in\mfa(Y)$, $t\in\bbR$, and $Y\subset \Lambda\subset\Lambda'$. Since $\tau_t^\Lambda(A) \in \mfa(\Lambda) \subset \mfa(\Lambda')$, 
we have for any $r>0$ such that $Y^r\subset\Lambda$, 
\begin{equation}
\left\Vert \Pi_{ Y^{r} }(\tau_t^\Lambda(A)) - \Pi_{ Y^{r} }(\tau_t^{\Lambda'}(A))\right\Vert
= \left\Vert \Pi_{ \Lambda',Y^{r} }\left(\tau_t^\Lambda(A) - \tau_t^{\Lambda'}(A)\right)\right\Vert 
\leq \left\Vert\tau_t^\Lambda(A)- \tau_t^{\Lambda'}(A)\right\Vert. \label{continuity-computation}
\end{equation}
Hence,
\begin{equation}
A'_{t,r} := \lim_{\Lambda\nearrow\Ga} \Pi_{ Y^{r} }(\tau_t^\Lambda(A)) \label{A-prime-x}
\end{equation}
exists and $A'_{t,r}\in\mfa(Y^r)$. Now we can write
\begin{equation}
\left\Vert \tau_t(A) - A'_{t,r} \right\Vert 
\leq \left\Vert \tau_t(A) - \tau_t^\Lambda(A) \right\Vert
+ \left\Vert \tau_t^\Lambda(A) - \Pi_{ Y^{r} }(\tau_t^\Lambda(A)) \right\Vert
+ \left\Vert \Pi_{ Y^{r} }(\tau_t^\Lambda(A)) - A'_{t,r} \right\Vert, \label{three-epsilon}
\end{equation}
keeping in mind that $r$ is restricted by $Y^r\subset\Lambda$. Let us now eliminate this restriction:  
The Lieb-Robinson bound (\ref{LRB}) yields
\begin{equation}
\left\Vert \left[\tau_t^\Lambda(A) , 1\otimes B\right]\right\Vert
\leq \frac{2 \|A\| \|B\|}{C_\lambda} \vert Y \vert \Vert F \Vert \e^{-\lambda (r-v_\lambda\vert t\vert)}
\end{equation}
for any $B\in\mfa(\Lambda\setminus Y^r)$. Therefore Lemma~\ref{Sven-lemma} gives 
\begin{equation}
\Vert \Pi_{Y^{r}} (\tau_t^\Lambda (A)) - \tau_t^\Lambda(A) \Vert \leq \frac{4 \|A\|}{C_\lambda} \vert Y \vert \Vert F \Vert \e^{-\lambda (r-v_\lambda\vert t\vert)},
\end{equation}
uniformly in $\Lambda$. We apply this bound to the second term on the r.h.s of  (\ref{three-epsilon}) and then take
the limit $\Lambda\nearrow \Ga$ which eliminates the remaining two terms and the restriction on $r$. Thus we are left with
\beqa
\left\Vert \tau_t(A) - A'_{t, r} \right\Vert \leq \frac{4 \|A\|}{C_\lambda} \vert Y \vert \Vert F \Vert\e^{-\lambda (r-v_\lambda\vert t\vert) },
\eeqa
where $r>0$ is arbitrary.  Now for any $\delta>0, t\in\real$  we set  $r$ equal to $r(\delta,t) = v_\lambda \vert t \vert +\delta$ 
and define $A_{\de,t}:=A'_{t,r(\delta,t)}$. This concludes the proof of (\ref{cor:localization-formula}).

The  piecewise continuity of $t\mapsto A_{\de,t}$, follows from  two observations: First,  $r\mapsto A'_{t,r}$, given by (\ref{A-prime-x}), is 
a step function for fixed $t$ (since $r\mapsto Y^r$ depends on  $r$ in a discrete manner). Second, $t\mapsto A'_{t,r}$ is continuous in 
norm for fixed $r$, which follows by replacing  $\tau_t^{\Lambda'}(A)$ with $\tau_{t'}^{\Lambda}(A)$  in  (\ref{continuity-computation}). \qed\\
Now we are ready to show that by smearing a local observable with a Schwartz class function we obtain an almost local observable.
\bel\label{almost-locality} Let $\La\in \P_{\fin}(\La)$, $A\in \mfa(\La)$ and $f\in S(\real\times \Ga)$.
Then $\tau_f(A)\in \mfa_{\aloc}$. 
\eel
\proof The fact that $\tau_f(A)\in \mfa$ follows from the strong continuity of
$\tau$ and properties of Bochner integrals, or from the approximation procedure below. Let us set
\beqa
A_{(r)}:=\int_{|t|\leq r} dt\sum_{x\in \Ga, |x|\leq r } \tau_{(t,x)}(A) f(t,x). \label{main-part-integral}
\eeqa 
Since $f\in S(\real\times \Ga)$, we have 
\beqa
\tau_f(A)=A_{(r)}+O(r^{-\infty}).
\eeqa
Now we fix some $\de>0$, depending on $r$, which will be specified later. Proposition~\ref{cor:localization} gives  observables 
$A_{t,\de}\in \mfa(\Lambda^{r_{\de}(t)})$, with $r_{\de}(t):= v_{\la}  t+ \delta$, such that
\begin{equation}
\left\Vert \tau_t(A) - A_{t,\delta} \right\Vert \leq C(A,\lambda) \e^{-\lambda \delta}. \label{approx-by-local}
\end{equation}
We note that $A_{t,\de}\in \mfa(\Lambda^{r_{\de}(r)})$, since $|t|\leq r$ in (\ref{main-part-integral}). Estimate~(\ref{approx-by-local}) 
gives
\beqa
A_{(r)}=\int_{|t|\leq r} dt\sum_{x\in \Ga, |x|\leq r }  \tau_{x}(A_{t,\de}) 
f(t,x)+O(\e^{-\la\de}), \label{upper-r}
\eeqa
where  we exploited the fact that the function $t\mapsto  A_{t,\de}$ is piecewise continuous in norm and
thus Bochner integrable (see  Proposition~\ref{cor:localization}). We denote by 
$A_r$ the integral on the r.h.s. of (\ref{upper-r})  and note that
$A_r\in \mfa(\La^{v_{\la}  r+ \delta+r})$.  We set  $\de=r$ and
obtain from (\ref{upper-r}) 
\beqa
A_{(r)}=A_r+O(r^{-\infty}).
\eeqa
Noting that $\La^{(v_{\la}  + 2)r }\subset \B^{2( v_{\la}  + 2) r}$ for $r\geq R_0$, where  $R_0$ depends on $\La$,
modifying $A_r$ for $r\leq  R_0$ and rescaling $r$ 
we conclude the proof. \qed\\
\bf Proof of Theorem~\ref{smearing-theorem}:\rm \   Let $A'\in \mfa_{\aloc}$ i.e. we have that there exist
$A'_{r}\in \mfa(\B^{r})$  s.t. 
$A'-A'_{r}=O(r^{-\infty})$. 

To show  (a), we note that for $(t,x)\in \real\times\Ga$
\beqa
\tau_{(t,x)}(A')-\tau_{(t,x)}(A'_{r})=O(r^{-\infty}).
\eeqa
Now by Proposition~\ref{cor:localization}, we obtain $A_r\in \mfa((\B^{r}+x)^{r})$ s.t.
\beqa
\tau_{(t,x)}(A'_{r})-A_r=O(r^{-\infty}).
\eeqa 
Using that $(\B^{r}+x)^{r}=\{x\}^{2r}\subset \B^{3r}$ for $r$ sufficiently large, we conclude part (a) as in the last step of the proof of Lemma~\ref{almost-locality}.

To show (b) we note that  for $f\in S(\real\times \Ga)$
we have
\beq
\|\tau_f(A')-\tau_f(A'_{r})\|\leq \|A'-A'_{r}\|\|f\|_1
\eeq
and therefore $\tau_f(A')-\tau_f(A'_{r})=O(r^{-\infty})$.
Now by Lemma~\ref{almost-locality} we obtain $A_r\in \mfa((\B^{r})^r)$ s.t.
\beqa
\tau_f(A'_{r})-A_r=O(r^{-\infty}).
\eeqa
Using that $(\B^{r})^r\subset \B^{2r}$,  we complete the proof in the case of $f\in S(\real\times \Ga)$.
The remaining cases follow similarly. \qed


\section{Notations}\label{Notations}
\setcounter{equation}{0}
For the convenience of the readers we here collect the main notations that we use and at the same time fix our normalisation constants for the Fourier transform. Note that we use a Minkowski sign convention, in that the time and spatial components carry an opposite sign in the Fourier transform.
\begin{enumerate}

\item For $f\in L^1(\real\times \Ga)$, $g\in L^1(\Ga)$, $ h\in L^1(\real)$  we define the following norms and Fourier transforms 
\beqa
 & \|f\|_1:=(2\pi)^{-\fr{(d+1)}{2}}\sum_{x\in \Ga}\int_\real dt\, |f(t,x)|,    \quad \wh f(E,p):=(2\pi)^{-\fr{(d+1)}{2}}\sum_{x\in \Ga}\int_\real dt\, \e^{\i  Et-\i px }f(t,x),  \label{spatialFT} \\
& \| g\|_1:= (2\pi)^{-\fr{d}{2}}\sum_{x\in \Ga} |g(x)|,    \quad \quad \quad \quad \quad \quad \wh g(p):= (2\pi)^{-\fr{d}{2}}\sum_{x\in \Ga} \e^{-\i px }g(x), \label{FT}\\
& \|h\|_1:=(2\pi)^{-\fr{1}{2}}\int_\real dt\, |h(t)|,    \quad\quad \quad \quad\quad \wh h(E):=(2\pi)^{-\fr{1}{2}}\int_\real dt\, \e^{\i  Et }h(t).  \label{timeFT} 
\eeqa

\item For $\wh f\in L^1(\real \times \wh \Ga)$, $\wh g\in L^1(\wh\Ga)$, $\wh h\in L^1(\real)$ we write
\beqa
&\|\wh f\|_1:=(2\pi)^{-\fr{(d+1)}{2}} \int_{\real\times \wh \Ga} dEdp\, |\wh f(E,p)|,  \quad  f(t,x)=(2\pi)^{-\fr{(d+1)}{2}} \int_{\real\times \wh \Ga} dEdp\, \e^{-\i  Et+\i px }\wh f(E,p), \label{FT-inv} \\
&\|\wh g\|_1:= (2\pi)^{-\fr{d}{2}} \int_{\wh \Ga} dp\, |\wh g(p)|, \quad \quad \quad \quad \quad\quad \ g(x)= (2\pi)^{-\fr{d}{2}} \int_{\wh \Ga} dp\, \e^{\i px }\wh g(p),  \label{spatialFT-inv} \\
&\|\wh h\|_1:=(2\pi)^{-\fr{1}{2}} \int_{\real} dE\, |\wh h(E)|,  \quad\quad \quad \quad \quad \ h(t)=(2\pi)^{-\fr{1}{2}} \int_{\real} dE\, \e^{-\i  Et }\wh h(E). \label{time-FT-inv} 
\eeqa

\item We write $\lan x\ran:=\sqrt{1+x^2}$.

\item We say that $g\in S(\Ga)$ if   $g(x)=O(\lan x \ran^{-\infty})$. Equivalently, $g\in S(\Ga)$ if $\wh g \in C^{\infty}(\wh\Ga)$.

\item We say that $f\in S(\real\times \Ga)$,  if  $f\in C^{\infty}(\real\times \Ga)$ and $\pa_t^n f(t,x)=O((\lan t \ran+\lan x\ran)^{-\infty})$
for any $n\in \nat$. Equivalently,  $f\in S(\real\times \Ga)$ if $\wh f\in C^{\infty}(\real\times \whGa)$ and $\pa_p^{\al}\pa_E^n\wh f(E,p)=O(\lan E \ran^{-\infty})$, uniformly in $p\in \whGa$, for any $n\in \nat$ and $\al\in \nat^d$.

\item Given $B\in \mathcal{B}(\hil)$ we set $B(t,x):=U(t,x)BU(t,x)^*, B(x):=B(0,x), B_t:=B(t,0)$. 

\item Given $B\in \mathcal{B}(\hil)$, $f\in L^1(\real\times  \Ga)$ and $g\in L^1(\Ga)$ we write
\beqa
B(f):=(2\pi)^{-\fr{(d+1)}{2}}\sum_{x\in \Ga} \int dt\,  B(t,x) f(t,x), \quad B(g):= (2\pi)^{-\fr{d}{2}}\sum_{x\in \Ga} B(x) g(x). \label{smearing}
\eeqa
Thus for $A\in\mfa$ we have $\pi(A)(f)=\pi(\tau_f(A))$, $\pi(A)(g)=\pi(\tau^{(d)}_g(A))$.

\item $O(x)$ denotes a term such that $\|O(x)\|\leq C|x|$
in some specified norms. $O(x^{-\infty})$ denotes a term which is $O(x^{-n})$ for any $n\in \nat$.

\end{enumerate}


\bibliography{Refs_Scattering}
\bibliographystyle{plain}   

\end{document}